\documentclass[11pt,a4wide]{article}
\pdfoutput=1
\usepackage{jheppub}

\usepackage{amsfonts}
\usepackage{amscd}
\usepackage{amssymb}
\usepackage{amsmath,bbm}
\usepackage{graphicx}
\usepackage{epsfig}
\usepackage{latexsym}
\usepackage{mathtools,verbatim}
\usepackage{hyperref}
\usepackage[vcentermath]{youngtab}
\usepackage{tikz}
\usepackage{float}
\usepackage[abs]{overpic}
\usepackage{relsize}
    
\def \be  {\begin{equation}}
\def \ee  {\end{equation}}
\def \bea {\begin{equation}\begin{aligned}}
\def \eea {\end{aligned}\end{equation}}
\def \ba  {\begin{eqnarray}}
\def \ea  {\end{eqnarray}}
\def \bb  {\begin{thebibliography}}
\def \eb  {\end{thebibliography}}
\def \lab #1 {\label{#1}}
\def\Re{{\rm Re}} \def\Im{{\rm Im}}

\newcommand\cL{\mathcal{L}}
\newcommand\cM{\mathcal{M}}
\newcommand\cN{\mathcal{N}}
\newcommand\cO{\mathcal{O}}

\newcommand\cS{\mathcal{S}}
\newcommand\cT{\mathcal{T}}

\newcommand\cW{\mathcal{W}}

\newcommand\al{\alpha}

\newcommand\q{\mathfrak{q}}
\newcommand\wf{w}

\newcommand\tf{t}

\newcommand\hf{h}

\newcommand\C{\mathbb{C}}
\newcommand\Z{\mathbb{Z}}
\newcommand\ep{\epsilon}

\newcommand\R{\mathbb{R}}

\newcommand\rd{\mathrm{d}}

\newcommand\ii{\mathrm{i}}

\newcommand\lb{\lambda}

\newcommand\la{\langle}
\newcommand\ra{\rangle}

\newcommand\tr{\mathrm{Tr}}
\newcommand{\da}[1]{\rd\nu_X(a_{#1})}

\newcommand\suf{\mathfrak{su}}
\newcommand\uf{\mathfrak{u}}

\newcommand\gf{\mathfrak{g}}

\definecolor{cardinal}{rgb}{0.6,0,0}
\definecolor{darkgreen}{rgb}{0,0.5,0}
\definecolor{golden}{rgb}{0.92, 0.7, 0}
\definecolor{midnight}{rgb}{0, 0, 0.5}
\definecolor{darkblue}{rgb}{0.2, 0, 0.8}
\definecolor{coolblue}{RGB}{0,76,153}

\newcommand{\bighash}{\mathop{\mathlarger{\#}}}

\usetikzlibrary{shapes.geometric,calc,shadows}

\def\centerarc[#1](#2)(#3,#4,#5,#6)
{\draw[#1] ($(#2)+({#5*cos(#3)},{#6*sin(#3)})$) arc [start angle=#3, end angle=#4, x radius=#5, y radius=#6]; }

\def\lineop at (#1,#2){\filldraw[coolblue,circular drop shadow={shadow yshift=-.15ex, shadow xshift=.15ex}] (#1,#2) circle (2.2pt); }

\def\interface at (#1,#2){\draw[line width=1pt, coolblue] (#1,#2-0.5) -- (#1,#2+0.5); }

\def\boundary at (#1,#2){\draw[line width=0.8pt, gray] (#1,#2-0.5) -- (#1,#2+0.5); }

\usepackage{amsmath}
\usepackage{amsfonts}
\usepackage{amssymb,tikz}

\DeclareSymbolFont{bbold}{U}{bbold}{m}{n}
\DeclareSymbolFontAlphabet{\mathbbold}{bbold}

\title{Refined 3d-3d Correspondence}
\author{Luis F. Alday,\,}
\author{Pietro Benetti Genolini,\,}
\author{Mathew Bullimore,\,}
\author{Mark van Loon}
\affiliation{Mathematical Institute, University of Oxford, Andrew Wiles Building, Radcliffe Observatory Quarter, Woodstock Road, Oxford, OX2 6GG, UK}

\emailAdd{luis.alday@maths.ox.ac.uk}
\emailAdd{pietro.benettigenolini@maths.ox.ac.uk}
\emailAdd{mathew.bullimore@maths.ox.ac.uk}
\emailAdd{mark.vanloon@maths.ox.ac.uk}

\abstract{We explore aspects of the correspondence between Seifert 3-manifolds and 3d $\mathcal{N}=2$ supersymmetric theories with a distinguished abelian flavour symmetry. We give a prescription for computing the squashed three-sphere partition functions of such 3d $\mathcal{N}=2$ theories constructed from boundary conditions and interfaces in a 4d $\mathcal{N}=2^*$ theory, mirroring the construction of Seifert manifold invariants via Dehn surgery. This is extended to include links in the Seifert manifold by the insertion of supersymmetric Wilson-'t Hooft loops in the 4d $\mathcal{N}=2^*$ theory. In the presence of a mass parameter for the distinguished flavour symmetry, we recover aspects of refined Chern-Simons theory with complex gauge group, and in particular construct an analytic continuation of the $S$-matrix of refined Chern-Simons theory.}

\begin{document}
\today
\maketitle


\section{Introduction}
\label{sec:intro}

An interesting class of quantum field theories with 3d $\cN=2$ supersymmetry arises from the twisted compactification of the 6d $\cN=(2,0)$ theory on three-manifolds. This leads to a `3d-3d correspondence' between the 3d $\cN=2$ theory denoted by $\cT(M_3)$ and the corresponding three-manifold $M_3$~\cite{Dimofte:2011ju,Dimofte:2011py,Cecotti:2011iy}. An important aspect of this correspondence is the equality between supersymmetric vacua of $\cT(M_3)$ on $S^1 \times \mathbb{R}^2$ and complex flat connections on the three-manifold $M_3$. Furthermore, the supersymmetric partition functions of $\cT(M_3)$, for example on squashed $S^3$ or $S^1 \times S^2$, can be identified with the partition function of Chern-Simons theory with complex gauge group on $M_3$. A derivation of the appearance of complex Chen-Simons theory using localization has appeared in~\cite{Yagi:2013fda,Cordova:2013cea,Lee:2013ida}. 
For a recent review of the 3d-3d correspondence we refer the reader to~\cite{Dimofte:2014ija}.

The purpose of this paper is to explore aspects of the 3d-3d correspondence for Seifert manifolds~\cite{Gadde:2013sca,Chung:2014qpa,Gukov:2016gkn,Gukov:2017kmk}. Seifert manifolds are circle fibrations over a Riemann surface and therefore admit a locally-free circle action. The corresponding 3d $\cN=2$ theory has a distinguished $\uf(1)_f$ flavour symmetry associated to this circle action, which can be incorporated into partition functions on squashed $S^3$ or $S^1 \times S^2$ by turning on a mass parameter or fugacity. Furthermore, the construction of Seifert manifolds via surgery on a torus is expected to have a counterpart in the construction of 3d $\cN=2$ theories using boundary conditions and interfaces implementing $SL(2,\Z)$ duality transformations in a 4d $\cN=2^*$ gauge theory.  

A natural question is how the additional parameter for the distinguished $\uf(1)_f$ flavour symmetry manifests itself as a `refinement' of complex Chern-Simons theory. Our goal is therefore to develop a concrete dicionary between the partition function of $\cT(M_3)$ on squashed $S^3$ with mass parameter for the $\uf(1)_f$ symmetry turned on and computations in Chern-Simons theory with complex gauge group. The mass parameter for the $\uf(1)_f$ symmetry corresponds to the presence of a particular network of defects in $M_3$, leading to a refinement of complex Chern-Simons theory. In particular, we will reproduce an analytic continuation of the $S$-matrix of refined Chern-Simons theory introduced in~\cite{Aganagic:2012ne,Aganagic:2011sg} from the partition functions of $\cT(S^3)$ in the presence of supersymmetric loop operators.

\subsection{Summary}

We will focus on twisted compactifications of the six-dimensional superconformal $\cN=(2,0)$ theory of type $\gf=\suf(N)$ on a compact Seifert manifold $M_3$. This leads to a 3d $\cN=2$ theory denoted by $\cT(M_3)$ with a distinguished $\uf(1)_f$ flavour symmetry corresponding to the circle action on $M_3$. 

Seifert manifolds can be constructed by Dehn surgery. The main step of this process takes a pair of 3-manifolds $M_3^\pm$ with torus boundary and constructs a new 3-manifold $M_3 = M_3^+ \cup_\phi M_3^-$ by identifying the torus boundaries through an element $\phi \in SL(2,\mathbb{Z})$ of the mapping class group. In order to understand the analogue of Dehn surgery for $\cT(M_3)$, it is therefore necessary to consider twisted compactifications of the $\cN=(2,0)$ theory on 3-manifolds with torus boundary.

As explained in~\cite{Dimofte:2013lba}, the twisted compactification on a 3-manifold with torus boundary should be regarded as a boundary condition in 4d $\cN=4$ gauge theory. Choosing a metric on $M_3$ such that the boundary region forms a semi-infinite cylinder $\R_+ \times T^2$ with complex structure $\tau$, compactification on $T^2$ in the asymptotic region leads to a 4d $\cN=4$ theory with gauge algebra $\gf$ on a half-line $\R_+$ with holomorphic gauge coupling $\tau$. The 3-manifold $M_3$ adjoined to this semi-infinite cylinder then defines a boundary condition in the 4d $\cN=4$ theory preserving 3d $\cN=4$ supersymmetry~\cite{Gaiotto:2008sd,Gaiotto:2008sa,Gaiotto:2008ak}.

\begin{figure}[htbp]
\centering
\includegraphics[height=4.5cm]{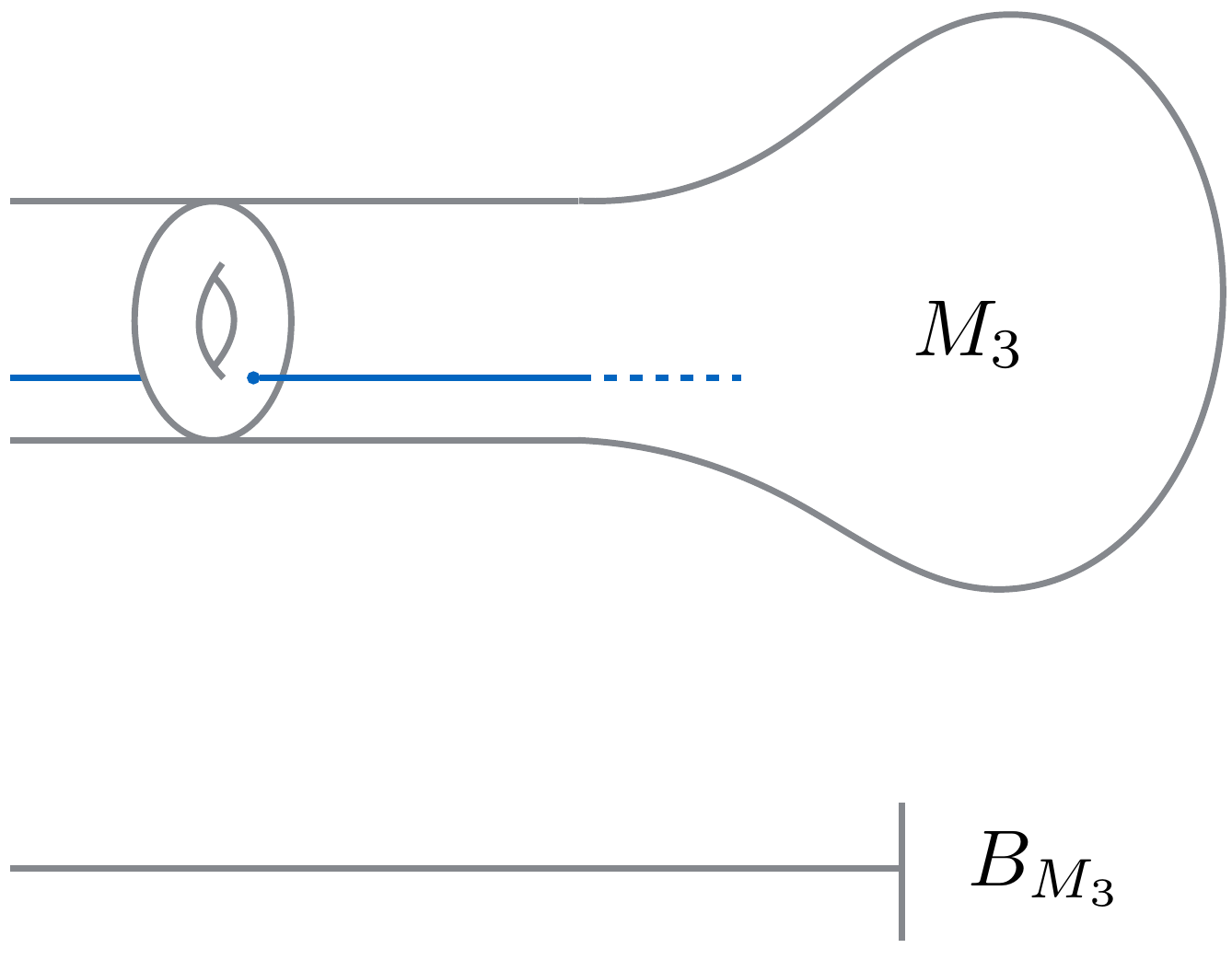}
\caption{A three-manifold $M_3$ with $T^2$ boundary and a codimension-2 defect intersecting the boundary $T^2$ at a point corresponds to a boundary condition $B_{M_3}$ in the 4d $\mathcal{N}=2^*$ theory.}
\label{fig:intro-bc}
\end{figure}

Turning on a mass parameter for the distinguished $\uf(1)_f$ flavour symmetry corresponds to adding a codimension-2 defect supporting the $\uf(1)_f$ flavour symmetry wrapping a curve in $M_3$ that intersects the boundary at a point $p \in T^2$. In particular, in the cylindrical region $\R_+ \times T^2$ the codimension-2 defect is wrapping $\R_+ \times \{\mathrm{pt}\}$. This is illustrated in the top of figure~\ref{fig:intro-bc}. This corresponds to turning on an $\cN=2^*$ deformation of the 4d $\cN=4$ gauge theory and the boundary condition now preserves 3d $\cN=2$ supersymmetry and flavour symmetry $\uf(1)_f$.

In many cases, a genuinely three-dimensional theory can be obtained from a boundary condition in the degeneration limit $\tau \to+ \ii \infty$, where the four-dimensional degrees of freedom are decoupled. In this limit, the boundary $T^2$ degenerates and we obtain a compact 3-manifold where the boundary is replaced by a maximal codimension-2 defect of the 6d $\cN=(2,0)$ theory supporting a flavour symmetry $\mathfrak{g}$. This flavour symmetry is then gauged in coupling to the 4d $\cN=2^*$ theory when the gauge coupling is turned back on.

\begin{figure}[htbp]
\centering
\begin{overpic}[height=5.8cm]{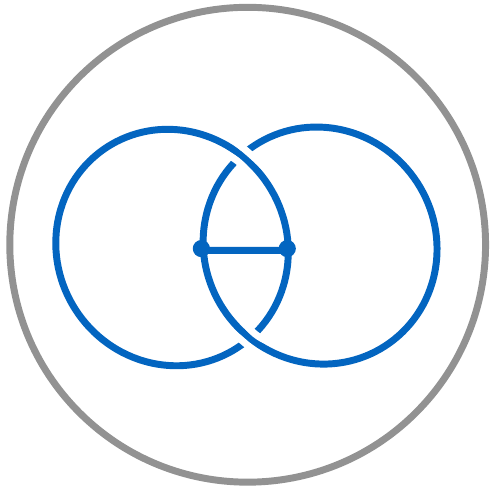}
\put(155,140){\large $S^3$}
\put(133,120){\small$\gf$}
\put(27,120){\small$\gf$}
\put(74,86){\small$\uf(1)$}
\end{overpic}
\caption{The $S$ generator corresponds to an $S^3$ with a Hopf network of defects with flavour symmetries as in the figure.}
\label{fig:S3HopfNetwork}
\end{figure}

Extending the discussion above, a 3-manifold with a pair of torus boundaries corresponds to an interface between 4d $\cN=2^*$ theories. For example, in the Dehn surgery $M_3^+ \cup_\phi M_3^-$, the mapping class element $\phi \in SL(2,\Z)$ corresponds to a mapping cylinder implementing the modular transformation on $T^2$. This corresponds to an interface implementing the corresponding $SL(2,\mathbb{Z})$ duality transformation of the 4d $\cN=2^*$ theory. Such interfaces can also viewed as 3d $\cN=2$ theories in their own right associated to compact 3-manifolds with a pair of codimension-2 defects supporting $\gf$ flavour symmetries. For example, the generator $\phi = S$ corresponds to a Hopf network of codimension-2 defects in $S^3$ supporting flavour symmetries $\gf$, $\gf$ and $\uf(1)_f$. This corresponds to the three-dimensional theory $T(\mathfrak{g})$ introduced in~\cite{Gaiotto:2008ak}. This is illustrated in figure \ref{fig:S3HopfNetwork}. 

A large class of Seifert manifolds known as Lens spaces can be constructed by starting from a mapping torus implementing an $SL(2,\Z)$ duality transformation and then capping off the torus boundaries with solid tori $D_2 \times S^1$. This corresponds to constructing the corresponding theory $\cT(M_3)$ by compactification of a 4d $\cN=2^*$ theory on an interval with boundary conditions at each end corresponding to the solid tori $D_2 \times S^1$ and a sequence of $SL(2,\Z)$ duality interfaces inserted in the intermediate region. For more general Seifert manifolds, one needs to consider boundary conditions and interfaces for a 4d $\cN=2^*$ theory with gauge algebra equal to a direct sum of several copies of $\gf$.

This setup can be further enriched by including codimension-4 defects of the 6d $\cN=(2,0)$ theory labelled by a dominant integral weight of $\mathfrak{g}$. We will focus on the case of codimension-4 defects labelled by the fundamental weights of $\gf$, or equivalently by the anti-symmetric tensor representations of $\suf(N)$. Adding a codimension-4 defect wrapping a knot $K \subset M_3$ corresponds to adding a supersymmetric line defect in the 3d $\cN=2$ theory $\cT(M_3)$. This can be incorporated into the surgery prescription such that, in an intermediate or asymptotic region where $M_3 \sim \mathbb{R} \times T^2$, the codimension-4 defects are supported at a point in $\R$ and a cycle in $T^2$. This will correspond to inserting supersymmetric Wilson-'t Hooft loops in the construction of $\cT(M_3)$ using boundary conditions and interfaces in the 4d $\cN=2^*$ theory.

In the course of this paper, we will implement the construction outlined above to compute the partition functions of theories $\cT(M_3)$ on the squashed three-sphere $S^3_b$~\cite{Hama:2011ea} (generalizing the round sphere introduced in~\cite{Jafferis:2010un,Kapustin:2009kz,Hama:2010av}) in the presence of a mass parameter for the distinguished $\uf(1)_f$ flavour symmetry.

\subsection{Outline}

We begin in section \ref{sec:setup} by summarizing our conventions for the 4d $\cN=2^*$ theory and describing the class of 3d $\cN=2$ boundary conditions and interfaces that will appear throughout the paper.

In section \ref{sec:S1timesR2vacua}, we consider the supersymmetric vacua of the 4d $\cN=2^*$ theory on $S^1\times \R^3$ and therefore the supersymmetric vacua of the theories $\cT(M_3)$ on $S^1\times \R^2$. We recall how the Coulomb branch has a description as the moduli space of $SL(N,\mathbb{C})$ flat connections on $T^2 / \{p\}$, and describe the Coulomb branch images of the aforementioned 3d $\cN=2$ boundary conditions and interfaces as holomorphic Lagrangian submanifolds.

In section \ref{sec:PartitionFunctionsOnS3b}, we consider boundary conditions and interfaces in the 4d $\cN=2^*$ theory on $S^3_b \times \mathbb{R}$ and how this is used to construct the partition functions of theories $\cT(M_3)$ on $S^3_b$. This corresponds to a quantization of the results in section~\ref{sec:S1timesR2vacua}, which are captured by a Chern-Simons theory with complex gauge group $SL(N,\mathbb{C})$. We discuss in detail the implementation of the general framework of boundary conditions and interfaces using results from localization of 3d $\cN=2$ supersymmetric field theories on $S^3_b$.

Having introduced the necessary tools, in section \ref{sec:CaseStudyTS3} we construct the partition function of $\cN=2$ theory $\cT(S^3)$ in a variety of ways from compactifying the 4d $\cN=2^*$ theory on an interval with appropriate boundary conditions. We then introduce codimension-4 defects labelled by anti-symmetric tensor representations of $\suf(N)$ using supersymmetric Wilson-'t Hooft loops in the 4d $\cN=2^*$ theory, corresponding to the unknot and Hopf link in $S^3$. In this way, we recover an analytic continuation of the $S$-matrix of refined Chern-Simons theory.

Finally, in section \ref{sec:Surgery} we construct the partition functions of $\cT(M_3)$ for more general Lens spaces and Seifert manifolds, and perform further checks of our proposal in various limits. We conclude in section \ref{sec:Conclusions} with directions for further study. Appendices~\ref{app:conv}-\ref{app:DetailsTwoDefects} provide some conventions, background and further details of our computations.


\section{Setup}
\label{sec:setup}


\subsection{The $\cN=2^*$ Theory}
\label{sec:bps-bound}

The 4d $\cN=2^*$ theory consists of an $\cN=2$ vectormultiplet together with a hypermultiplet in the adjoint representation of the gauge algebra $\gf$, which we will assume to be $\suf(N)$\footnote{We use conventions where adjoint fields take the form $ \Phi = \sum_A \Phi_A t^A$, where $t^A$ are antihermitian matrices and the covariant derivative is $D_\mu = \partial_\mu + A_\mu$.}. In addition to the standard R-symmetry $\mathfrak{u}(1)_r\oplus \mathfrak{su}(2)_R$, the theory has a $\uf(1)_f$ flavour symmetry acting on the adjoint hypermultiplet. The mass parameter for the adjoint hypermultiplet is obtained by coupling to a background vectormultiplet for $\uf(1)_f$ and turning on a background expectation value $m$ for the scalar component.

We will denote the complex scalar in the dynamical vectormultiplet by $\phi$ and decompose the adjoint hypermultiplet scalars into a pair of complex scalars $(X,Y)$. The charges of these fields under the Cartan generators of the R- and flavour symmetries are given in table~\ref{tab:ChargesComplexScalars}.

\begin{table}[htb]
\centering
\begin{tabular}{ c | c | c | c }
& $T_r$ & $T_R$ & $T_f$ \\ \hline
$\phi$ & $+2$ & $0$ & $0$ \\ 
$X$ & $0$ & $+1$ & $+1$ \\
$Y$ & $0$ & $+1$ & $-1$ \\ 
\end{tabular}
\caption{Charges of the complex scalars in the $\cN=2$ vectormultiplet and hypermultiplet under the Cartan generators of the R-symmetry $\uf(1)_r \oplus \suf(2)_R$ and flavour symmetry $\uf(1)_f$}
\label{tab:ChargesComplexScalars}
\end{table}


\subsection{Boundary Conditions}
\label{sec:boundary-def}

We will consider boundary conditions preserving a 3d $\cN=2$ supersymmetry with unbroken R-symmetry and $\uf(1)_f$ flavour symmetry. We introduce a coordinate $s$ normal to the boundary and coordinates $x^j=\{x^1,x^2,x^3\}$ parallel to the boundary. In general there is an $S^1 \times \mathbb{CP}^1$ family of such boundary conditions corresponding to a choice of breaking pattern $\uf(1)_r \oplus \suf(2)_R \rightarrow \{ \mathrm{pt}\} \oplus \uf(1)_R$. We choose the phase such that $(A_j,\Re(\phi))$ and $(A_s,\Im(\phi))$ transform as a 3d $\cN=2$ vectormultiplet and chiral multiplet respectively, and $\uf(1)_R$ is generated by $T_R$ from table~\ref{tab:ChargesComplexScalars} such that $X$ and $Y$ transform as chiral multiplets. 

The basic boundary conditions for the vectormultiplet correspond to a choice of Neumann boundary conditions for $(A_j,\Re(\phi))$ and Dirichlet boundary conditions $(A_s,\Im(\phi))$ or vice versa~\cite{Dimofte:2013lba}. In more detail, the boundary conditions are defined by
\bea
& \mathrm{Neumann}  \quad && : \quad F_{sj}|  = 0 \qquad D_s \Re(\phi)|  = 0 \qquad \Im(\phi)|=0 \, , \\
& \mathrm{Dirichlet} \quad && : \quad F_{ij}| = 0 \qquad D_s \Im(\phi)| = 0 \qquad  \Re(\phi)|=a
\label{eq:neu-dir-def}
\eea
and $a$ is a valued in a Cartan subalgebra of $\mathfrak{g}$. Neumann boundary conditions preserve the full gauge symmetry $\mathfrak{g}$, whereas Dirichlet boundary conditions break the gauge symmetry but inherit a global symmetry equal to the subalgebra of $\gf$ commuting with $a$. For Neumann boundary conditions $(A_j,\Re(\phi))$ transform as a 3d $\cN=2$ vectormultiplet at the boundary, whereas for Dirichlet boundary conditions $(A_s,\Im(\phi))$ transform as a chiral multiplet.

The boundary conditions for the $\cN=2$ hypermultiplet correspond to a choice of Neumann boundary conditions for $X$ and Dirichlet for $Y$ or vice versa. We will therefore consider the following `Neumann' boundary conditions
\bea
& N_X &&: \quad \mathrm{Neumann} \qquad +  \qquad D_s X | = 0 \qquad Y|=0  \\
& N_Y &&: \quad  \mathrm{Neumann} \qquad + \qquad D_s Y | = 0 \qquad X|=0
\label{eq:neu-def}
\eea
and `Dirichlet' boundary conditions
\bea
& D_X &&: \quad \mathrm{Dirichlet} \qquad +  \qquad D_s Y | = 0 \qquad X|=0  \\
& D_Y &&: \quad  \mathrm{Dirichlet} \qquad + \qquad D_s X | = 0 \qquad Y|=0 \, .
\label{eq:dir-def}
\eea
Note that $X$ has Neumann boundary conditions in $N_X$ and $D_Y$ and becomes a chiral multiplet on the boundary, whereas $Y$ has Neumann boundary conditions in $N_Y$ and $D_X$ and becomes a chiral multiplet on the boundary, with charges as in table~\ref{tab:ChargesComplexScalars}. If we want to emphasize the dependence on the boundary expectation value $a$, we will write Dirichlet boundary conditions as $D_X(a)$, $D_Y(a)$.

These basic boundary conditions can be modified by coupling to boundary degrees of freedom~\cite{Dimofte:2013lba}. For example, the Neumann boundary condition $N_X$ can be modified by coupling to a 3d $\cN=2$ theory with unbroken R-symmetry $\uf(1)_R$ and flavour symmetry at least $\uf(1)_f \oplus \gf$ by coupling to the dynamical vectormultiplet at the boundary. We can also add a boundary superpotential $W(X|,\cO)$ depending on additional boundary chiral operators $\cO$, which modifies a right boundary condition to
\be
 Y| =  \frac{\partial W}{\partial X|} \qquad 0 = \frac{\partial W}{\partial \cO} \, ,
 \label{eq:bdryW-right}
\ee
and a left boundary condition to
\be
| Y = - \frac{\partial W}{\partial | X} \qquad 0 = \frac{\partial W}{\partial \cO} \, .
 \label{eq:bdryW-left}
\ee
In the paper we use the notation $\cdot \, |$ and $| \, \cdot$ to denote the expectation values of bulk operators at right and left boundary conditions respectively.

An important example is to deform the right Neumann boundary condition $N_X$ by a boundary chiral multiplet $\cO_Y$ with the same $T_R$ and $T_f$ charges as $Y$ and a boundary superpotential
\be
W = \tr (X|\cO_Y) \, .
\ee
From equations~\eqref{eq:bdryW-right}, it is straightforward to see that this boundary condition flows to $N_Y$ with $Y|=\cO_Y$, and similarly one can convert the boundary condition $N_Y$ back to $N_X$. There is an essentially identical construction for Dirichlet boundary conditions. Following~\cite{Dimofte:2012pd,Dimofte:2013lba}, we will refer to this operation as a `flip'.


\subsection{Interfaces}
\label{sec:sl2z}

We will also  consider interfaces preserving a 3d $\cN=2$ supersymmetry with unbroken R-symmetry $\uf(1)_R$ and flavour symmetry $\uf(1)_f$. A variety of such interfaces can be constructed by coupling the basic boundary conditions introduced above to additional three-dimensional degrees of freedom by gauging and/or adding a boundary superpotential.

An important class of interfaces are those that flow to the identity interface. For example, let us first impose Dirichlet boundary conditions $D_Y(a)$ on the left and $D_Y(a')$ on the right of the interface. We then identify the boundary flavour symmetry on each of these boundary conditions and gauge it by coupling to a dynamical 3d $\cN=2$ vectormultiplet. Finally, we add a boundary chiral multiplet $\cO$ and a boundary superpotential
\be
W = \tr \left( X| \, \cO  - \cO \, |X ' \right)\,.
\ee
The boundary superpotential requires
\be
Y| = \frac{\partial W}{\partial X|} = \cO  \qquad |Y'= - \frac{\partial W}{\partial |X'} = \cO  \qquad 0 = \frac{\partial W}{\partial \cO} = X| - |X' \, ,
\ee
ensuring that the interface identifies the chiral multiplets on each side. There is an identical construction starting from $D_X$ boundary conditions by exchanging the role of $Y$ and $X$. Such interfaces will be used to `cut' the path integral in our computations in section~\ref{sec:PartitionFunctionsOnS3b}.

Another important class of interfaces are those that implement $SL(2,\mathbb{Z})$ duality transformations\footnote{Provided it is simply-laced, $SL(2,\Z)$ transformations do not change the gauge algebra $\gf$. However, there are distinct physical theories on $\mathbb{R}^4$ with the same $\mathfrak{g}$ but different sets of mutually compatible line operators, on which $SL(2,\Z)$ transformations act in an intricate way~\cite{Aharony:2013hda}. We will generally omit this distinction, mentioning it explicitly when needed.}.
$SL(2,\mathbb{Z})$ duality transformations are generated by $S$ and $T$ satisfying
\be
S^2 = P\, , \qquad (ST)^3 = P\, , 
\ee
where $P$ is a central element such that $P^2=\mathbb{I}$.
The corresponding interfaces were introduced in~\cite{Gaiotto:2008ak}.

\begin{figure}[htp]
\centering
\includegraphics[height=3.25cm]{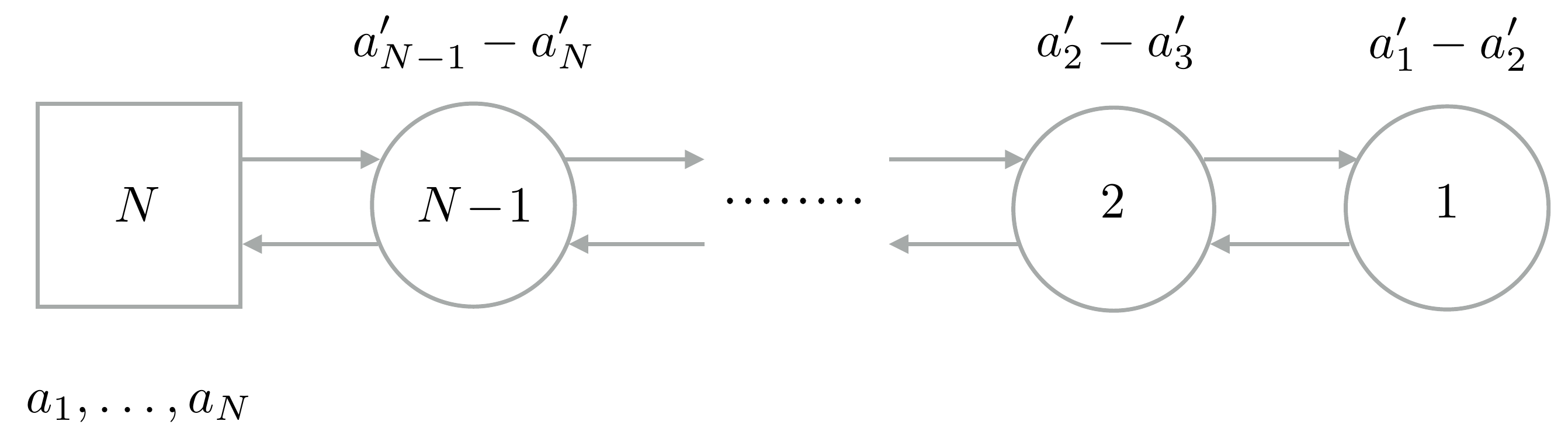}
\caption{A 3d $\mathcal{N}=2$ quiver description of $T(\mathfrak{g})$ with mass parameters $(a_1,\ldots,a_N)$ and FI parameters $\left( a_1'-a_2', \ldots, a_{N-1}'-a_N'\right) $.}
\label{fig:TG}
\end{figure}

The interface generating the action of $T$ on boundary conditions is constructed by adding an $\cN=2$ supersymmetric Chern-Simons term at level $+1$. To construct an $S$-duality interface at $s=0$, we deform a right $N_X$ boundary condition on $s\leq 0$ and a left $N_Y$ boundary condition on $s\geq 0$ by coupling to the three-dimensional theory $T(\gf)$ at $s=0$ and gauging the flavour symmetry $\gf \oplus \gf $~\cite{Gaiotto:2008ak}.

There is a description of $T(\gf)$ as a triangular quiver with gauge algebras $\uf(j)$ for $j=1,\ldots,N-1$. The $\gf$ symmetry that rotates the $N$ pairs of chiral at the final node is manifest, while the second one is an enhancement of the $\uf(1)^{N-1}$ topological symmetry in the infrared. Sandwiching the $S$ interface between Dirichlet boundary conditions $D_X(a)$ on the left and $D_Y(a')$ on the right isolates the three-dimensional degrees of freedom in $T(\mathfrak{g})$. In particular, $a = (a_1,\ldots,a_N)$ are identified with the mass parameters and $a'=(a'_1,\ldots,a_N')$ with the FI parameters of $T(\gf)$ - as shown in figure~\ref{fig:TG}.


\section{Supersymmetric Vacua on $S^1 \times \R^2$}
\label{sec:S1timesR2vacua}

Upon compactification on a circle, the Coulomb branch of supersymmetric vacua of the 4d $\cN=2^*$ theory coincides with the Hitchin moduli space on a punctured torus $T^2 / \{p\}$ with boundary conditions at $p$ determined by the hypermultiplet mass $m$. This is a hyper-K\"ahler moduli space $\cM$. 

Our choice of boundary conditions and interfaces fixing a point $\{\mathrm{pt}\} \subset \uf(1)_f$ are compatible with a complex structure in which $\cM$ is the moduli space of complex flat connections on $T^2/\{p\}$ with fixed monodromy around the puncture $p$ determined by the mass parameter $m$. The moduli space is then parameterized by the traces of the holonomy around the cycles of $T^2$, which are the expectation values of supersymmetric loop operators in the 4d $\cN=2^*$ theory wrapping the circle. 

The Coulomb branch image of a 3d $\cN=2$ boundary condition is a holomorphic Lagrangian submanifold in $\cM$ cut out by the additional `boundary Ward identities' imposed upon supersymmetric loop operators at the boundary. Similarly, interfaces determine holomorphic Lagrangian submanifolds in the product of Coulomb branch moduli spaces on each side of the interface. Our task in this section is to determine the Coulomb branch images of the 3d $\cN=2$ boundary conditions and interfaces constructed in section~\ref{sec:setup}.


\subsection{$SL(N,\C)$ Flat Connections}
\label{sec:flat-conn}

For definiteness, let us compactify the 4d $\cN=2^*$ theory on a circle by identifying $x^1 \sim x^1+2\pi R$. As explained above, in the complex structure compatible with our choice of 3d $\cN=2$ boundary conditions, the Coulomb branch moduli space $\cM$ can be identified with the moduli space of $SL(N,\C)$ flat connections on $T^2/\{p\}$. This is parameterized by holonomy matrices $W$, $H$ around the $(1,0)$, $(0,1)$ cycles which obey
\be
WHW^{-1}H^{-1} = E \, .
\label{eq:torus-puncture}
\ee
modulo conjugation by $SL(N,\mathbb{C})$ matrices. The holonomy $E$ around the puncture at $p$ has fixed eigenvalues $\{ \tf^{-1},\ldots,\tf^{-1},\tf^{N-1} \}$, where $\tf=e^{2\pi R m}$ and the hypermultiplet mass parameter $m$ is in this section complexified by a background Wilson loop for the $\uf(1)_f$ flavour symmetry wrapping the circle

At a generic point on the moduli space $\cM$, the gauge symmetry is broken to a Cartan subalgebra and the  eigenvalues $\{\wf_1,\ldots,\wf_N\}$ of $W$ can be identified with the expectation values of abelian supersymmetric Wilson loops obeying $\prod_j\wf_j=1$. We have $\wf_j = e^{2\pi Ra_j}$, where $a = (a_1,\ldots,a_N)$ is the expectation value of the scalar field $\Re(\phi)$, complexified by the holonomy of the gauge field around the circle.

By an $SL(N,\mathbb{C})$ transformation, we can diagonalize the holonomy matrix $W$ and introduce the following convenient parameterization of the holonomy matrix $H$,
\be
W^i{}_j = \delta^i{}_j \wf_j \qquad  H^i{}_j = \frac{ \prod_{k \neq j} ( \tf^{1/2} \wf_i  - \tf^{-1/2} \wf_k  )}{\prod_{k \neq i} ( \wf_i  - \wf_k )} \hf_j \, ,
\label{eq:holonomy-sol}
\ee
where the coordinates $\{\hf_1,\ldots,\hf_N\}$ are the expectation values of supersymmetric abelian 't Hooft loops, and obey $\prod_j \hf_j=1$. With these coordinates, the holomorphic symplectic form is given by
\be
\Omega = \sum_{j=1}^N \rd \log \wf_j \wedge \rd \log \hf_j \, .
\ee
Note that removing the puncture, $\tf\to1$, the holonomy matrix $H$ also becomes diagonal with eigenvalues $\{\hf_1,\ldots,\hf_N\}$. However, we emphasize that the coordinates $\{\hf_1,\ldots,\hf_N\}$ are not in general the eigenvalues of $H$.

The holonomy matrix $H$ can be identified with the Lax matrix of the complex $N$-body trigonometric Ruijsenaars-Schneider model~\cite{Gaiotto:2013bwa,Bullimore:2014awa} and therefore methods from classical integrable systems are very useful. In particular, a convenient set of invariant functions on $\cM$ is obtained by expanding the Lax determinants
\bea
\det(z-W) & = \sum_{r=1}^N (-1)^r z^{N-r} W^{(r)} \\
\det(z-H) & =\sum_{r=1}^N (-1)^r z^{N-r} H^{(r)} \, ,
\eea
where
\be
W^{(r)} =  \sum_{|I|=r} \wf_{I} 
\label{eq:non-abloops1}
\ee
and 
\be
H^{(r)}  = \sum_{|I|=r} \; \hf_I\; \prod_{  i \in I, j \notin I   } \frac{\tf^{1/2}\wf_i  -\tf^{-1/2} \wf_j  }{\wf_i  - \wf_j  } 
\label{eq:non-abloops2}
\ee
are the traces of the holonomy matrices $\tr_{\Lambda^r} (W)$ and $\tr_{\Lambda^r} (H)$ respectively in the antisymmetric tensor representations $\Lambda^r$ of $SL(N,\mathbb{C})$ of rank $r = 1,\ldots,N-1$. In these expressions, we use the notation $I = \{ i_1,\ldots,i_r\} \subset \{1,\ldots,N\}$ and introduce the convenient shorthand $\wf_I = \wf_{i_1} \ldots \wf_{i_r}$ and $\hf_I = \hf_{i_1} \ldots \hf_{i_r}$. The functions~\eqref{eq:non-abloops1} and~\eqref{eq:non-abloops2} are the Coulomb branch expectation values of non-abelian supersymmetric Wilson and 't Hooft loops respectively wrapping the circle.

Since the holonomy matrices are valued in $SL(N,\C)$, they have unit determinant and traces in conjugate representations $\Lambda^r$ and $\Lambda^{N-r}$ are obtained by inverting the holonomy matrix. For example, we have $H^{(N-r)} = \tr_{\Lambda^r}(H^{-1})$. Traces in conjugate representations can be expressed nicely in terms of $\{ \tilde h_1, \ldots,\tilde h_N\}$ defined by 
\be
\hf_i \, \widetilde \hf_i = \prod_{ j \neq i} \frac{\tf^{-1/2}\wf_i  - \tf^{1/2} \wf_j  }{\tf^{1/2} \wf_i - \tf^{-1/2}\wf_j } \, .
\ee
For example,
\be
\tr_{\Lambda^r}(H^{-1}) = \sum_{|I|=r} \; \tilde h_I \; \prod_{  i \in I, j \notin I   } \frac{\tf^{1/2} \wf_i  -  \tf^{-1/2} \wf_j}{\wf_i  - \wf_j  } \,.
\label{eq:inverse-thooft}
\ee
It is also straighforward to compute the trace of the holonomy around other cycles of $T^2/\{p\}$ in terms of these coordinates,
\bea
\tr_{\Lambda^r} (WH) & = \sum_{|I|=r} \; \wf_I \hf_I \; \prod_{ i \in I , j \notin I }  \frac{\tf^{1/2} \wf_i  - \tf^{-1/2} \wf_j }{\wf_i  - \wf_j  } \\
\tr_{\Lambda^r} (W^{-1}H) & = \sum_{|I|=r} \; \frac{ \hf_I}{\wf_I} \; \prod_{ i \in I , j \notin I }  \frac{\tf^{1/2} \wf_i  - \tf^{-1/2} \wf_j }{\wf_i  - \wf_j  } \\
\tr_{\Lambda^r} (WH^{-1}) & = \sum_{|I|=r} \; \wf_I \tilde\hf_I \; \prod_{ i \in I , j \notin I }  \frac{\tf^{1/2} \wf_i  - \tf^{-1/2} \wf_j }{\wf_i  - \wf_j  } \\
\tr_{\Lambda^r} (W^{-1}H^{-1}) & = \sum_{|I|=r} \; \frac{ \tilde\hf_I}{\wf_I} \; \prod_{ i \in I , j \notin I }  \frac{\tf^{1/2} \wf_i  - \tf^{-1/2} \wf_j }{\wf_i  - \wf_j  } \, .
\label{eq:mixed-ev}
\eea
These expressions are identified with the Coulomb branch expectation values of supersymmetric mixed Wilson-'t Hooft loops. 


\subsection{Boundary Conditions}
\label{sec:bc-hollag}

\begin{figure}[htp]
\centering
\includegraphics[height=3.2cm]{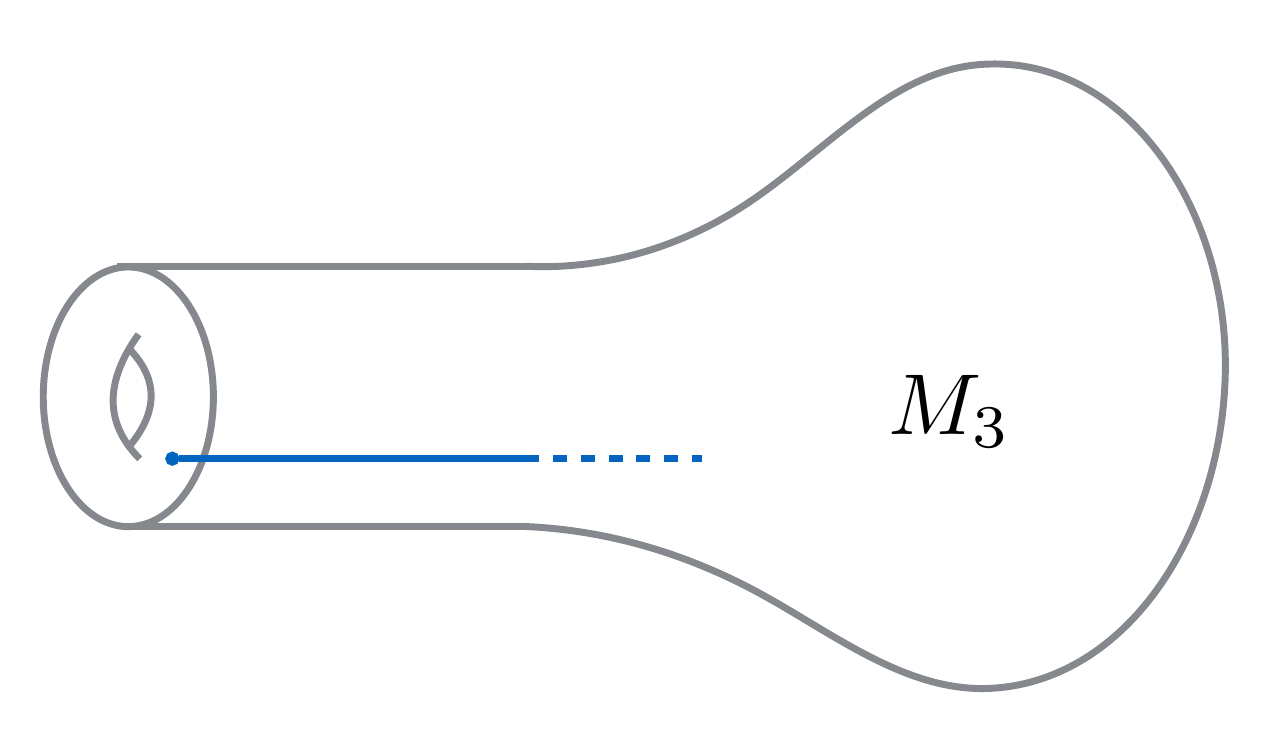}
\caption{Three-manifold $M_3$ with defect of monodromy eigenvalues $\{ t^{-1},\ldots,t^{-1},t^{N-1} \}$ ending on the boundary $T^2/\{p\}$.}
\label{fig:torus-puncture-boundary}
\end{figure}

The image of a boundary condition preserving 3d $\cN=2$ supersymmetry is a holomorphic Lagrangian submanifold $\cL \subset \cM$ encoding the boundary Ward identities for supersymmetric loop operators brought to the boundary. This Lagrangian describes a choice of three-manifold with boundary $T^2 / \{ p\} $ and defect with holonomy $\{ \tf^{-1},\ldots,\tf^{-1},\tf^{N-1} \}$ - as shown in figure~\ref{fig:torus-puncture-boundary}. The holomorphic Lagrangian $\cL$ consists of those $SL(N,\C)$ flat connections on the boundary that extend into the three-manifold.

In order to describe the holomorphic Lagrangians $\cL \subset \cM$ associated to the basic boundary conditions in section~\ref{sec:setup}, it is convenient to introduce a new set of variables 
\bea
h_i^+ & = h_i \prod_{j\neq i} \frac{t^{1/2} w_i - t^{-1/2}w_j}{w_i-w_j} \\
h_i^- & = h^{-1}_i \prod_{j\neq i} \frac{t^{-1/2} w_i - t^{1/2}w_j}{w_i-w_j}
\label{eq:newh}
\eea
with
\be
h_i^+h_i^- = \prod_{j\neq i} \frac{(t^{1/2} w_i - t^{-1/2}w_j)}{w_i - w_j} \frac{(t^{-1/2} w_i - t^{1/2}w_j)}{w_i-w_j}\, .
\label{eq:ring-rel}
\ee
The parameters $\{h_1^+,\ldots,h_N^+\}$ and $\{h_1^-,\ldots,h_N^-\}$ are the four-dimensional lift of the abelian monopole operators introduced in~\cite{Bullimore:2015lsa} to describe the Coulomb branch of 3d $\cN=4$ gauge theories and further used in~\cite{Bullimore:2016nji} to find the Coulomb branch images of 2d $\cN=(2,2)$ boundary conditions. We can therefore uplift these results to compute the Coulomb branch images of 3d $\cN=2$ boundary conditions in the 4d $\cN=2^*$ theory. 


\subsubsection{Neumann}

Let us first consider Neumann boundary conditions. The holomorphic Lagrangians for right Neumann boundary conditions $N_X$ and $N_Y$ are 
\bea
& N_X \quad : \quad h_i^+ | = \prod_{j\neq i} \frac{t^{1/2} w_i - t^{-1/2}w_j}{w_i-w_j} | \qquad h_i^-| = \prod_{j\neq i} \frac{t^{-1/2} w_i - t^{1/2}w_j}{w_i-w_j} | \\
& N_Y \quad : \quad h_i^+ | = \prod_{j\neq i} \frac{t^{-1/2} w_i - t^{1/2}w_j}{w_i-w_j} | \qquad h_i^-| = \prod_{j\neq i} \frac{t^{1/2} w_i - t^{-1/2}w_j}{w_i-w_j} | \, .
\label{eq:neu-right}
\eea
In terms of the original variables, the Neumann boundary condition $N_X$ is described by $h_i = 1$ whereas $N_Y$ is described by $\tilde h_i=1$. 

It is straightforward to check that both Neumann boundary conditions $N_X$ and $N_Y$ in fact describe the same holomorphic Lagrangian, which can be defined invariantly by fixing the eigenvalues of the holonomy matrix $H$ to be $t^\rho$, where
\be
\rho = \left( \tfrac{N-1}{2} ,  \tfrac{N-3}{2} , \ldots,  \tfrac{1-N}{2} \right)
\ee
is the Weyl vector. 

In terms of supersymmetric non-abelian 't Hooft loops, the right $N_X$ boundary condition has the property that
\be
H^{(r)} \, | = \sum_{|I|=r} \; \prod_{  i \in I, j \notin I   } \Big. \frac{\tf^{1/2}\wf_i  -\tf^{-1/2} \wf_j  }{\wf_i  - \wf_j  } \, \Big| \, .
\label{eq:thooft-at-neu}
\ee
This expression is in fact independent of $w_j$ and sums to 
\be
\dim_t( \Lambda^r) = W^{(r)}(\wf \to \tf^\rho) \, ,
\ee
which is the quantum dimension of the representation $\Lambda^r$ with quantum parameter $t$. Since the quantum dimension is invariant under $t \to t^{-1}$, we obtain the same result for $N_Y$. This reproduces the localization computation of the $S^1$ partition function of an $\cN=4$ gauged quantum mechanics that flows to a sigma model onto the Grassmannian $Gr(r,N)$ \cite{Hori:2014tda}. This can be interpreted as the $S^1$ partition function of the one-dimensional degrees of freedom supported on the 't Hooft loop.

It will also be important to note the expectation values of mixed Wilson-'t Hooft loops at the Neumann boundary condition $N_X$,
\bea \label{eq:WHonNeumann}
\tr_{\Lambda^r} (WH) \, | & = \tf^{r(N-r)/2} \, W^{(r)}\, | \\
 \tr_{\Lambda^r} (W^{-1}H) \, |  & = \tf^{-r(N-r)/2} \, W^{(N-r)}\, | \\
\tr_{\Lambda^r} (WH^{-1}) \, | & = \tf^{-r(N-r)/2} \, W^{(r)}\, | \\
\tr_{\Lambda^r} (W^{-1}H^{-1}) \, |  & = \tf^{r(N-r)/2} \, W^{(N-r)}\,| \, .
\eea

Removing the puncture by turning off the mass parameter for the $\uf(1)_f$ symmetry sends $\tf \to 1$, and therefore the holomorphic Lagrangian for a Neumann boundary condition becomes
\be
\tr_{\Lambda^r} ( H )\, | = \dim(\Lambda^r) \, .
\ee
This shows that the holonomy around the $(0,1)$ cycle becomes trivial. The 3-manifold corresponding to this holomorphic Lagrangian is therefore a solid torus $S^1 \times D_2$ obtained by contracting the $(0,1)$ cycle. 

Turning back on the mass parameter for the $\uf(1)_f$ symmetry, the holomorphic Lagrangian still describes a solid torus $S^1 \times D_2$ obtained by collapsing the $(0,1)$ cycle, but now punctured by a monodromy defect at the origin of the disk $D_2$ with fixed holonomy eigenvalues $t^\rho$. We will simply refer to this as the solid torus $S^1 \times D_2$ obtained by collapsing the $(0,1)$ cycle, with the presence of the monodromy defect understood.

Finally, the boundary Ward identities for left Neumann boundary conditions are found by exchanging the roles of $h_i^+$ and $h^-_i$ in the above formulae, which define the same holomorphic Lagrangian in this example.


\subsubsection{Generalized Neumann}
\label{sec:gen-neu}

We now briefly consider the generalized Neumann boundary conditions $N_X[T]$ and $N_Y[T]$ obtained by coupling Neumann boundary conditions $N_X$ or $N_Y$ to a 3d $\cN=2$ gauge theory $T$ with unbroken R-symmetry $\uf(1)_R$ and flavour symmetry at least $\gf \oplus \uf(1)_f$.

Let us denote the effective twisted superpotential of the three-dimensional theory $T$ by $\widetilde \cW(w_j,t,s_a)$, where $s_a$ are the abelian Wilson loops for the three-dimensional gauge symmetry\footnote{In order to simplify our notation, we multiply the effective twisted superpotential $\widetilde \cW$ by a factor of $(2\pi R)^2$ compared to the standard conventions, for example~\cite{Gaiotto:2013bwa}.}. The boundary Ward identities generalizing those for pure Neumann boundary conditions~\eqref{eq:neu-right} are
\bea
& N_X[T] &&: \quad h_i^+ | = e^{\frac{\partial \widetilde \cW }{ \partial \log w_i}} \prod_{j\neq i} \frac{t^{1/2} w_i - t^{-1/2}w_j}{w_i-w_j}  | \qquad h_i^-| = e^{ -\frac{\partial \widetilde \cW }{ \partial \log w_i}}\prod_{j\neq i} \frac{t^{-1/2} w_i - t^{1/2}w_j}{w_i-w_j} | \\
& N_Y[T] &&: \quad h_i^+ | = e^{ \frac{\partial \widetilde \cW }{ \partial \log w_i}} \prod_{j\neq i} \frac{t^{-1/2} w_i - t^{1/2}w_j}{w_i-w_j} | \qquad h_i^-| =e^{ -\frac{\partial \widetilde \cW }{ \partial \log w_i}} \prod_{j\neq i} \frac{t^{1/2} w_i - t^{-1/2}w_j}{w_i-w_j} | \, ,
\label{eq:neu-gen-right}
\eea
which are supplemented by the vacuum equations
\be
e^{ \frac{\partial \widetilde \cW }{ \partial \log s_a}} = 1 \, .
\ee
As above, the boundary Ward identities for left boundary conditions are found by exchanging $h^+_i$ and $h^-_i$ in the above.

This result allows us to check the compatibility of the boundary Ward identities for pure Neumann boundary conditions~\eqref{eq:neu-right} with the flip. As explained in section~\ref{sec:boundary-def}, the flip corresponds to coupling the Neumann boundary condition $N_X$ to a boundary chiral multiplet $\cO_Y$ with the same charges as $Y$ with superpotential $W=\tr(X|\cO_Y)$. The boundary chiral multiplet $\cO_Y$ has effective twisted superpotential
\be
\widetilde\cW = \sum_{i \neq j} f( w_i / t w_j) + \mathrm{const}
\ee
where the function $f(w)$ satisfies 
\be
e^{\frac{\partial f}{\partial \log w}} = w^{1/2}-w^{-1/2} \, .
\ee 
It is straightforward to check using equation~\eqref{eq:neu-gen-right} that the boundary Ward identities for  $N_X(\cO_Y)$ are equivalent to those for $N_Y$, up to a sign that can be absorbed in the definition of the abelian 't Hooft loop operators. A similar derivation shows that the boundary condition $N_Y(\cO_X)$ is equivalent to $N_Y$.


\subsubsection{Dirichlet}

Let us now consider the Dirichlet boundary conditions $D_X$. The holomorphic Lagrangian is defined by setting the eigenvalues of $W$ equal to fixed values $\{\wf_1^0,\ldots,\wf_N^0\}$, or equivalently by fixing the expectation values of supersymmetric Wilson loops $W^{(r)}$ for all $r=1,\ldots,N-1$.

The corresponding three-manifold is therefore the solid torus $S^1 \times D_2$ obtained by contracting the $(1,0)$ cycle, punctured by a monodromy defect at the origin of the disk $D_2$ with eigenvalues $\{ \wf_1^0,\ldots,\wf^0_N\}$.


\subsection{Interfaces}
\label{sec:int}

An interface corresponds to a holomorphic Lagrangian submanifold in the product $\cM \times \cM'$ of Coulomb branch moduli spaces on each side of the interface, with holomorphic symplectic form $\Omega - \Omega'$. We now describe the holomorphic Lagrangians corresponding to the interfaces generating $SL(2,\Z)$ transformations that were described in section~\ref{sec:sl2z}.

In preparation for our discussion of the $T$ interface, let us first consider a class of interfaces generalizing $N_X[T]$, which are constructed by coupling to a 3d $\cN=2$ gauge theory with unbroken R-symmetry $\uf(1)_R$ and flavour symmetry at least $\gf \oplus \uf(1)_f$. As above, we denote the effective twisted superpotential of this theory by $\widetilde\cW(\wf_j,t,s_a)$. This interface defines the holomorphic Lagrangian
\be
\wf_i \, | = | \, \wf'_i\,, \qquad \hf_i^+ \, | = | \, e^{\frac{ \partial \widetilde\cW }{ \partial \log \wf'_j }} h_i'^{-} \, ,
\ee
where in the second equation we have cancelled a factor of 
\be
\prod_{j\neq i} \frac{t^{1/2} w_i - t^{-1/2}w_j}{w_i-w_j}
\ee
on each side since $w_i| =|w_i'$ from the first equation. This is again supplemented by the vacuum condition
\be
e^{\frac{\partial \widetilde\cW}{\partial \log s_a}} = 1 \, .
\ee

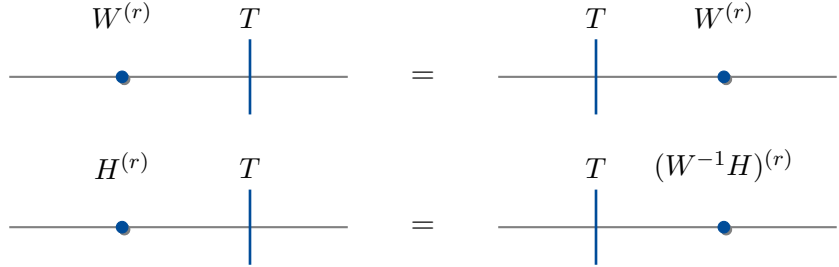
\begin{figure}[htbp]
\begin{center}
\begin{tikzpicture}
\draw[line width=0.8pt, gray] (-5.5,1) -- (-1,1);
\draw[line width=0.8pt, gray] (1,1) -- (5.5,1);
\interface at (-2.3,1)
\interface at (2.3,1)
\lineop at (-4,1)
\lineop at (4,1)
\node at (-2.3,1.78) {$T$};
\node at (2.3,1.78) {$T$};
\node at (-4,1.82) {$W^{(r)}$};
\node at (4,1.82) {$W^{(r)}$};
\node at (0,1) {{\large $=$}};
\draw[line width=0.8pt, gray] (-5.5,-1) -- (-1,-1);
\draw[line width=0.8pt, gray] (1,-1) -- (5.5,-1);
\interface at (-2.3,-1)
\interface at (2.3,-1)
\lineop at (-4,-1)
\lineop at (4,-1)
\node at (-2.3,-0.22) {$T$};
\node at (2.3,-0.22) {$T$};
\node at (-4,-0.18) {$H^{(r)}$};
\node at (4,-0.18) {$(W^{-1}H)^{(r)}$};
\node at (0,-1) {{\large $=$}};
\end{tikzpicture}
\end{center}
\vspace{-0.5cm}
\caption{A Wilson loop commutes with the $T$ interface, whereas an 't Hooft loop becomes a mixed Wilson-'t Hooft loop.}
\label{fig:Tinterface}
\end{figure}

The $T$ interface is now a special case of the above construction where we couple to a supersymmetric Chern-Simons term at level $+1$, with effective twisted superpotential
\be
\widetilde\cW(w_j) = - \frac{1}{2} \sum_{j=1}^N (\log \wf_j)^2 \, .
\ee
It therefore corresponds to the holomorphic Lagrangian 
\be
\wf_i \, | = | \, \wf'_i \, ,\qquad \hf^+_i  \, | = | \, w_i'^{-1}  \hf'^-_i \, ,
\ee
which can be written more invariantly as
\be
\tr_{\Lambda_r}(W) \, | = | \, \tr_{\Lambda^r}(W') \, ,\qquad \tr_{\Lambda_r}(H) \, | = | \, \tr_{\Lambda^r}(W'^{-1}H'^{-1}) \, .
\ee
In what follows, we will introduce a graphical notation where supersymmetric loop operators are always denoted acting on right boundary conditions. With this convention, the translation of supersymmetric loop operators through the $T$ interface is shown in figure~\ref{fig:Tinterface}.

\begin{figure}[htbp]
\begin{center}
\begin{tikzpicture}
\draw[line width=0.8pt, gray] (-5.5,1) -- (-1,1);
\draw[line width=0.8pt, gray] (1,1) -- (5.5,1);
\interface at (-2.3,1)
\interface at (2.3,1)
\lineop at (-4,1)
\lineop at (4,1)
\node at (-2.3,1.78) {$S$};
\node at (2.3,1.78) {$S$};
\node at (-4,1.82) {$H^{(r)}$};
\node at (4,1.82) {$W^{(r)}$};
\node at (0,1) {{\large $=$}};
\draw[line width=0.8pt, gray] (-5.5,-1) -- (-1,-1);
\draw[line width=0.8pt, gray] (1,-1) -- (5.5,-1);
\interface at (-2.3,-1)
\interface at (2.3,-1)
\lineop at (-4,-1)
\lineop at (4,-1)
\node at (-2.3,-0.22) {$S$};
\node at (2.3,-0.22) {$S$};
\node at (-4,-0.18) {$W^{(r)}$};
\node at (4,-0.18) {$H^{(N-r)}$};
\node at (0,-1) {{\large $=$}};
\end{tikzpicture}
\end{center}
\vspace{-0.5cm}
\caption{Under $S$ duality, a Wilson loop becomes an 't Hooft loop.}
\label{fig:S-interface}
\end{figure}
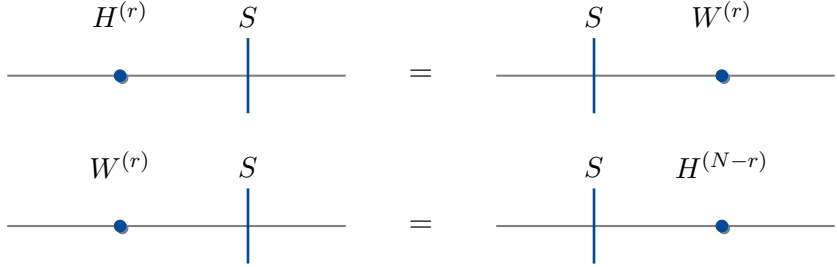

Let us now consider the $S$ transformation. Recall that in the construction of section~\ref{sec:sl2z}, the 3d $\cN=2$ theory $T(\gf)$ is isolated by sandwiching the $S$ interface in between Dirichlet boundary conditions $D_X(a)$ and $D_Y(a')$. This has the inconvenient feature that it interpolates between flat connections on $T^2/\{p\}$ with the monodromy eigenvalues at the puncture inverted, $\tf \to \tf^{-1}$. It is therefore convenient to combine this interface with a flip and denote by $\widetilde\cW(w_j,w_j',s_a)$ the effective twisted superpotential of the degrees of freedom obtained by sandwiching the $S$ interface between boundary conditions $D_X(a)$ and $D_X(a')$. With this understood, the holomorphic Lagrangian is a generalization of that for the boundary condition $N_X[T]$ to
\bea
h_i^+ \, | & = e^{\frac{\partial \widetilde \cW }{ \partial \log w_i}} \prod_{j\neq i} \frac{t^{1/2} w_i - t^{-1/2}w_j}{w_i-w_j}  | \\
| \, h_i'^- & = | e^{\frac{\partial \widetilde \cW }{ \partial \log w'_i}} \prod_{j\neq i} \frac{t^{1/2} w'_i - t^{-1/2}w'_j}{w'_i-w'_j} \, ,
\eea
together with
\be
e^{\frac{\partial\widetilde W}{\partial \log s_a}} = 1\, .
\ee
From the detailed computations in \cite{Bullimore:2014awa,Gaiotto:2013bwa}, this holomorphic Lagrangian can be written invariantly as
\be
H^{(r)} \, | = | \, W^{(r)} \, \qquad W^{(r)} \, | = | \, H^{(r)} \, .
\ee
In diagrammatic conventions, with the understanding that all operators act on right boundary conditions, the action of the $S$ interface on supersymmetric loop operators is shown in figure \ref{fig:S-interface}.

The $S$-dual of the Neumann boundary conditions $N_X$ and $N_Y$ will play an important r\^ole later. We denote them by Nahm pole boundary conditions $NP_X$ and $NP_Y$. Given that Neumann boundary conditions of all types correspond to setting the eigenvalues of $H$ equal to $t^\rho$, the Nahm pole boundary conditions correspond to setting the eigenvalues $\{ w_i,\ldots,w_i\}$ of $W$ to $t^\rho$. Equivalently, we have
\be
W^{(r)} | = \dim_t(\Lambda^r) \qquad | W^{(r)}  = \dim_t(\Lambda^r)
\ee
for Nahm pole boundary conditions.


\section{Squashed $S^3$ Partition Function}
\label{sec:PartitionFunctionsOnS3b}

In this section, we will replace $S^1\times \R^2$ parallel to the boundary conditions and interfaces by a squashed three-sphere $S^3_b$. This will lead to a quantization of the Coulomb branch moduli space $\cM$ of $SL(N,\C)$ flat connections on $T^2/\{p\}$, which is captured by a Chern-Simons theory with complex gauge group $SL(N,\C)$. Such a quantization is specified by a pair levels $(k,\sigma)$ where $k \in \mathbb{Z}$ is quantized and $\sigma\in \C$ is continuous~\cite{Witten:1989ip}. From supersymmetric localization of the six-simensional $\cN=(2,0)$ theory~\cite{Cordova:2013cea}, the expected levels for the complex Chern-Simons theory corresponding to $S^3_b$ partition functions are
\be 
k = 1\, ,  \qquad \sigma = \frac{1-b^2}{1+b^2} \, .
\ee 
Our approach will be to utilize results from supersymmetric localization of 3d $\cN=2$ theories on $S^3_b$ to construct partition functions of $SL(N,\C)$ Chern-Simons theory on Seifert manifolds by surgery on $T^2/\{p\}$.
 

\subsection{Setup}

A 4d $\cN=2$ theory on $\mathbb{R} \times S^3_b$ can be viewed as an infinite-dimensional supersymmetric quantum mechanics on $\mathbb{R}$ with a pair of supercharges $Q,Q^\dagger$, which coincide with the supercharges used in the localization of 3d $\cN=2$ theories on $S^3_b$. A compatible boundary condition that preserves 3d $\cN=2$ supersymmetry in flat space can be represented as a `boundary state' in the space of supersymmetric ground states annihilated by $Q,Q^\dagger$. Instead of attempting to describe this supersymmetric quantum mechanics directly, for example as in~\cite{Assel:2015nca}, we will perform computations using known localization results for 3d $\cN=2$ theories on $S^3_b$. 

Our conventions regarding contributions to the $S^3_b$ partition functions are summarized in appendix~\ref{app:conv}. In particular, we have imaginary mass parameters $(a_1,\ldots,a_N)$ obeying $\sum_j a_j = 0$, in keeping with our choice of anti-hermitian Lie algebra generators, and an imaginary hypermultiplet mass parameter $m$ associated to the $T_f$ symmetry. It will also be convenient to also introduce the combination $\ep = \tfrac{Q}{2} - m$, where $Q = b+b^{-1}$, such that $\ep^* = \tfrac{Q}{2}+m$.

With this notation, the contribution of a 3d $\cN=2$ vectormultiplet is
\be
\nu(a)  = \prod ^N_{\substack{i, j=1\\i\neq j}}\frac{1}{S_b(a_i-a_j)} 
\ee
The contributions from chiral multiplets in the adjoint representation with the same $T_R$ and $T_f$ charges charges as $X$ and $Y$ (shown in table~\ref{tab:ChargesComplexScalars}) are
\bea
K_X(a) & = \frac{1}{S_b(\epsilon)} \prod_{i,j=1}^N S_b(\ep+a_i-a_j) \\
K_Y(a) & = \frac{1}{S_b(\epsilon^*)} \prod_{i,j=1}^N S_b(\ep^*+a_i-a_j) 
\eea
respectively. An important consequence of the identity $S_b(x)S_b(Q-x)=1$ is that these partition functions obey $K_X(a)K_Y(a)=1$. The physical reason is the existence of the superpotential $\tr(XY)$ allowing both chiral multiplets to be integrated out. As we will see momentarily, it also ensures consistency of the flip.

It is also convenient to introduce the notation
\be\label{eq:nuX}
\nu_X(a) = \nu(a)K_X(a)\,, \qquad \nu_Y(a) = \nu(a) K_Y(a)\,,
\ee
which combine a 3d $\cN=2$ vectormultiplet and an adjoint chiral multiplet with the same charges as $X$ or $Y$. These combinations correspond to the contributions from 3d $\cN=4$ vectormultiplets or twisted vectormultiplets, deformed to 3d $\cN=2$ supersymmetry by the mass parameter $m$ associated to $T_f$.


\subsection{Basic Overlaps}
\label{sec:basic-overlaps}

The basic computation we want to perform is the parition function of the 4d $\cN=2^*$ theory on $S^3_b$ times an interval with 3d $\cN=2$ boundary conditions at each end. This corresponds to the overlap of boundary states in the putative supersymmetric quantum mechanics. A standard but crucial observation is that the momentum generator $P_s \propto \{ Q,Q^\dagger \}$ is exact with respect to both supercharges, and therefore acts trivially on the boundary states that are annihilated by $Q, Q^\dagger$. The correlation functions of boundary conditions are therefore independent of the position on the $s$-axis, and we can perform computations by reducing the distance between boundary conditions to zero and applying known localization computations for 3d supersymmetric gauge theories on $S^3_b$. To gain some familiarity with such computations, we will compute the correlation functions of the  Neumann and Dirichlet boundary conditions introduced in section~\ref{sec:boundary-def}. 

Let us first consider the overlap of a Neumann boundary condition and a Dirichlet boundary condition. For the overlap of $D_X(a)$ with $N_X$ or $D_Y(a)$ with $N_Y$, after sending the distance between the boundary conditions to zero, it is straightforward to see from the definitions~\eqref{eq:neu-def} and~\eqref{eq:dir-def} that there are no fluctuating degrees of freedom remaining on $S^3_b$ and therefore the partition functions are `$1$'. We write this as
\be
\langle D_X(a), N_X \rangle = 1\,,\qquad \langle D_Y(a), N_Y \rangle = 1  \, .
\ee
However, for the boundary conditions $D_Y(a)$ and $N_X$, the chiral multiplet $X$ has Neumann boundary conditions at both ends and therefore contributes to the correlation function. Similarly, $Y$ contributes to the correlation function of $D_X(a)$ and $N_Y$. We therefore have
\be
\la D_Y(a), N_X \rangle = K_X(a) \,,\qquad \la D_X(a), N_Y \rangle = K_Y(a) \, .
\ee
This is summarized in figure \ref{fig:basic-overlaps}.

Next consider the correlation function a pair of Dirichlet boundary conditions $D_X(a)$ and $D_Y(a')$. If $a \neq a'$, the boundary conditions are incompatible and the partition function should vanish. If $a = a'$, from equation~\eqref{eq:dir-def} we expect to get contributions from an adjoint 3d $\cN=2$ chiral multiplet of $T_R$ charge $0$ and $T_f$ charge $0$, which has Neumann boundary conditions at both ends. This would lead to the contribution
\be
S_b(0)^{N-1} \prod_{i \neq j}^N S_b(a_i-a_j) .
\ee
However, this expression is singular with a pole of order $N-1$ from the contribution $S_b(0)^{N-1}$ of the neutral scalars, indicating that a more careful analysis is needed. Note that there is a simple pole for each independent parameter, since $\sum_j a_j=0$. Further, recall that the $a_j$ are imaginary: $a_j = \ii r_j$, and that the residue of $S_b(\ii r)$ at $r=0$ is $\frac{1}{2\pi\ii}$.
We therefore replace the singular contribution by a Weyl invariant delta function,
\be
\Delta(a,a') = \frac{1}{N!} \sum_{\sigma \in S_N} \prod_{j=1}^N \delta(a_j - a_{\sigma(j)})\ ,
\label{eq:ND}
\ee
where $S_N$ is the set of permutations of $\{1,\ldots,N\}$. 
This delta function should be considered as a contour prescription around the aforementioned pole.
Using the identity $S_b(x)=1/S_b(Q-x)$, we therefore find
\be
\langle D_X(a), D_Y(a')\rangle  = \frac{1}{\nu(a)} \Delta(a,a')  \, .
\label{eq:DD}
\ee

This argument extends immediately to 
\be
\langle D_X(a), D_X(a')\rangle  = \frac{1}{\nu_X(a)} \Delta(a,a')\, , \qquad \langle D_Y(a), D_Y(a')\rangle  = \frac{1}{\nu_Y(a)} \Delta(a,a') \, ,
\label{eq:DD2}
\ee
where the additional contributions come respectively from the chiral multiplets $Y$ and $X$. It is straightforward to check that equations~\eqref{eq:DD} and~\eqref{eq:DD2} are compatible with the partition functions of other boundary conditions and the `cutting' construction introduced in section \ref{sec:cutting}.

Finally, let us consider the correlation function of a pair of Neumann boundary conditions. For $N_X$ with $N_Y$ we have a dynamical 3d $\cN=2$ vectormultiplet with partition function
\be
\langle N_X,N_Y \rangle = \int \frac{\mathrm{d}^{N-1}a}{\ii^{N-1} N!} \, \nu(a)  \, ,
\label{eq:NN}
\ee
where we defined $\rd^{N-1} a \equiv \rd a_1 \cdots \rd a_N ~\delta(a_1+\dots+a_N)$.
For a pair of $N_X$ or $N_Y$ boundary conditions we have additional adjoint chiral multiplets $X$ and $Y$ on the boundary, so that
\be
\langle N_X,N_X \rangle = \int \frac{\mathrm{d}^{N-1}a}{\ii^{N-1}N!} \, \nu_X(a)  \,,
\qquad \langle N_X,N_X \rangle = \int \frac{\mathrm{d}^{N-1}a}{\ii^{N-1}N!} \, \nu_Y(a) \, .
\label{eq:NN2}
\ee
These correspond to the partition functions of `bad' theories in the terminology of~\cite{Gaiotto:2008ak} and therefore formally diverge due to the presence of unitarity violating monopole operators~\cite{Kapustin:2010xq}. They can nevertheless be defined by analytic continuation, as explained in~\cite{Yaakov:2013fza}. 

\begin{figure}[htp]
\begin{center}
\begin{tikzpicture}
\draw[line width=0.8pt, gray] (-4,4) -- (-0.5,4);
\boundary at (-4,4)
\boundary at (-0.5,4)
\node at (-4,4.75) {$D_X$};
\node at (-0.5,4.75) {$N_X$};
\node[right] at (0.5,4) {{$= \qquad 1$}};
\draw[line width=0.8pt, gray] (-4,2) -- (-0.5,2);
\boundary at (-4,2)
\boundary at (-0.5,2)
\node at (-4,2.75) {$D_Y$};
\node at (-0.5,2.75) {$N_X$};
\node[right] at (0.5,2) {{$= \qquad K_X(a)$}};
\draw[line width=0.8pt, gray] (-4,0) -- (-0.5,0);
\boundary at (-4,0)
\boundary at (-0.5,0)
\node at (-4,0.75) {$D_X$};
\node at (-0.5,0.75) {$D_X$};
\node[right] at (0.5,0) {{$\displaystyle = \qquad\frac{\Delta(a,a')}{\nu_X(a)}$}};
\draw[line width=0.8pt, gray] (-4,-2) -- (-0.5,-2);
\boundary at (-4,-2)
\boundary at (-0.5,-2)
\node at (-4,-1.25) {$N_X$};
\node at (-0.5,-1.25) {$N_X$};
\node[right] at (0.5,-2) {{$\displaystyle = \qquad \int \frac{\rd^{N-1}a}{\ii^{N-1}N!}\, \nu_X(a)$}};
\end{tikzpicture}
\end{center}
\vspace{-0.5cm}
\caption{A sampling of the correlation functions of Neumann and Dirichlet boundary conditions.
There is an isomorphic set of functions obtained by interchanging $X \leftrightarrow Y$.} 
\label{fig:basic-overlaps}
\end{figure}

Finally, we note that these correlation functions are compatible with the `flip'. For example, the Dirichlet boundary condition $D_X$ is obtained from $D_Y$ coupled to a boundary chiral multiplet $\cO_Y$ with the same charges as $Y$ with the boundary superpotential $W = \tr (X|\cO_Y)$. Since the partition functions are independent of boundary superpotential couplings, we would therefore expect correlation functions of $D_X(a)$ to be obtained from those of $D_Y(a)$ by multiplying by the contribution $K_Y(a)$ from $\cO_Y$. Using the identity $K_X(a)K_Y(a)=1$, it is straightforward to verify that this is the case in the above examples. 


\subsection{Cutting the Interval} \label{sec:cutting}

Our strategy for computing a general correlation function $\la B_1,B_2\ra$ is to `cut' the path integral at an intermediate point and express the result in terms of the `wave functions' $\la B_1,D_X(a)\ra$ and $\la D_X(a),B_2\ra$ associated to the boundary conditions $B_1$ and $B_2$. It is therefore convenient to introduce a shorthand notation
\be
Z_{X,B}(a) = \la D_X(a) , B \ra \qquad Z_{Y,B}(a) = \la D_Y(a) ,B \ra \, .
\ee
The cutting construction can be performed using $D_X(a)$ or $D_Y(a)$ or a mixture of both, leading to considerable flexibility in notation.

Let us briefly recall the construction of the `identity' interface from section~\ref{sec:sl2z}. First, cut the interval at some intermediate point and impose the boundary condition $D_X(a)$ on the left and the boundary condition $D_X(a')$ on the right of the cut. Next, identify the boundary flavour symmetry on each side of the cut, forcing $a=a'$, and introduce a dynamical 3d $\cN=2$ vectormultiplet, together with a chiral multiplet $\cO_X$ and the boundary superpotential
\be
W = \tr \left( Y| \, \cO_X  - \cO_X \, |Y ' \right)
\ee
which identifies the chiral multiplets $X$ and $Y$ across the interface.

This construction is straightforward to implement at the level of partition functions: the boundary superpotential is exact and therefore makes no contribution. Hence, the result is
\be
\int  \rd \nu_X(a)\,  Z_{X,B_1}(a)  Z_{X,B_2}(a)
\label{eq:inner-prod}
\ee
where we introduce the shorthand notation
\be
\rd\nu_X(a) = \frac{\mathrm{d}^{N-1}a}{\ii^{N-1}N!} \nu_X(a)
\label{eq:meas-short}
\ee
for the measure of integration. This is illustrated in figure~\ref{fig:gluing}.

Although we will mostly concentrate on cutting the path integral using $D_X(a)$ boundary conditions, it is straightforward to provide a similar construction using $D_Y(a)$ boundary conditions, leading to the following equivalent expressions
\bea
\la B_1,B_2\ra  & = \int \rd \nu(a)\,  Z_{X,B_1}(a)  Z_{Y,B_2}(a)  \\
			& = \int \rd \nu_Y(a)\,  Z_{Y,B_1}(a)  Z_{Y,B_2}(a) \, ,
\label{eq:inner-prod-2}
\eea
where we introduce shorthand notations for the measures analogous to equation~\eqref{eq:meas-short}. These expressions are of course compatible since
\be
Z_{X,B}(a) = K_Y(a) Z_{Y,B}(a)\,, \qquad Z_{Y,B}(a) = K_X(a) Z_{X,B}(a) 
\ee
by performing a flip.

\begin{figure}[tp]
\begin{center}
\begin{tikzpicture}
\draw[line width=0.8pt, gray] (-7,0) -- (-4,0);
\boundary at (-7,0)
\boundary at (-4,0)
\node at (-7,0.75) {$B_-$};
\node at (-4,0.75) {$B_+$};
\node[right] at (-3.5,0) {{$\displaystyle = \quad \int \rd\nu_X(a)$}};
\draw[line width=0.8pt, gray] (-0.5,0) -- (2.5,0);
\boundary at (-0.5,0)
\boundary at (2.5,0)
\node at (-0.5,0.75) {$B_1$};
\node at (2.5,0.75) {$D_X$};
\draw[line width=0.8pt, gray] (3.2,0) -- (6.2,0);
\boundary at (3.2,0)
\boundary at (6.2,0)
\node at (3.2,0.75) {$D_X$};
\node at (6.2,0.75) {$B_2$};
\end{tikzpicture}
\end{center}
\vspace{-0.5cm}
\caption{The construction of a general correlation function $\la B_1,B_2\ra$ by inserting cutting the path integral and expressing the result in terms of the wave functions $Z_{X,B_1}(a)$ and $Z_{X,B_2}(a)$.}
\label{fig:gluing}
\end{figure}

Finally, it is straightforward to check that all of the correlation functions of Neumann and Dirichlet boundary conditions in section~\eqref{sec:basic-overlaps} are compatible with this procedure.


\subsection{Loop Operators}

Supersymmetric Wilson-'t Hooft operators can be inserted at points in the interval and on Hopf linked circles $S^1$ and $\tilde S^1$ of length $2\pi b$ and $2\pi/b$ in the squashed three-sphere $S^3_b$. This corresponds to the insertion of operators in the putative supersymmetric quantum mechanics annihilated by $Q$ or $Q^\dagger$. As before, their correlation functions are independent of the position on the $s$-axis. We will focus on supersymmetric loop operators wrapping $S^1$. 

It will be sufficient to determine the correlation function of a supersymmetric loop operator inserted between a Dirichlet boundary condition $D_X(a)$ or $D_Y(a)$ and a general boundary condition $B$. Results from supersymmetric localization imply this will act as a difference operator on the wave functions $Z_{X,B}(a)$ or $Z_{Y,B}(a)$. From these ingredients, more general correlation functions can be computed by cutting the path integral.

\subsubsection{Wilson Loops}

Let us first consider a supersymmetric Wilson loop in the representation $\Lambda^r$ inserted between a Dirichlet boundary condition $D_X(a)$ or $D_Y(a)$ and another boundary condition $B$. Moving the supersymmetric Wilson loop operator to the Dirichlet boundary condition, it is evaluated on the vacuum expectation value $A_j=0$ and $\Re(\phi) = a$. We therefore find
\be
W^{(r)}(a) Z_{X,B}(a) \qquad W^{(r)}(a) Z_{Y,B}(a)
\ee
where
\be
W^{(r)}(a)  = \sum_{|I|=r} e^{2\pi \ii b a_I} 
\ee
is the character of the representation $\Lambda^r$ and we write $a_I = \sum_{i\in I}a_i$. Note that if we define exponentiated variables $w_j = e^{2\pi \ii ba_j}$ this contribution concides with the expectation value of a supersymmetric Wilson loop from section~\ref{sec:flat-conn}. This is summarized in figure~\ref{fig:wl}.

\begin{figure}[htp]
\begin{center}
\begin{tikzpicture}
\draw[line width=0.8pt, gray] (-4.5,1) -- (-1,1);
\lineop at (-2.75,1)
\boundary at (-4.5,1)
\boundary at (-1,1)
\node at (-2.75,1.82) {$W^{(r)}$};
\node at (-4.5,1.75) {$D_X$};
\node at (-1,1.78) {$B$};
\node [right] at (0,1) {{$\displaystyle = \qquad W^{(r)}(a)\, Z_{X,B}(a)$}};
\draw[line width=0.8pt, gray] (-4.5,-1) -- (-1,-1);
\boundary at (-4.5,-1)
\boundary at (-1,-1)
\lineop at (-2.75,-1)
\node at (-2.75,-0.18) {$W^{(r)}$};
\node at (-4.5,-0.22) {$D_Y$};
\node at (-1,-0.25) {$B$};
\node [right] at (0,-1) {{$\displaystyle = \qquad W^{(r)}(a)\, Z_{Y,B}(a)$}};
\end{tikzpicture}
\end{center}
\vspace{-0.5cm}
\caption{Correlation functions with supersymmetric Wilson loops inserted.}
\label{fig:wl}
\end{figure}

The correlation function of a supersymmetric Wilson loop between any pair of boundary conditions $B_1$ and $B_2$ is then
\be
\int  \rd\nu_X(a) ~  Z_{X,B_1}(a) W^{(r)}(a)  Z_{X,B_2}(a) \, ,
\label{eq:inner-wl}
\ee
by cutting the path integral on either side of the supersymmetric Wilson loop insertion. As in equation~\eqref{eq:inner-prod-2}, there are equivalent expressions involving $D_Y(a)$ boundary conditions using $\rd\nu(a)$ and $\rd\nu_Y(a)$.

\subsubsection{'t Hooft loops}

Let us now move to supersymmetric 't Hooft loops. We first consider an 't Hooft loop in the antisymmetric tensor representation $\Lambda^r$ inserted between $D_X(a)$ or $D_Y(a)$ on the left and a boundary condition $B$ on the right. This correlation function is given by a difference operator acting on the original wave function,
\be
H^{(r)}_X(a) \la D_X(a) , B\ra\,, \qquad H^{(r)}_Y(a) \la D_Y(a) , B\ra \, .
\ee
The form of these difference operators can be determined from supersymmetric localization~\cite{Gomis:2011pf}. The result takes the following form\footnote{The localization results in~\cite{Gomis:2011pf} are for supersymmetric 't Hooft loops on $S^4$ supported on a circle $S^1\subset S^3$ where $S^3$ is the equator. In the neighbourhood of the equator, the background looks like our $\mathbb{R} \times S^3$. Since the contributions to the difference operator arise from 1-loop contributions localized at the equator, we expect these expressions to be correct also for our computation. A further conjugation is required to bring these operators into the form shown here~\cite{Bullimore:2014nla,Bullimore:2013xsa}.}
\bea
H^{(r)}_X(a) & =
\sum_{|I|=r} \, \prod_{i \in I , j\notin I} \frac{\sin\pi b(\ep+a_i-a_j)}{\sin\pi b(a_i-a_j)} h_I \,, \\
H^{(r)}_Y(a) & =
\sum_{|I|=r} \, \prod_{i \in I , j\notin I} \frac{\sin\pi b(\ep^*+a_i-a_j)}{\sin\pi b(a_i-a_j)}  h_I \, ,
\label{eq:thooft-op}
\eea
where
\be
h_i : a_j \mapsto a_j+ b(\delta_{ij} -\tfrac{1}{N})
\ee
are elementary difference operators preserving the constraint $\sum_ja_j=0$ and we have used the shorthand notation $h_{I} = h_{i_1}\cdots h_{i_r}$ for $I=\{i_1,\ldots,i_r\}$. The contributions in the numerators of these difference operators arise from 1-loop contributions from the chiral fields $X$ and $Y$ in the background of an 't Hooft loop, explaining the relative dependence on the combinations $\ep$ and $\ep^*$.

If we define exponentiated parameters
\be
w_j = e^{2\pi \ii ba_j} \qquad t = e^{2\pi \ii b \ep} \qquad q = e^{2\pi \ii b^2} \, ,
\ee
the difference operators become
\bea
H^{(r)}_X(a) & =
\sum_{|I|=r} \, \prod_{i \in I , j\notin I } \frac{t^{1/2}w_i-t^{-1/2}w_j}{w_i-w_j} h_I \,, \\
H^{(r)}_Y(a) & =
\sum_{|I|=r} \, \prod_{i \in I , j\notin I } \frac{(q/t)^{1/2}w_i-(q/t)^{-1/2}w_j}{w_i-w_j}  h_I \, .
\eea
In the `classical' limit $b\to0$, the difference operators $H^{(r)}_X(a)$ coincide with the Coulomb branch expectation values of supersymmetric 't Hooft loops in section~\ref{sec:flat-conn}, where the eigenvalues of the holonomy around the puncture are $\{t^{-1},\ldots,t^{-1},t^{N-1}\}$. On the other hand, the difference operators $H^{(r)}_Y(a)$ coincide with the expectations values of supersymmetric 't Hooft loops in a setup where the eigenvalues of the holonomy around the puncture are inverted to $\{t,\ldots,t,t^{1-N}\}$. 

This means that choosing to construct wave functions with $D_X(a)$ or $D_Y(a)$ correspond to quantizations of $SL(2,\Z)$ flat connections on $T^2/\{p\}$ with the holonomy eigenvalues at $p$ inverted. In what follow, we focus on constructing wave functions with $D_X(a)$, so that our formulae reduce directly to those in section~\ref{sec:flat-conn} in the `classical' limit $b^2\to0$.

\begin{figure}[htp]
\begin{center}
\begin{tikzpicture}
\draw[line width=0.8pt, gray] (-4.5,1) -- (-1,1);
\lineop at (-2.75,1)
\boundary at (-4.5,1)
\boundary at (-1,1)
\node at (-2.75,1.82) {$H^{(r)}$};
\node at (-4.5,1.75) {$D_X$};
\node at (-1,1.78) {$B$};
\node [right] at (0,1) {{$\displaystyle = \qquad H^{(r)}_X(a)Z_{X,B}(a)$}};
\draw[line width=0.8pt, gray] (-4.5,-1) -- (-1,-1);
\boundary at (-4.5,-1)
\boundary at (-1,-1)
\lineop at (-2.75,-1)
\node at (-2.75,-0.18) {$H^{(r)}$};
\node at (-4.5,-0.23) {$D_Y$};
\node at (-1,-0.25) {$B$};
\node [right] at (0,-1) {{$\displaystyle = \qquad H^{(r)}_Y(a)Z_{Y,B}(a)$}};
\end{tikzpicture}
\end{center}
\vspace{-0.5cm}
\caption{The insertion of a supersymmetric 't Hooft loop between a boundary conditions $D_X(a)$ and $B$ acts as a difference operator on the wave function $Z_{X,B}(a)$. }
\label{fig:tl}
\end{figure}

Let us now compute the partition function of a supersymmetric 't Hooft loop between any pair of boundary conditions $B_1$ and $B_2$ by cutting the interval to the left of the supersymmetric  't Hooft loop with $D_X(a)$ boundary conditions, 
\be
\int \rd\nu_X(a)~ Z_{X,B_1}(a)  \left[ \, H^{(r)}_X(a) Z_{X,B_2}(a) \, \right] \, .
\ee
Provided the wave functions $Z_{X,B_1}(a)$ and $Z_{X,B_2}(a)$ have no poles inside the region $|  \Re(a_j) |< b(1-\tfrac{1}{N})$, the difference operators obey the following conjugation property,
\be
\begin{split}
\int \rd\nu_X(a)~ & Z_{X,B_1}(a)  \left[ \, H^{(r)}_X(a) Z_{X,B_2}(a) \, \right] = \\
& \int \rd\nu_X(a) ~\left[ \, H^{(r)}_X(-a) Z_{X,B_1}(a) \, \right]  Z_{X,B_2}(a)  \, ,
\end{split}
\label{eq:conj}
\ee
which can be shown by suitably deforming the contour of integration and using the functional properties of the double sine function~\cite{Bullimore:2014awa}. The difference operator appearing on the right coincides with that of the 't Hooft loop in the conjugate representation,
\be
H^{(r)}_X(-a) = H^{(N-r)}_X(a) \, .
\ee

Compatibility with the freedom to cut the path integral at any point now requires that the partition function of an 't Hooft loop in the representation $\Lambda^r$ between a general boundary condition $B$ on the left and a Dirichlet boundary condition $D_X(a)$ or $D_Y(a)$ on the right is
\be
H^{(r)}_X(-a) \la B, D_X(a) \ra \, ,\qquad H^{(r)}_Y(-a) \la  B, D_Y(a)  \ra \, .
\ee
In other words, the 't Hooft loop acts on a left boundary condition by the difference operator for the conjugate representation. This is compatible with the prescription for left / right boundary conditions in the limit $b \to 0$ in section~\ref{sec:bc-hollag}.

Finally, the difference operators acting on wave functions constructed using $D_X(a)$ and $D_Y(a)$ are intertwined by the contribution from chiral multiplets $X$ and $Y$,
\be\label{eq:intertwined}
H^{(r)}_X(a) K_Y(a) = K_Y(a) H^{(r)}_Y(a)\, \qquad H^{(r)}_Y(a) K_X(a) = K_X(a) H^{(r)}_X(a)\, ,
\ee
which is a consequence of the identity
\be
K_Y(a) \, \left[ h_i  K_X(a) \right]= \prod_{\substack{j=1\\j \neq i}}^N\frac{\sin \pi b (\ep+a_i-a_j)}{\sin\pi b(\ep^* +a_i-a_j)} \, .
\ee
This ensures compatibility of the action of the difference operators with the flip: we can consistently cut the path integral using $D_X(a)$, $D_Y(a)$ or a mixture of both, even in the presence of supersymmetric 't Hooft loop insertions.

It is interesting to compute the correlation function of an 't Hooft loop between $D_X(a)$ and $N_Y$. In the absence of the 't Hooft loop, we have the wave function $\la D_X(a) , N_X \ra =1$. Therefore, we expect to reproduce the partition function of a supersymmetric quantum mechanics on $S^1$ for the degrees of freedom supported on the 't Hooft loop. Indeed, by the same computation as in equation~\eqref{eq:thooft-at-neu}, we find
\bea
\label{eq:PartFuncGaugedQM}
H^{(r)}_X(a) \cdot 1 = \sum_{\substack{ I \subset \{ 1,\ldots,N \} \\ |I|=r} } \prod_{\substack{i \in I \\ j\notin I} } \frac{\sin\pi b(\ep+a_i-a_j)}{\sin\pi b(a_i-a_j)}  = W^{(r)}(\rho\ep) \, .
\eea
As in section~\ref{sec:bc-hollag}, this coincides with the partition function of a gauged $\cN=4$ supersymmetric quantum mechanics on $S^1$ that flows to a sigma model to the Grassmannian $Gr(r,N)$, and gives the quantum dimension $\dim_{\tf}(\Lambda^r)$ of the representation $\Lambda^r$, where now $\tf=e^{2\pi \ii b \ep}$. 


\subsection{$SL(2,\Z)$ Interfaces}
\label{sec-duality-interface}

Let us first consider the $T$ transformation.
As discussed in section \ref{sec:sl2z}, this corresponds to the addition of a supersymmetric Chern-Simons term at level $+1$.
Moving the $T$ interface onto a Dirichlet boundary condition $D_X(a)$ of $D_Y(a)$ evaluates the supersymmetric Chern-Simons term at the expectation value $A_j=0$ and $\Re(\phi) = a$, leading to an insertion of
\be
T(a) = \exp\Big( - \ii \pi \sum_j a_j^2 \Big) \, .
\label{eq:cs-contribution}
\ee
The insertion of the $T$ interface between a pair of Dirichlet boundary conditions $D_X(a)$ and $D_X(a')$ is summarized in figure~\ref{fig:Tsandwich}. 

\begin{figure}[htp]
\begin{center}
\begin{tikzpicture}
\draw[line width=0.8pt, gray] (-4.5,0) -- (-1,0);
\interface at (-2.75,0)
\boundary at (-4.5,0)
\boundary at (-1,0)
\node at (-2.75,0.75) {$T$};
\node at (-4.5,0.75) {$D_X$};
\node at (-1,0.75) {$D_X$};
\node [right] at (0,0) {{$\displaystyle = \qquad \frac{\Delta(a,a')}{\nu_X(a)}T(a)$}};
\end{tikzpicture}
\end{center}
\vspace{-0.6cm}
\caption{The correlation function of the $T$ duality interface between a pair of Dirichlet boundary conditions $D_X(a)$ and $D_X(a')$.}
\label{fig:Tsandwich}
\end{figure}

As in section~\ref{sec:int}, this interface is characterized by Ward identities for supersymmetric loop operators, which translate into difference equations for the function $T(a)$. Wilson loops act multiplicatively and therefore commute with the interface. On the other hand, an 't Hooft loop becomes a mixed Wilson-'t Hooft loop upon translation through the interface. For the supersymmetric 't Hooft loop in the representation $\Lambda^r$, we find
\be\label{eq:HWdef}
H^{(r)}_X(a) \; T(a)  = q^{-\frac{r(N-r)}{2N}}  T(a) (W^{-1}H)^{(r)}_X(a)  
\ee
where
\be\label{eq:HWform}
(W^{-1}H)^{(r)}_X= \sum_{\substack{ I \subset \{ 1,\ldots,N \} \\ |I|=r} }  \left[ \prod_{i \in I , j\notin I } \frac{\sin\pi b(\ep+a_i-a_j)}{\sin\pi b(a_i-a_j)} \right]  e^{-2\ii\pi b a_I} h_I\, .
\ee
This difference operator corresponds to the expectation value of the mixed Wilson-'t Hooft loop given by $\tr_{\Lambda^r}(W^{-1}H)$ from section~\ref{sec:flat-conn}.

Analogously, we find
\be\label{eq:HWdef2}
H^{(r)}_X(a) \; T^{-1}(a)  = q^{\frac{r(N-r)}{2N}}  T^{-1}(a) (WH)^{(r)}_X(a)  
\ee
where
\be\label{eq:HWform2}
(WH)^{(r)}_X= \sum_{\substack{ I \subset \{ 1,\ldots,N \} \\ |I|=r} }  \left[ \prod_{i \in I , j\notin I } \frac{\sin\pi b(\ep+a_i-a_j)}{\sin\pi b(a_i-a_j)} \right]  e^{2\ii\pi b a_I} h_I\, .
\ee

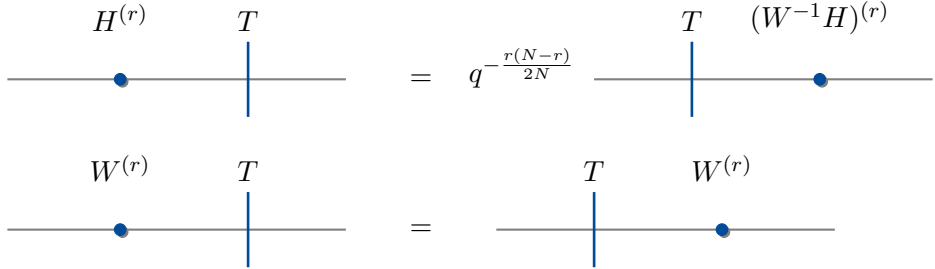
\begin{figure}[htp]
\begin{center}
\begin{tikzpicture}
\draw[line width=0.8pt, gray] (-5.5,1) -- (-1,1);
\draw[line width=0.8pt, gray] (2.3,1) -- (6.8,1);
\interface at (-2.3,1)
\interface at (3.6,1)
\lineop at (-4,1)
\lineop at (5.3,1)
\node at (-2.3,1.78) {$T$};
\node at (3.6,1.78) {$T$};
\node at (-4,1.82) {$H^{(r)}$};
\node at (5.3,1.82) {$(W^{-1}H)^{(r)}$};
\node at (0,1) {{$=$}};
\node [right] at (0.5,1.15) {$q^{-\frac{r(N-r)}{2N}} $};
\draw[line width=0.8pt, gray] (-5.5,-1) -- (-1,-1);
\draw[line width=0.8pt, gray] (1,-1) -- (5.5,-1);
\interface at (-2.3,-1)
\interface at (2.3,-1)
\lineop at (-4,-1)
\lineop at (4,-1)
\node at (-2.3,-0.22) {$T$};
\node at (2.3,-0.22) {$T$};
\node at (-4,-0.18) {$W^{(r)}$};
\node at (4,-0.18) {$W^{(r)}$};
\node at (0,-1) {{$=$}};
\end{tikzpicture}
\end{center}
\vspace{-0.5cm}
\caption{Translation of a supersymmetric 't Hooft loop through a $T$ interface generates a supersymmetric mixed Wilson-'t Hooft loop.}
\label{fig:Mixed_WH}
\end{figure}

We now consider the interface implementing the $S$ transformation.
As discussed in section \ref{sec:sl2z}, this is done by coupling to the theory $T(\mathfrak{g})$ at the interface.
Since the overlap between Neumann and Dirichlet boundary conditions is `$1$', the definition in section~\ref{sec:sl2z} makes it clear that the correlation function of the interface between Dirichlet boundary conditions $D_X(a)$ and $D_Y(a')$ reproduces the $S^3_b$ partition function $Z(a,a',\ep)$ of the theory $T(\gf)$ - as shown in figure~\ref{fig:Ssandwich}.

\begin{figure}[htp]
\centering
\includegraphics[height=3cm]{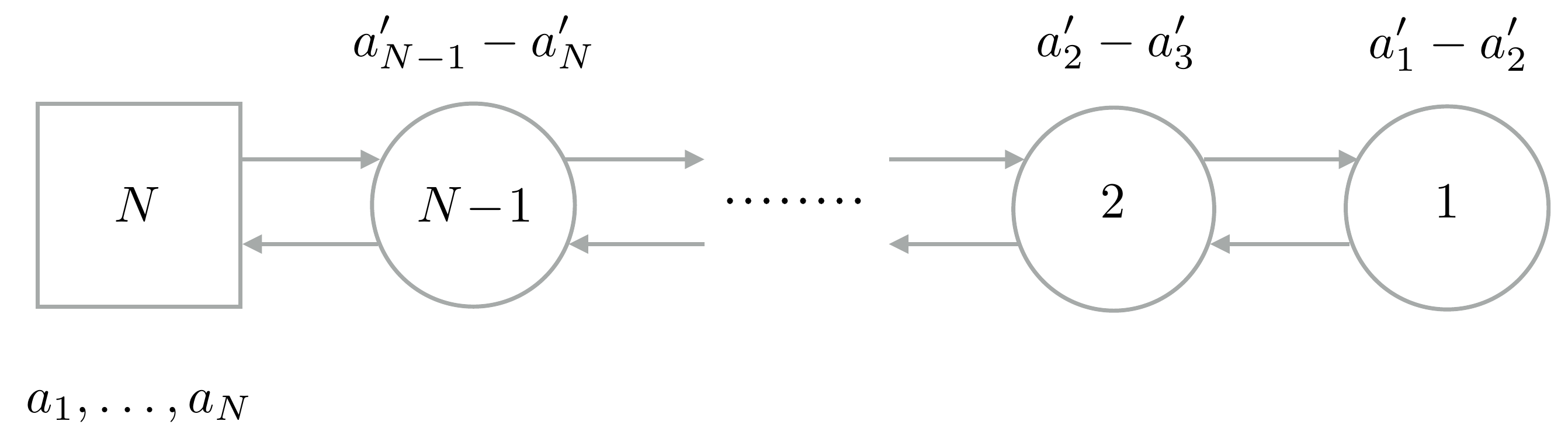}
\caption{A quiver description of $T(\mathfrak{g})$ with hypermultiplet mass parameters $(a_1,\ldots,a_N)$ and FI parameters labelled $\left( a_1^{'}-a_2', \ldots, a_{N-1}'-a_N'\right) $.}
\label{fig:TG2}
\end{figure}

The partition function $Z(a,a',\ep)$ can be constructed from the Lagrangian description of $T(\gf)$ shown in figure \ref{fig:TG2}. This leads to the following integral formula, 
\be
Z(a,a',\ep) = \int \prod_{n=1}^{N-1}  \rd\nu_X\left(a^{(n)}\right) Q_{n+1,n}\left(a^{(n+1)},a^{(n)}\right)e^{2\pi \ii\left(a'_n-a'_{n+1}\right)\left(a^{(n)}_1+\cdots+a^{(n)}_n\right)} \, .
\label{eq:s-integral}
\ee
Here we have introduced parameters $\{ a^{(n)}_1,\ldots,a^{(n)}_n \}$ valued in the Cartan subalgebra of $\uf(n)$ for $n=0,\ldots,N-1$, and by convention we define $\{ a_1,\ldots,a_N \} = \{ a^{(N)}_1,\ldots, a^{(N)}_N \}$ to be mass parameters at the final node. The FI parameter at the $n$-th node is $a'_n-a'_{n+1}$. Finally
\be
Q_{n+1,n}\left(a^{(n+1)},a^{(n)}\right) =\prod_{i=1}^{n+1}\prod_{j=1}^nS_b\left(\tfrac{\ep^*}{2}+a^{(n+1)}_i-a^{(n)}_j\right)S_b\left(\tfrac{\ep^*}{2}-a^{(n+1)}_i+a^{(n)}_j\right)
\ee
is the one-loop contribution to the partition function from the hypermultiplet in the bifundamental representation of $\uf(n+1) \oplus \uf(n)$.

The integral~\eqref{eq:s-integral} may be evaluated as a series expansion in $e^{2\pi \ii(a'_n-a'_{n-1})}$ by summing the contributions from the poles of the hypermultiplet contributions, see for example~\cite{Bullimore:2014awa}. However, the resulting expression is rather unwieldy. An exception is the limit $b=1$ and $\ep = 1$, in which the partition function reduces to a product of simple trigonometric functions~\cite{Benvenuti:2011ga}. Nevertheless, using the integral representation~\eqref{eq:s-integral}, it is possible to show that the partition function obeys the following properties:
\begin{itemize}
\item Mirror symmetry
\be
 Z(a,a',\ep) = Z(a',a,\ep^*) \, .
 \label{eq-mirror}
\ee
\item It has an analytic continuation away from imaginary $a,a'$ with simple poles at
\bea
 a_i - a_j  & = -\left(\ep^*+n_1 b+n_2 b^{-1}\right)\, , \\
 a_i' - a_j' & = -\left(\ep\,\,+n_1 b+n_2 b^{-1}\right)\, ,
  \label{eq-poles}
\eea
for all $i<j$ and $n_1,n_2 \in \mathbb{Z}_{\geq 0}$.
\item It is a simultaneous eigenfunction of 't Hooft loop difference operators
\bea
H^{(r)}_X(a) \cdot Z(a,a',\ep) & = W^{(r)}(a') \, Z(a,a',\ep)\, , \\ 
H^{(r)}_Y(a') \cdot Z(a,a',\ep)  & = W^{(r)}(a) \, Z(a,a',\ep)\, ,
\label{eq-diffeq}
\eea
with identical equations for supersymmetric loop operators wrapping the circle of length $2\pi /b$.
\end{itemize}

The first symmetry property~\eqref{eq-mirror} reflects the expectation that $T(\gf)$ is self-dual under three-dimensional mirror symmetry. This property has been proved in the case $N=2$ using the integral representation in reference~\cite{Hosomichi:2010vh}.

The analytic structure~\eqref{eq-poles} in the mass parameters $(a_1,\ldots,a_N)$ can be determined from the integral representation~\eqref{eq:s-integral} by analysing where the poles from the hypermultiplet contributions to the integrand collide and pinch the contour. The analytic structure in the FI parameters $(a_1', \ldots, a_N')$ is not simple to determine directly from the integral representation~\eqref{eq:s-integral} but can be determined from the analytic structure in $(a_1,\ldots,a_N)$ using the mirror symmetry property~\eqref{eq-mirror}.

\begin{figure}[htbp]
\begin{center}
\begin{tikzpicture}
\draw[line width=0.8pt, gray] (-5.5,1) -- (-1,1);
\draw[line width=0.8pt, gray] (1,1) -- (5.5,1);
\interface at (-2.3,1)
\interface at (2.3,1)
\lineop at (-4,1)
\lineop at (4,1)
\node at (-2.3,1.78) {$S$};
\node at (2.3,1.78) {$S$};
\node at (-4,1.82) {$H^{(r)}$};
\node at (4,1.82) {$W^{(r)}$};
\node at (0,1) {{\large $=$}};
\draw[line width=0.8pt, gray] (-5.5,-1) -- (-1,-1);
\draw[line width=0.8pt, gray] (1,-1) -- (5.5,-1);
\interface at (-2.3,-1)
\interface at (2.3,-1)
\lineop at (-4,-1)
\lineop at (4,-1)
\node at (-2.3,-0.22) {$S$};
\node at (2.3,-0.22) {$S$};
\node at (-4,-0.18) {$W^{(r)}$};
\node at (4,-0.18) {$H^{(N-r)}$};
\node at (0,-1) {{\large $=$}};
\end{tikzpicture}
\end{center}
\vspace{-0.5cm}
\caption{Under $S$ duality, a Wilson loop becomes an 't Hooft loop.}
\label{fig:WilsonHooft_Exchange}
\end{figure}
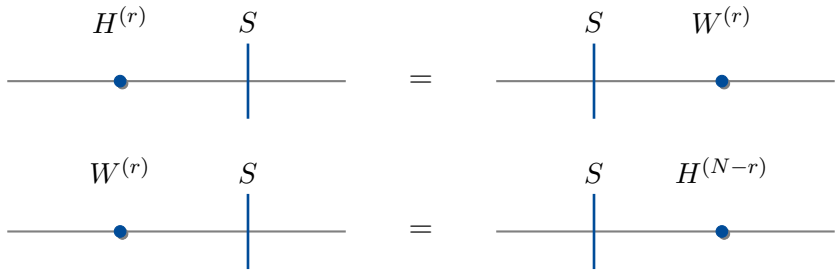

Finally, the difference equations encode the transformation properties of supersymmetric Wilson and 't Hooft loops under $S$-duality. This property can be proved by induction on $N$ using the various properties of the 't Hooft loop difference operators as shown in~\cite{Bullimore:2014nla,Bullimore:2014awa}.

\subsection{$SL(2,\Z)$ Relations}

We now want to check that above interfaces generate an action of $SL(2,\mathbb{Z})$ on the wave functions associated to boundary conditions. 

For this purpose, it is convenient to choose a uniform convention for cutting the path integral using the Dirichlet boundary conditions $D_X(a)$ and integrating using the measure $\rd\nu_X(a)$. The partition function $Z(a,a',\ep)$ obtained from sandwiching $S$ between the boundary conditions $D_X(a)$ and $D_Y(a)$ is therefore inconvenient with this choice. Instead, we will work with the partition function
\be\label{eq:curlySdef}
S_X(a,a') := Z(a,a',\ep) K_Y(a')
\ee
obtained from sandwiching the interface $S$ between boundary conditions $D_X(a)$ and $D_X(a)$. (For consistency, we could also define a function $S_Y(a,a')$ by sandwiching the interface in between boundary conditions $D_Y(a)$ and $D_Y(a')$, although we will not need it.) The origin of the two functions is summarized in figure \ref{fig:Ssandwich}.


\begin{figure}[htp]
\begin{center}
\begin{tikzpicture}
\draw[line width=0.8pt, gray] (-4.5,1) -- (-1,1);
\interface at (-2.75,1)
\boundary at (-4.5,1)
\boundary at (-1,1)
\node at (-2.75,1.78) {$S$};
\node at (-4.5,1.75) {$D_X$};
\node at (-1,1.75) {$D_Y$};
\node [right] at (0,1) {{$\displaystyle = \qquad Z(a,a',\ep)$}};
\draw[line width=0.8pt, gray] (-4.5,-1) -- (-1,-1);
\boundary at (-4.5,-1)
\boundary at (-1,-1)
\interface at (-2.75,-1)
\node at (-2.75,-0.23) {$S$};
\node at (-4.5,-0.25) {$D_X$};
\node at (-1,-0.25) {$D_X$};
\node [right] at (0,-1) {{$\displaystyle = \qquad S_X(a,a')$}};
\end{tikzpicture}
\end{center}
\vspace{-0.5cm}
\caption{The partition function of $S$ duality interface between a pair of Dirichlet boundary conditions $D_X(a)$ and $D_Y(a')$.}
\label{fig:Ssandwich}
\end{figure}


Using the analytic properties of the functions $K_X(a)$, $K_Y(a)$ and their intertwining property with respect to the difference operators $H^{(r)}_X(a)$, $H^{(r)}_Y(a)$, we find that the function $S_X(a,a')$ has the following properties:
\begin{itemize}
\item Mirror symmetry
\be
S_X(a,a') = S_X(a',a) \, .
\label{eq-mirror-2}
\ee
\item It has an analytic continuation away from imaginary $a,a'$ with simple poles at
\bea
 a_i - a_j  & = -\left(\ep^*+n_1 b+n_2 b^{-1}\right)\, , \\
 a_i' - a_j' & = -\left(\ep^*+n_1 b+n_2 b^{-1}\right)\, ,
  \label{eq-poles-2}
\eea
for all $i<j$ and $n_1,n_2 \in \mathbb{Z}_{\geq 0}$.
\item Simultaneous eigenfunction of 't Hooft loop difference operators
\bea
H^{(r)}_X(a) \cdot S_X(a,a') & = W^{(r)}(a') S_X(a,a')\, , \\ 
H^{(r)}_X(a') \cdot S_X(a,a')  & = W^{(r)}(a) S_X(a,a')\, ,
\label{eq-diffeq-2}
\eea
with identical equations for supersymmetric loop operators wrapping the circle of length $2\pi /b$. 
\end{itemize}

We now want to show that the concatenation of our kernels $S_X(a,a')$ and $T(a)$ with respect to the measure $\nu_X(a)$ defines a representation of $SL(2,\mathbb{Z})$. The standard relations $S^2=P$ and $(ST)^3=P$ correspond to 
the following equations 
\be
\int \rd\nu_X(a) \,  S_X(a,a') S_X(a',a'') = \frac{1}{N} \frac{\Delta(a,-a'')}{ \nu_X(a)}
\label{eq-S^2}
\ee
and
\be
\int  \rd\nu_X(a')\,  S_X(a,a')  T(a')  S_X(a',a'')  =  \zeta T^{-1}(a) S_X(a,a'')  T^{-1}(a'') \, . 
\label{eq-STcubed}
\ee
At the level of partition functions, $P$ corresponds to the replacement $a \to -a$. There is an additional constant contribution
\be
\zeta = \frac{1}{\sqrt{N}}e^{\frac{\ii \pi}{4}((N-1)+N(N-1)\ep\ep^*) } \, ,
\ee 
which is expected to be the contribution of a decoupled topological sector. This is a familiar feature from the $SL(2,\mathbb{Z})$ action of three-dimensional quantum field theories with abelian flavour symmetries~\cite{Witten:2003ya}.

We can prove the relation $S^2=P$ by inserting a supersymmetric 't Hooft loop in between the $S$ transformation interfaces. 
Using the eigenfunction property~\eqref{eq-diffeq-2} and the conjugation property~\eqref{eq:conj} we find that
\be
\left(W^{(r)}(a) - W^{(r)}(-a'')\right) \int\rd\nu_X(a)  \, S_X(a,a')  S_X(a',a'') = 0.
\ee
A similar equations applies for supersymmetric Wilson loops wrapping the circle of radius $2\pi/b$. This implies that the integral vanishes unless $a=-a''$ and is therefore proportional to a Weyl-invariant delta function. A simple way to determine the particular normalization in~\eqref{eq-S^2} is to examine the limit $b \to1$ and $m\to0$, where everything reduces to trigonometric functions~\cite{Benvenuti:2011ga}.

In section~\ref{sec:S3part}, we will perform an explicit check of the relation $(ST)^3=P$ for the specific values $a = \rho \ep$ and $a' = \rho \ep$ by equating two different ways to compute the partition function associated to the 3-manifold $M_3=S^3$ by surgery. In particular, by analysing the asymptotics of this formula as $\ep\to\infty$ with $\Im\ep>0$, this will allow us to determine the additional factor $\zeta$.

The extraneous factors of $\zeta$ and $\sqrt{N}$ can always be removed from the formulae~\eqref{eq-S^2}-\eqref{eq-STcubed} by rescaling the transformation functions $T(a)$ and $S_X(a,a')$. In particular, we can define the `dressed' functions
\bea
\cT(a)  = e^{-\frac{\ii\pi}{12}(N-1+N(N+1) \ep \ep^*)} T(a) \qquad
\cS_X(a,a')  = \sqrt{N}S_X(a,a') 
\eea
such that
\be
\int \rd\nu_X(a) \,  \cS_X(a,a') \cS_X(a',a'') = \frac{\Delta(a,-a'')}{ \nu_X(a)}
\label{eq-S^2-2}
\ee
and
\be
\int  \rd\nu_X(a')\,  \cS_X(a,a')  \cT(a')  \cS_X(a',a'')  = \cT^{-1}(a) \cS_X(a,a'')  \cT^{-1}(a'') \, . 
\label{eq-STcubed-2}
\ee
We will work in the rest of the paper with the functions $\cS_X(a,a')$ and $\cT(a)$, which satisfy the $SL(2,\mathbb{Z})$ relations exactly.

In particular, the dressed transformation $\cT(a)$ can be written in terms of quantities that are particularly natural in Toda conformal field theory of type $A_{N-1}$,
\be\label{eq:DressedT}
\cT(a) = \exp\Big( \Delta(\al) - \frac{c}{24} - \frac{\Delta_\ep}{12} \Big)
\ee
where
\begin{itemize}
\item $\Delta(\al) = ( \al , 2 Q\rho - \al)/2$ is the conformal dimension of a non-degenerate representation of the $W_N$-algebra corresponding to momentum $\al = Q\rho-a$ around the $(1,0)$ cycle of $T^2/\{p\}$.
\item $\Delta_\ep = \Delta(N\ep\omega_{N-1})$ is the conformal dimension of a semi-degenerate representation of the $W_N$-algebra associated to the puncture on $T^2/\{p\}$ with momentum $N \ep \omega_{N-1}$ where $\omega_j$ are the fundamental weights of $\gf=\suf(N)$.
\item $c = (N-1)+N(N^2-1)Q^2 $ is the standard parameterization of the central charge of the $W_N$-algebra.
\end{itemize}
The appearance of $A_{N-1}$ Toda conformal field theory is consistent with the proposal that we are constructing partition functions of Chern-Simons theory with complex gauge group $SL(N,\C)$ on 3-manifolds with boundary. It would be interesting to understand how to provide a concrete justification for the addition of these factors from the viewpoint of correlation functions of 3d $\cN=2$ boundary conditions and interfaces.

The $SL(2,\Z)$ relations allow us to derive how mixed Wilson-'t Hooft loops defined in \eqref{eq:HWform} and \eqref{eq:HWform2} transform under $S$ transformations.
An 't Hooft loop $H^{(r)}$ through the combination of interfaces $SPT^{-1}ST^{-1}$ becomes $q^{\frac{r(N-r)}{2N}}(WH)^{(r)}$.
On the other hand, the combination of interfaces above simply corresponds to $TS$, which leads to $q^{\frac{-r(N-r)}{2N}}(W^{-1}H)^{(r)}$ acting on $S$. The relation just found corresponds to the following equation
\begin{equation}\label{eq:BehaviourMixedWilsontHooft}
(W^{-1}H)^{(r)}(a)\cS_X(a,a')=\cS_X(a,a')(WH)^{(r)}(a') q^{\frac{r(N-r)}{N}} \, .
\end{equation}
Similarly, moving an 't Hooft loop $H^{(r)}$ through the $SL(2,\Z)$ interface $STST=T^{-1}S$, we find the following relation
\begin{equation}\label{eq:BehaviourMixedWilsontHooft2}
(WH)^{(r)}(a)\cS_X(a,a')=\cS_X(a,a')(W^{-1}H)^{(N-r)}(a') q^{-\frac{r(N-r)}{N}} \, .
\end{equation}
These formulae will be important for computing the parition function associated to an unkot and Hopf link in $S^3$ in section \ref{sec:CaseStudyTS3}.


\subsection{Boundary Conditions Revisited} \label{sec:boundrevis}

Now that we have constructed the partition functions of interfaces generating $SL(2,\mathbb{Z})$ duality transformations, we can in principle compute the partition functions involving boundary conditions in the $SL(2,\mathbb{Z})$ orbits of the basic Neumann and Dirichlet boundary conditions introduced in section~\ref{sec:bps-bound}.

In particular, we will define the Nahm pole boundary condition such that the Neumann boundary condition $N_X$ is the $S$ transformation of $NP_X$. The wave functions $Z_{X,NP_X}(a) = \la D_X(a),NP_X \ra $ for the Nahm pole boundary condition and $Z_{X,N_X}(a) = 1$ are then related by
\be
1 = \int \rd\nu_X(a)\, \cS_X(a',a) Z_{X,NP_X}(a) \, ,
\label{eq:nahmtoneum}
\ee
and its inverse
\be
Z_{X,NP_X}(a) = \int \rd\nu_X(a')\, \cS_X(-a,a') \, .
\label{eq:neumto nahm}
\ee
We will not need an explicit expression for the Nahm pole wave function $Z_{X,NP_X}$, as we can rely the following property. By inserting a supersymmetric 't Hooft loop between the interface and the Neumann boundary condition and using the conjugation property~\eqref{eq:conj}, the eigenfunction equation~\eqref{eq-diffeq-2} and $H_X^{(r)}(a')\cdot 1 = W^{(r)}(\rho\ep)$, we find 
\be
(W^{(r)}(a) - W^{(r)}(\rho\ep))Z_{X,NP_X}(a) = 0 \, ,
\label{eq:nahm-delta}
\ee
with an identical equation for supersymmetric loop operators wrapping the circle of length $2\pi /b$. This implies $Z_{X,NP_X}(a)$ vanishes in the physical regime where $a$ is imaginary. We can define the wave function by analytic continuation, although its detailed form will not be needed. The important point is that, due to~\eqref{eq:nahm-delta}, we can replace $a\to\rho\ep$ in any invariant function $f(a)$ multiplying the wave function $Z_{X,NP_X}(a)$.


\section{Case Study: $\cT\left(S^3\right)$}
\label{sec:CaseStudyTS3}

We will now apply the results of the previous section to the computation of the $S^3_b$ partition function of the 3d $\cN=2$ theory associated to $S^3$, $\cT\left(S^3\right)$. In addition, we compute the partition function of $\cT(S^3)$ in the presence of loop operators corresponding to the unknot and the Hopf link in $S^3$ labelled by antisymmetric tensor representations of $SL(N,\C)$ by adding supersymmetric Wilson-'t Hooft loops in the surgery construction. In this way, we will recover an analytic continuation of the $S$-matrix of refined Chern-Simons theory introduced in~\cite{Aganagic:2012ne,Aganagic:2011sg}.


\subsection{$S^3$ Partition Function} \label{sec:S3part}

The simplest way to construct the three-manifold $S^3$ by surgery is to identify the boundaries of two solid tori $D^2\times S^1$ by an $SL(2,\Z)$ transformation $\phi = S$. Using solid tori obtained by contracting the $(1,0)$ cycle of the boundary $T^2$, this corresponds to computing the correlation function of the $S$ interface between a pair of Nahm pole boundary conditions $NP_X$. Equivalently, it corresponds to the correlation function of a Nahm pole boundary condition $NP_X$ and a Neumann boundary condition $N_X$. The partition function of $\cT(S^3)$ can therefore be expressed as
\bea
Z_{\cT(S^3)} & = \int \rd \nu_X(a) \rd \nu_X(a') ~ Z_{X,NP_X}(a) \cS_X(a,a') Z_{X,NP_X}(a')  \\
& = \int \rd \nu_X(a)\, Z_{X,NP_X}(a) \, .
\eea

We can evaluate the integral in the second line without requiring the form of the Nahm pole wave function $Z_{X,NP_X}(a)$. We start from the relation between the Neumann and Nahm pole wave functions~\eqref{eq:nahmtoneum} and consider the limit as $a'\to \rho \ep$. We claim that the function $\cS_X(a',a)$ remains finite in this limit and is independent of $a$. In particular, from the eigenfunction equation~\eqref{eq-diffeq-2}, we find
\be
H^{(r)}_X(a) \cS_X(\rho\ep,a) = W^{(r)}(\rho\ep) \cS_X(\rho\ep,a) \, .
\ee
for all $r=1,\ldots,N-1$ with a similar equation for supersymmetric loop operators wrapping the circle of length $2\pi/b$. This implies that the function $\cS_X(\rho\ep,a)$ is independent of $a$. An explicit computation using the perturbative expansion of the function $\cS_X(a,a')$ in powers of $e^{2\pi \ii(a'_n-a'_{n-1})}$ is consistent with this argument and demonstrates that in fact
\be
\cS_X(\rho\ep,a) =  \sqrt{N}\prod_{j=2}^N S_b(j\ep)^{-1} \, .
\ee
The computation is performed in Appendix~\ref{app:DetailsTwoDefects}. Therefore, we find
\be
Z_{\cT(S^3)} = \frac{1}{\sqrt{N}}\prod_{j=2}^N S_b(j\ep) \, .
\label{eq:ZS3withoutdefects}
\ee
Apart from the $1 / \sqrt{N}$ factor out front, this expression coincides with the partition function of $(N-1)$ chiral multiplets with $T_R$ charges $2,\ldots,N$ and $T_f$ charges $2,\ldots,N$.

There is an alternative surgery construction of the partition function of $\cT(S^3)$, which is related to the computation above by following the sequence of operations shown in figure~\ref{fig:ChainS3}. The starting point for this computation is the correlation function of the $S$ interface between a pair of Nahm poles $NP_X$. The next step is to note that the interface $T$ acts on the Nahm pole wave function $Z_{X,NP_X}(a)$ by multiplying by
\be 
\cT(\rho\ep) = \exp\left(-\frac{\pi\ii }{12} N(N^2-1)\ep^2 - \frac{\pi\ii}{12}(N-1)(1+N\ep^*\ep)\right)
\ee
as a consequence of equation~\eqref{eq:nahm-delta}. We can therefore insert a pair of $T^{-1}$ interfaces at the expense of a framing factor $\cT(\rho\ep)^2$. Next, applying the relation~\eqref{eq-STcubed-2} and using the resulting $S$ interfaces to convert the Nahm pole boundary condition to Neumann boundary conditions, we arrive at the final line in figure~\ref{fig:ChainS3}.

\begin{figure}[tbp]
\begin{center}
\begin{tikzpicture}
\draw[line width=0.8pt, gray] (-6,1) -- (-1.5,1);
\interface at (-3.75,1)
\boundary at (-6,1)
\boundary at (-1.5,1)
\node at (-3.75,1.78) {$\cS$};
\node at (-6,1.75) {$NP_X$};
\node at (-1.5,1.75) {$NP_X$};
\node [right] at (-0.9,1) {{$=$}};
\node [right] at (-0.25,1.1) {$\cT(\rho\ep)^2$};
\draw[line width=0.8pt, gray] (1.5,1) -- (6,1);
\boundary at (1.5,1)
\boundary at (6,1)
\interface at (2.625,1)
\interface at (4.875,1)
\interface at (3.75,1)
\node at (2.63,1.78) {$\cT^{-1}$};
\node at (3.75,1.78) {$\cS$};
\node at (4.875,1.78) {$\cT^{-1}$};
\node at (1.5,1.75) {$NP_X$};
\node at (6,1.75) {$NP_X$};
\draw[line width=0.8pt, gray] (-6,-1) -- (-1.5,-1);
\interface at (-4.875,-1)
\interface at (-2.625,-1)
\interface at (-3.75,-1)
\boundary at (-6,-1)
\boundary at (-1.5,-1)
\node [right] at (-8.4,-1) {{$=$}};
\node [right] at (-7.75,-0.9) {$\cT(\rho\ep)^2$};
\node at (-4.875,-0.22) {$\cS$};
\node at (-2.625,-0.22) {$\cS$};
\node at (-3.75,-0.21) {$\cT$};
\node at (-6,-0.25) {$NP_X$};
\node at (-1.5,-0.25) {$NP_X$};
\node [right] at (-0.9,-1) {{$=$}};
\node [right] at (-0.25,-0.9) {$\cT(\rho\ep)^2$};
\draw[line width=0.8pt, gray] (1.5,-1) -- (6,-1);
\boundary at (1.5,-1)
\boundary at (6,-1)
\interface at (3.75,-1)
\node at (3.75,-0.21) {$\cT$};
\node at (1.5,-0.25) {$N_X$};
\node at (6,-0.25) {$N_X$};
\end{tikzpicture}
\end{center}
\vspace{-0.5cm}
\caption{The sequence of moves relating the different surgery constructions of $\cT\left(S^3\right)$.}
\label{fig:ChainS3}
\end{figure}

Therefore, modulo framing, $\cT(S^3)$ can also be constructed from the $T$ interface sandwiched between a pair of Neumann boundary conditions $N_X$, leading to a description in terms of a supersymmetric Chern-Simons theory at level $+1$ and a chiral multiplet with the same charges as $X$. The sequence of moves shown in figure~\ref{fig:ChainS3} translates into concrete expressions at the level of partition functions, 
\bea
\label{eq:ZS3NTinvN}
Z_{\cT(S^3)} = \cT(\rho\ep)^2 Z'_{\cT(S^3)} \qquad Z'_{\cT(S^3)} = \int \da{}\, \cT(a) \, .
\eea
In Appendix~\ref{app:DetailsTwoDefects}, we check agreement of the asymptotic behaviour of both sides of this equation in the limit $\ep \to \infty$ with $\Im(\ep)>0$. In particular, this asymptotic analysis determines the framing factor $T(\rho\ep)^2$ in equation~\eqref{eq:ZS3NTinvN} exactly, which furthermore determines the coefficient $\zeta$ in the $SL(2,\Z)$ relations~\eqref{eq-STcubed}.

We therefore find that $\cT(S^3)$ is a supersymmetric $SU(N)$ Chern-Simons theory at level $+1$ together with an a chiral multiplet in the adjoint representation, as proposed in \cite{Gukov:2015sna}. In our construction, the adjoint chiral multiplet comes naturally with the same $T_R$ charge as $X$, namely $+1$. However, at the level of partition functions this can be modified by analytic continuation in the mass parameter $m$ for the $T_f$ symmetry.
The equivalence with $(N-1)$ chiral multiplets together with a decoupled topological sector is a known three-dimensional duality~\cite{Jafferis:2011ns,Kapustin:2011vz}. 

Let us briefly consider the special case $N=2$. The equivalence between the supersymmetric Chern-Simons and $(N-1)$ chiral multiplet descriptions~\eqref{eq:ZS3NTinvN} is equivalent to the following integral identity,
\be
\label{eq:ZS3A1claim}
\int_{\gamma_{\ep}} \frac{\rd a}{2\ii} S_b(\epsilon) \frac{S_b(\ep + 2a) S_b(\ep-2a)}{S_b(2a)S_b(-2a)}e^{-2\pi \ii a^2} 
 = \frac{1}{\sqrt{2}}e^{\frac{\pi\ii}{4}}e^{\frac{\pi\ii}{2}\epsilon(Q+\epsilon)}S_b(2\epsilon)
\ee
where $\gamma_{\ep}$ is a suitably deformed contour from supersymmetric localization, which satisfies $\gamma_{\ep}= \ii\R$ in the physical region, where $\Re(\ep)>0, \Im(\ep)>0$. In the limit $b\to 1$ of a round three-sphere, this reproduces the result checked numerically in \cite{Jafferis:2011ns}, with $\epsilon$ analytically continued from $\Delta \in (0,\infty)$. 

Finally, in the limit that we remove the mass parameter $m\to 0$ for $\uf(1)_f$ and set the $T_R$ charge $r$ to an even integer, the partition function \eqref{eq:ZS3NTinvN} vanishes.
This supports the expectation that, due to the absence of $SL(N,\C)$ flat connections on $S^3$ without monodromy defects, supersymmetry is spontaneously broken in $\cT\left(S^3\right)$. 


\subsection{Unknot in $S^3$}

Let us now consider adding a single codimension-4 defect of the $\cN=(2,0)$ theory, labelled by an antisymmetric tensor representation of rank $r$, wrapping $S^1 \subset S^3$. In the construction of $S^3$ by gluing two solid tori $S^1 \times D_2$ with an $S$ transformation, this corresponds to adding a codimension-4 defect at the origin of $D_2$ in one of the solid tori. 

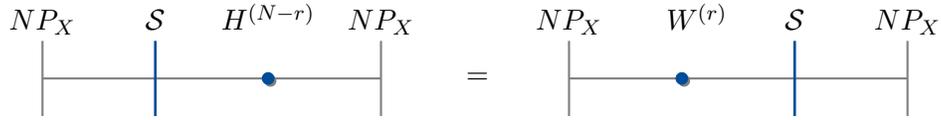
\begin{figure}[tbp]
\begin{center}
\begin{tikzpicture}
\draw[line width=0.8pt, gray] (-5.5,1) -- (-1,1);
\interface at (-4,1)
\boundary at (-5.5,1)
\boundary at (-1,1)
\lineop at (-2.5,1)
\node at (-2.5,1.82) {$H^{(N-r)}$};
\node at (-4,1.78) {$\cS$};
\node at (-5.5,1.75) {$NP_X$};
\node at (-1,1.75) {$NP_X$};
\node [right] at (0,1) {{$=$}};
\draw[line width=0.8pt, gray] (1.5,1) -- (6,1);
\boundary at (1.5,1)
\boundary at (6,1)
\lineop at (3,1)
\interface at (4.5,1)
\node at (3.2,1.82) {$W^{(r)}$};
\node at (4.5,1.78) {$\cS$};
\node at (1.5,1.75) {$NP_X$};
\node at (6,1.75) {$NP_X$};
\end{tikzpicture}
\end{center}
\vspace{-0.5cm}
\caption{The computations leading to an unknot in $S^3$.}
\label{fig:OneDefectS3setup}
\end{figure}

This corresponds to the correlation function of a supersymmetric 't Hooft loop in the representation $\Lambda^{N-r}$ in between a Neumann boundary condition $N_X$ and a Nahm pole boundary condition $NP_X$. This can be evaluated by moving the supersymmetric 't Hooft loop through the $S$ interface to become a supersymmetric Wilson loop, as shown on the right of figure~\ref{fig:OneDefectS3setup}. This contributes $W^{(r)}(a)$ to the integrand, which should be evaluated at $a = \rho \ep$ since it multiplies the Nahm pole wave function:
\be
\frac{Z_{\cT(S^3)}(\omega_r)}{Z_{\cT(S^3)} }  = W^{(r)}(\rho\ep)  = W^{(N-r)}(\rho\ep)\, .
\ee
In terms of the exponentiated variable, $t = e^{2\pi \ii b \epsilon}$, we have
\begin{equation}
\frac{Z_{S^3}(\omega_r)}{Z_{S^3}}  = \dim_t(\Lambda^r) = \dim_t(\Lambda^{N-r}) \,,
\end{equation}
which is the quantum dimension of the representation $\Lambda^r$ or $\Lambda^{N-r}$, with quantum parameter $t$. We also recognize this result as an analytic continuation of $S_{r,0}$, where $S_{r,s}$ is the $S$-matrix of the refined Chern-Simons theory from \cite{Aganagic:2011sg}.

\begin{figure}[tbp]
\begin{center}
\begin{tikzpicture}
\draw[line width=0.8pt, gray] (-7.5,1) -- (-3,1);
\interface at (-4.5,1)
\lineop at (-6,1)
\boundary at (-7.5,1)
\boundary at (-3,1)
\node at (-4.5,1.78) {$\cS$};
\node at (-5.7,1.82) {$W^{(r)}$};
\node at (-7.5,1.75) {$NP_X$};
\node at (-3,1.75) {$NP_X$};
\node [right] at (-2.4,1) {{$=$}};
\node [right] at (-1.75,1.1) {$\cT(\rho\ep)^2$};
\draw[line width=0.8pt, gray] (0,1) -- (5.8,1);
\boundary at (0,1)
\boundary at (5.8,1)
\lineop at (1.25,1)
\interface at (2.5,1)
\interface at (3.6,1)
\interface at (4.7,1)
\node at (1.27,1.82) {$W^{(r)}$};
\node at (2.5,1.82) {$\cT^{-1}$};
\node at (3.6,1.78) {$\cS$};
\node at (4.7,1.82) {$\cT^{-1}$};
\node at (0,1.75) {$NP_X$};
\node at (5.8,1.75) {$NP_X$};
\draw[line width=0.8pt, gray] (-1,-1) -- (-6.8,-1);
\boundary at (-1,-1)
\boundary at (-6.8,-1)
\lineop at (-5.55,-1)
\interface at (-4.3,-1)
\interface at (-3.2,-1)
\interface at (-2.1,-1)
\node [right] at (-9.2,-1) {{$=$}};
\node [right] at (-8.55,-0.9) {$\cT(\rho\ep)^2$};
\node at (-5.45,-0.18) {$W^{(r)}$};
\node at (-4.3,-0.22) {$\cS$};
\node at (-3.2,-0.22) {$\cT$};
\node at (-2.1,-0.22) {$\cS$};
\node at (-1,-0.25) {$NP_X$};
\node at (-6.8,-0.25) {$NP_X$};
\node [right] at (-0.4,-1) {{$=$}};
\node [right] at (0.35,-0.9) {$\cT(\rho\ep)^2$};
\draw[line width=0.8pt, gray] (2,-1) -- (5.7,-1);
\boundary at (2,-1)
\boundary at (5.7,-1)
\lineop at (3.23,-1)
\interface at (4.56,-1)
\node at (3.23,-0.18) {$H^{(N-r)}$};
\node at (4.56,-0.21) {$\cT$};
\node at (2,-0.25) {$N_X$};
\node at (5.7,-0.25) {$N_X$};
\draw[line width=0.8pt, gray] (1.7,-3) -- (-2.5,-3);
\boundary at (1.7,-3)
\boundary at (-2.5,-3)
\interface at (-1.57,-3)
\lineop at (-0.19,-3)
\node [right] at (-6.2,-3) {{$=$}};
\node [right] at (-5.55,-2.9) {$\displaystyle \cT(\rho\ep)^2\, q^{-\frac{r(N-r)}{2N}}$};
\node at (-1.57,-2.18) {$\cT$};
\node at (0,-2.21) {$(W^{-1}H)^{(N-r)}$};
\node at (1.7,-2.25) {$N_X$};
\node at (-2.5,-2.25) {$N_X$};
\node [right] at (-6.6,-5) {{$=$}};
\node [right] at (-5.95,-4.9) {$\cT(\rho\ep)^2\, q^{-\frac{r(N-r)}{2N}}\, t^{-\frac{r(N-r)}{2}}$};
\draw[line width=0.8pt, gray] (-1.5,-5) -- (2.5,-5);
\boundary at (-1.5,-5)
\boundary at (2.5,-5)
\lineop at (1.16,-5)
\interface at (-0.17,-5)
\node at (1.16,-4.18) {$W^{(r)}$};
\node at (-0.17,-4.21) {$\cT$};
\node at (-1.5,-4.25) {$N_X$};
\node at (2.5,-4.25) {$N_X$};
\end{tikzpicture}
\end{center}
\vspace{-0.5cm}
\caption{Sequence of moves to evaluate $Z_{S^3}(\omega_r)$.}
\label{fig:OneDefectS3eval}
\end{figure}

As before, we can express the same result in the alternative framing of $S^3$ by performing the sequence of operations shown in figure~\ref{fig:OneDefectS3eval}. At the final stage, modulo a factor $q^{-\frac{r(N-r)}{2N}}$ from equation~\eqref{eq:HWdef} from translating a supersymmetric 't Hooft loop through $T$, we find the correlation function of $T$ and $(W^{-1}H)^{(N-r)}$ in between a pair of Neumann boundary conditions $N_X$. The action of the mixed Wilson-'t Hooft loop on $N_X$ is
\begin{equation}
\left(W^{-1}H\right)^{(N-r)}(a) \cdot 1 = t^{-\frac{r(N-r)}{2}}W^{(r)}(a)\, .
\end{equation}
Thus we conclude that
\be
Z_{S^3}(\omega_r) = \theta_r^{-1}\cT(\rho\ep)^2 Z'_{S^3}(\omega_r) = \cT(\rho\ep)\cT(\rho\ep+b\omega_r) Z'_{S^3}(\omega_r)
\ee
where
\be
Z'_{S^3}(\omega_r) = \int \da{}\, W^{(r)}(a) \cT(a) \, ,  
\ee
and
\be
\theta_r = \theta_{N-r} =  q^{\frac{r(N-r)}{2N}}  t^{\frac{r(N-r)}{2}}\,,
\ee
which satisfies
\begin{equation}
\theta_r^{-1} \cT(\epsilon\rho) = \cT(\epsilon\rho+b\omega_r)\,.
\end{equation}
The insertion of the defect can therefore be interpreted as the insertion of a Wilson loop in the representation $\Lambda^r$ in the supersymmetric Chern-Simons description of $\cT(S^3)$.


\subsection{Hopf Link in $S^3$}

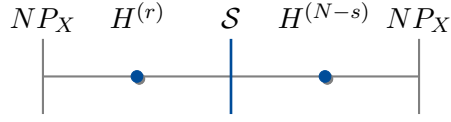
\begin{figure}[tbp]
\begin{center}
\begin{tikzpicture}
\draw[line width=0.8pt, gray] (-2.5,0) -- (2.5,0);
\lineop at (-1.25,0)
\interface at (0,0)
\lineop at (1.25,0)
\boundary at (2.5,0)
\boundary at (-2.5,0)
\node at (-1.24,0.82) {$H^{(r)}$};
\node at (1.24,0.82) {$H^{(N-s)}$};
\node at (0,0.78) {$\cS$};
\node at (-2.5,0.75) {$NP_X$};
\node at (2.5,0.75) {$NP_X$};
\end{tikzpicture}
\end{center}
\caption{The partition function for $\cT\left(S^3\right)$ with two defects corresponds to the insertion of 't Hooft loop operators. 
This partition function is symmetric in $r$ and $s$ because in our conventions operators in diagrams act to the right.}
\label{fig:TwoDefectsS3}
\end{figure}

We now consider two codimension-4 defects labelled by anti-symmetric tensor representations of rank $r$ and $s$ wrapping two Hopf-linked circles in $S^3$. 
In the first surgery construction of $S^3$ by gluing two solid tori $S^1 \times D_2$ with an $S$ transformation, this corresponds to inserting a pair of codimension-4 defects at the origin of each $D_2$. 

This corresponds to inserting two supersymmetric 't Hooft loops on the two sides of the interface, as depicted in figure \ref{fig:TwoDefectsS3}. 
By cutting the path integral at both sides of the $S$ interface, we find the following integral representation of this correlation function,
\begin{equation}
Z_{\cT(S^3)}(\omega_r,\omega_s) = \int \rd\nu_X(a)\rd\nu_X(a') ~ Z_{X,NP_X}(a) Z_{X,NP_X}(a') H_X^{(s)}(a') \cdot \left( H_X^{(r)}(a) \cS_X(a,a')\right)\,.
\end{equation}
The evaluation of the 't Hooft loop difference operators on the $S$ interface kernel yields the following expression,
\begin{equation} \label{eq:twoH-oneS}
H_X^{(s)}(a') \cdot \left( H_X^{(r)}(a) \cS_X(a,a')\right) = \sum_{|I|=s}  B_I(a') \,  W^{(r)}(a'+b\delta_I) \, \cS_X(a,a'+b\delta_I)\, ,
\end{equation}
where
\begin{equation}
B_I(a) = \prod_{\substack{i \in I \\ j\notin I} } \frac{\sin\pi b(\epsilon+a_i-a_j)}{\sin\pi b(a_i-a_j)} \, ,
\end{equation}
and by a slight abuse of notation, we have defined $\delta_I$ to be the vector whose elements satisfy $\left(\delta_I\right)_j = \chi_I(j)-\tfrac{|I|}{N}$, with $\chi_I$ the indicator function of $I$.
Now we make use of the results for the Nahm pole in section \ref{sec:boundrevis} to see that we need to evaluate equation \eqref{eq:twoH-oneS} at $a,a'= \rho\ep$.
In this case, the only contribution in the sum in equation \eqref{eq:twoH-oneS} is from the set $I = \{ 1,\dots,s\}$.
Since
\begin{equation}
\cS_X (\rho\ep,a')= 1/ Z_{\cT(S^3)}
\end{equation}
we therefore find that
\begin{equation}
\left. H_X^{(s)}(a') \cdot \left( H_X^{(r)}(a) \cS_X(a,a')\right) \right|_{(a,a')=(\rho\ep,\rho\ep)}= \frac{1}{Z_{\cT(S^3)}} W^{(s)}(\rho\ep) W^{(r)}\left(\rho\ep + b\omega_s\right)\, ,
\end{equation}
where $\omega_s = \delta_{\{1,\dots,s\}}$ is the highest weight of the rank $s$ fundamental representation of $\mathfrak{su}(N)$.
Finally, we evaluate
\begin{equation}
\int \rd\nu_X(a)\rd\nu_X(a') ~ Z_{X,NP_X}(a) Z_{X,NP_X}(a') = \left(\frac{1}{\cS_X(\rho\ep,\rho\ep)}\right)^2 = \left(Z_{\cT(S^3)}\right)^2\,.
\end{equation}

Putting everything together, we find that
\begin{equation}
\frac{ Z_{\cT(S^3)}(\omega_r,\omega_s) }{Z_{\cT(S^3)}} =  W^{(s)}(\rho\ep) W^{(r)}\left(\rho\ep + b\omega_s\right)\,,
\end{equation}
or in terms of the exponentiated variables $\tf = e^{2\pi \ii b \epsilon}$, $q = e^{2\pi \ii b^2}$:
\begin{equation} \label{eq:MainClaimS3TwoDefects}
\frac{Z_{\cT(S^3)}\left( \omega_r, \omega_s \right) }{Z_{\cT(S^3)}}=  W^{(r)}\left(\tf^\rho\right)W^{(s)}\left(\tf^\rho q^{\omega_r} \right)\, .
\end{equation}
This precisely reproduces an analytic continuation of the $S$-matrix $S_{r,s}$ for a pair of anti-symmetric tensor representations $\Lambda^r$ and $\Lambda^s$ in refined Chern-Simons theory \cite{Aganagic:2012ne,Aganagic:2011sg}.

\begin{figure}[htb]
\begin{center}
\begin{tikzpicture}
\draw[line width=0.8pt, gray] (-8,1) -- (-3.5,1);
\interface at (-5.75,1)
\lineop at (-6.87,1)
\lineop at (-4.62,1)
\boundary at (-8,1)
\boundary at (-3.5,1)
\node at (-5.75,1.78) {$\cS$};
\node at (-6.87,1.82) {$H^{(r)}$};
\node at (-4.62,1.82) {$H^{(N-s)}$};
\node at (-7.88,1.75) {$NP_X$};
\node at (-3.38,1.75) {$NP_X$};
\node [right] at (-2.9,1) {{$=$}};
\node [right] at (-2.25,1.1) {$\cT(\rho\ep)^2$};
\draw[line width=0.8pt, gray] (-0.5,1) -- (5.8,1);
\boundary at (-0.5,1)
\boundary at (5.8,1)
\interface at (0.55,1)
\lineop at (1.6,1)
\interface at (2.65,1)
\lineop at (3.7,1)
\interface at (4.75,1)
\node at (0.55,1.8) {$\cT^{-1}$};
\node at (1.6,1.82) {$H^{(r)}$};
\node at (2.65,1.78) {$\cS$};
\node at (3.7,1.82) {$H^{(N-s)}$};
\node at (4.82,1.82) {$\cT^{-1}$};
\node at (-0.5,1.75) {$NP_X$};
\node at (5.8,1.75) {$NP_X$};
\node [right] at (-7.7,-1) {{$=$}};
\node [right] at (-7.05,-0.9) {$\cT(\rho\ep)^2\, q^{-\frac{r(N-r)}{2N}}q^{\frac{s(N-s)}{2N}}$};
\draw[line width=0.8pt, gray] (-2.7,-1) -- (4.7,-1);
\boundary at (-2.7,-1)
\boundary at (4.7,-1)
\lineop at (-1.42,-1)
\interface at (0.1,-1)
\interface at (1,-1)
\interface at (1.9,-1)
\lineop at (3.45,-1)
\node at (-1.36,-0.22) {$(W^{-1}H)^{(r)}$};
\node at (0.1,-0.18) {$\cT^{-1}$};
\node at (1,-0.22) {$\cS$};
\node at (1.9,-0.18) {$\cT^{-1}$};
\node at (3.45,-0.22) {$(WH)^{(N-s)}$};
\node at (-2.8,-0.22) {$NP_X$};
\node at (4.9,-0.25) {$NP_X$};
\node [right] at (-7.7,-3) {{$=$}};
\node [right] at (-7.05,-2.9) {$\cT(\rho\ep)^2\, q^{-\frac{r(N-r)}{2N}}q^{\frac{s(N-s)}{2N}}$};
\draw[line width=0.8pt, gray] (-2.7,-3) -- (4.7,-3);
\boundary at (-2.7,-3)
\boundary at (4.7,-3)
\lineop at (-1.42,-3)
\interface at (0.1,-3)
\interface at (1,-3)
\interface at (1.9,-3)
\lineop at (3.45,-3)
\node at (-1.36,-2.22) {$(W^{-1}H)^{(r)}$};
\node at (0.1,-2.22) {$\cS$};
\node at (1,-2.22) {$\cT$};
\node at (1.9,-2.22) {$\cS$};
\node at (3.45,-2.22) {$(WH)^{(N-s)}$};
\node at (-2.8,-2.25) {$NP_X$};
\node at (4.9,-2.25) {$NP_X$};
\node [right] at (-7,-5) {{$=$}};
\node [right] at (-6.25,-4.9) {$\displaystyle \cT(\rho\ep)^2\, \, q^{\frac{r(N-r)}{2N}}q^{-\frac{s(N-s)}{2N}}$};
\draw[line width=0.8pt, gray] (-1.8,-5) -- (4.2,-5);
\interface at (1.2,-5)
\lineop at (-0.3,-5)
\lineop at (2.7,-5)
\boundary at (-1.8,-5)
\boundary at (4.2,-5)
\node at (1.2,-4.22) {$\cT$};
\node at (-0.3,-4.22) {$(WH)^{(r)}$};
\node at (2.8,-4.22) {$(W^{-1}H)^{(N-s)}$};
\node at (-1.8,-4.25) {$N_X$};
\node at (4.3,-4.25) {$N_X$};
\draw[line width=0.8pt, gray] (-2,-7) -- (2.5,-7);
\interface at (0.25,-7)
\lineop at (-0.7,-7)
\lineop at (1.48,-7)
\boundary at (-2,-7)
\boundary at (2.5,-7)
\node at (0.25,-6.22) {$\cT$};
\node at (-0.7,-6.18) {$W^{(r)}$};
\node at (1.48,-6.18) {$W^{(s)}$};
\node at (-2,-6.25) {$N_X$};
\node at (2.5,-6.25) {$N_X$};
\node [right] at (-5.9,-7) {{$=$}};
\node [right] at (-5.15,-6.9) {$\displaystyle \cT(\rho\ep)^2\, \, \theta_r^{-1}\, \theta_s^{-1}$};
\end{tikzpicture}
\end{center}
\vspace{-0.5cm}
\caption{An evaluation of $Z_{\cT(S^3)}(\omega_r,\omega_s)$}
\label{fig:S3twodefects1}
\end{figure}

Again, we can make contact with the alternative framing of $S^3$ by the sequence of operations shown in figure~\ref{fig:S3twodefects1}.
We begin with the same setup as before and treat symmetrically the operators on either sides of the interface, using the property that $T$ acts as multiplication by a constant on a Nahm pole boundary condition to insert a $T$ interface, as represented in the first step of figure \ref{fig:S3twodefects1}. 
Then, we move the $T$ interfaces towards the center using \eqref{eq:HWdef2} and the relation
\be
(W^{-1}H)^{(r)}_X(a)\cT^{-1}(a) = q^{\frac{r(N-r)}{2N}}\cT^{-1}(a)H^{(r)}(a)\, .
\ee
Now we use the $SL(2,\Z)$ relations to get to the third line to figure \ref{fig:S3twodefects1}. Recalling the transformation of supersymmetric Wilson-'t Hooft loops~\eqref{eq:BehaviourMixedWilsontHooft}, we end up at the fourth line.
For the supersymmetric Wilson-'t Hooft loop on the right of the $T$ interface, the action of the difference operator on the Neumann boundary condition is
\begin{equation}
\left(W^{-1}H\right)^{(N-s)}(a) \cdot 1 = t^{-\frac{s(N-s)}{2}}W^{(s)}(a) \, .
\end{equation}
However, for the supersymmetric Wilson-'t Hooft loop on the left of the $T$ interface, we first need to use the conjugation property
\be
\int d\nu_X(a) f(a) \left[(WH)^{(r)}(a')g(a')\right] = q^{-\frac{r(N-r)}{N}} \int d\nu_X(a) \left[(W^{-1}H)^{(N-r)}(a)f(a) \right] g(a')\, .
\ee
This allows us to conclude that
\bea
\label{eq:TwoDefectsTS}
Z_{\cT(S^3)}(\omega_r,\omega_s) &= \theta_r^{-1}\theta_s^{-1}\cT(\rho\ep)^2Z_{\cT(S^3)}'(\omega_r, \omega_s) \\
&= \cT(\rho\ep+b\omega_r)\cT(\rho\ep+b\omega_s)Z_{\cT(S^3)}'(\omega_r, \omega_s)\, ,
\eea
where
\be
\label{eq:two-defects-answer2}
Z_{\cT(S^3)}'(\omega_r, \omega_s) = \int\da{}\, W^{(r)}(a)W^{(s)}(a)\cT(a) \, .
\ee
This corresponds to the insertion of a pair of supersymmetric Wilson loops in the anti-symmetric tensor representations $\Lambda^r$ and $\Lambda^s$ in the supersymmetric Chern-Simons description of $\cT(S^3)$. This can be interpreted as a complex version of the Cherednik-Macdonald-Mehta identity~\cite{Etingof:1998}.


\section{Surgery} \label{sec:Surgery}

Closed orientable three-manifolds have the property that they can be constructed by Dehn surgery along links in $S^3$. This is determined by an element of the mapping class group $SL(2,\Z)$ of the torus boundaries of both the knot exterior in $S^3$ and the tubular neighbourhood of the knot. In this section we consider the Dehn surgery construction of Seifert manifolds $M_3$, and the corresponding construction of the partition function of the theory $\cT(M_3)$.


\subsection{Lens Spaces} \label{sec:Lens}

The Lens space $L(p,1)$ can be constructed by gluing a pair of $(1,0)$ solid tori by the $SL(2,Z)$ transformation $ST^{-p}S$, or equivalently two $(0,1)$ solid tori by the $SL(2,\Z)$ transformation $T^{-p}$. This corresponds to the partition function of the interface $T^{-p}$ in between a pair of Neumann boundary conditions $N_X$ or $N_Y$. Sending the size of the interval to zero, this leaves a supersymmetric Chern-Simons theory for $\gf$ at level $-p$ together with an adjoint chiral multiplet\footnote{The choice of supersymmetric Chern-Simon term at level $p$ and $-p$ correspond to the Lens spaces $L(-p,1)$ and $L(p,1)$, which differ only by a change of orientation.}.
Applying our considerations from section \ref{sec:PartitionFunctionsOnS3b}, the partition function is given by the following integral
\be
Z_{\cT(L(p,1))} =  \int \da{} \,
\cT^{-p}(a) \, .
\ee

For a general Lens space $L(p,q)$, we expand $-p/q$ as a continued fraction 
\be
-p/q = [r_1,\dots,r_m] = r_1 - \frac{1}{r_2- \frac{1}{r_3 - \dots}} \ .
\ee
The Lens space $L(p,q)$ is then constructed using rational surgery by gluing two solid tori $(0,1)$ with the $SL(2,\mathbb{Z})$ transformation $T^{r_1}ST^{r_2} S\dots S T^{r_m}$. This corresponds to a series of $SL(2,\Z)$ duality interfaces between a pair of Neumann boundary conditions $N_X$. The partition function of $\cT(L(p,q))$ is
\be 
\label{eq:generalLensPart}
\begin{split}
Z_{\cT(L(p,q))} = \int \da{1}\cdots \da{m}& \, 
\cT^{r_1}(a_1)  \cS_X\left(a_1,a_2\right) \cT^{r_2}(a_2) \cS_X\left(a_2,a_3\right) \\
 & \cdots \cT^{r_{m-1}}(a_{m-1})\cS_X\left(a_{m-1},a_m\right) \cT^{r_m}(a_m) \ .
\end{split}
\ee

Note that continued fraction expansions are not unique, for instance $1 = [1] = [0,-1] = [2,1]= [0,0,2,1] = \dots$.
The difference in the constructions of the same Lens space $L(p,q)$ through different continued fraction expansions for $-p/q$ is the resulting framing of the manifold. 
However, the framing only affects the partition function by an overall constant factor, and we indeed find that different choices of continued fraction expansions in \eqref{eq:generalLensPart} yield partition functions that only differ by a framing factor.

Furthermore, equation \eqref{eq:generalLensPart} respects known homeomorphisms of Lens spaces, namely if $qq'\equiv 1 \mod p$, then $L(p,q) \cong L(p,q')$.
The continued fraction expansions of the two pairs of coefficients are in 1-1 correspondence: if $-p/q = [\rho_1,\dots,\rho_k]$, then $-p/q' = [\rho_k,\dots,\rho_1]$ (and vice versa).
Since the expression for the partition function is explicitly invariant under reversal of the sequence $(\rho_1,\dots,\rho_k)$, it indeed respects this homeomorphism.
%
%

\subsection{Seifert Manifolds}

Seifert manifolds are $S^1$-orbibundles; they can be realized using surgery on various solid tori and they are described by a collection of pairs of integer numbers $(p_i,q_i)\in \Z\oplus\Z$, as described in appendix \ref{app:3-surgery}.
To compute the partition function of the 3d $\cN=2$ theory associated to a general Seifert manifold $\cM\left((m_1,n_1),\dots,(m_k,n_k)\right)$ we must now consider the 4d $\cN=2^*$ theory with gauge algebra $\gf^{\oplus k}$.

The boundary condition on the right is `unentangled': it is a product of $(m_j,n_j)$ type boundary conditions for each factor of the gauge group separately. 
After expanding each $m_j/n_j$ as a continued fraction: $m_j/n_j = \left[r^j_1,\dots,r^j_{l_j}\right]$, the wave function associated to this boundary condition is given by
\be
\phi(a_1,\dots , a_k) = \phi_1(a_1)\cdots \phi_k(a_k)\,.
\ee
where
\begin{align} \label{eq:SeifertArm}
\phi_i(a_i) &= 
\int \rd\nu_X\left(a_i^{(2)}\right)\dots \rd\nu_X\left(a_i^{(l_i)}\right) \, \cT^{r^i_1}\left(a_i\right)  \cS_X\left(a_i,a^{(2)}_i\right) \cT^{r^i_2}\left(a^{(2)}_i\right) \nonumber\\
&\quad \quad \cdots \cS_X\left(a_i^{(l_{i-1})},a_i^{(l_{i})}\right) \cT^{r^i_{l_i}}\left(a^{(l_i)}_i\right)\, 
\end{align}
encodes all the information down each fibre of the plumbing tree.

However, on the left  we must introduce an `entangled' boundary condition corresponding to the manifold $S^2 \times S^1 \setminus \left(\cup_{i=1}^k N_i\right)$, where the $N_i$ are $k$ unlinked solid tori. This is defined by starting from Neumann boundary conditions $N_X$ for each factor $\gf$ in the gauge algebra, and deforming it by coupling to the dimensional reduction of the class $\cS$ theory corresponding to $S^2$ with $k$ full punctures and flavour symmetry $\gf^{\oplus k}$. This has a mirror description as a star-shaped quiver~\cite{Benini:2010uu}, leading to the wave function
\be
\psi(a_1,\dots, a_k) = \int \rd \nu_X(a) \, \cS_X(a,a_1) \ldots \cS_X(a,a_k) \, .
\ee
The partition function corresponding to the Seifert manifold is now
\bea
Z_{\cT(M)} & = \int \da{1}\cdots \da{k} \,  \psi(a_1,\dots,a_k) \phi(a_1,\dots,a_k) \\
& = \int \rd\nu_X(a) \prod_{i=1}^k\da{i}\,  \cS_X(a,a_i) \phi_i(a_i)\, .
\label{eq:general-seifert}
\eea
This expression mirrors the standard surgery construction for Seifert manifolds in regular Chern-Simons theory.

\begin{figure}[htbp]
\begin{center}
\begin{tikzpicture}
\node at (-2,0) (0) [circle, draw, minimum size=1mm, inner sep=0.5mm, fill=black, label={left:\small{$0$}}]{};
\node at (0,1.5) (11) [circle, draw, minimum size=1mm, inner sep=0.5mm, fill=black, label={[yshift=-0.1cm]\small{$r_1^1$}}] {};
\node at (0,0.6) (21) [circle, draw, minimum size=1mm, inner sep=0.5mm, fill=black, label={[yshift=-0.1cm]\small{$r_1^2$}}] {};
\node at (0,-1.3) (n1) [circle, draw, minimum size=1mm, inner sep=0.5mm, fill=black, label={[yshift=-0.1cm]\small{$r_1^n$}}] {};
\node at (1.5,1.5) (12) [circle, draw, minimum size=1mm, inner sep=0.5mm, fill=black, label={[yshift=-0.1cm]\small{$r_2^1$}}] {};
\node at (1.5,0.6) (22) [circle, draw, minimum size=1mm, inner sep=0.5mm, fill=black, label={[yshift=-0.1cm]\small{$r_2^2$}}] {};
\node at (1.5,-1.3) (n2) [circle, draw, minimum size=1mm, inner sep=0.5mm, fill=black, label={[yshift=-0.1cm]\small{$r_2^n$}}] {};
\node at (6,1.5) (1m1) [circle, draw, minimum size=1mm, inner sep=0.5mm, fill=black, label={[yshift=-0.1cm]\small{$r_{l_1}^1$}}] {};
\node at (6,0.6) (2m2) [circle, draw, minimum size=1mm, inner sep=0.5mm, fill=black, label={[yshift=-0.1cm]\small{$r_{l_2}^2$}}] {};
\node at (6,-1.3) (nmn) [circle, draw, minimum size=1mm, inner sep=0.5mm, fill=black, label={[yshift=-0.1cm]\small{$r_{l_n}^n$}}] {};
\draw (0) -- (11) -- (12) -- (2.3,1.5);
\draw (0) -- (21) -- (22) -- (2.3,0.6);
\draw (0) -- (n1) -- (n2) -- (2.3,-1.3);
\draw (5.2,1.5) -- (1m1);
\draw (5.2,0.6) -- (2m2);
\draw (5.2,-1.3) -- (nmn);
\draw[dotted] (0,0.3) -- (0,-0.5);
\draw[dotted] (1.5,0.3) -- (1.5,-0.5);
\draw[dotted] (6,0.3) -- (6,-0.5);
\draw[dotted] (3,1.5) -- (4.5,1.5);
\draw[dotted] (3,0.6) -- (4.5,0.6);
\draw[dotted] (3,-1.3) -- (4.5,-1.3);
\end{tikzpicture}
\end{center}
\vspace*{-0.4cm}
\caption{Plumbing tree for a general Seifert manifold.}
\label{fig:generalSeifertPlumbing}
\end{figure}
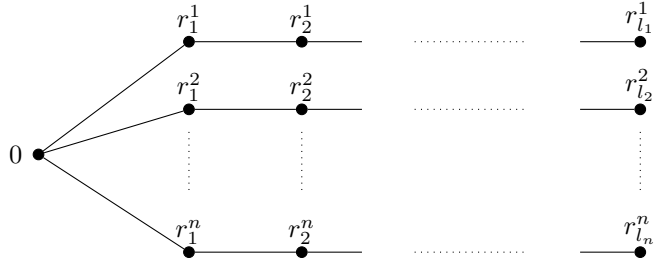

The structure of the result~\eqref{eq:general-seifert} is manifest in the plumbing diagram for the Seifert manifold, represented in figure \ref{fig:generalSeifertPlumbing}, where to each node we associate an integral and a $\cT$-function, and to each edge we associate an $\cS_X$-function:
\begin{align}
\text{node $j$ with label $r_j$} \quad &\leftrightarrow \quad \int \rd\nu_X(a_j)~\cT^{r_j}(a_j)\,, \\
\text{edge joining nodes $i$ and $j$} \quad &\leftrightarrow \qquad \cS_X(a_i,a_j)\,.
\end{align} 

We can check that this reproduces the result \eqref{eq:generalLensPart} for a Lens space in two different ways, First, using the representation of the Lens space $L(p,q)$ as $\cM\left((q,p)\right)$, we write $q/p = [r_1,\dots, r_l]$.
Then $L(p,q)$ has the following partition function,
\bea
Z_{\cT(L(p,q))} & = \int \da{}\, \da{1}\dots\da{l} ~ \cS_X(a,a_1) \cT^{r_1}(a_1) \cdots \cS_X(a_{l-1},a_l) \cT^{r_l}(a_l). \\
& = \int \da{1}\dots\da{k} ~ \cT^{\rho_1}(a_1) \cS_X(a_1,a_2) \cT^{\rho_2}(a_2) \cdots \cS_X(a_{k-1},a_k) \cT^{\rho_k}(a_k) \, ,
\eea
where in the second line we have trivially re-written the partition function in terms of the expansion $-p/q = [0,q/p] = [0,r_1,\dots,r_l]=[\rho_1,\dots,\rho_k]$, where $k=l+1$. This reproduces the result \eqref{eq:generalLensPart}.

We can alternatively construct the Lens space $L(p,q)$ as the Seifert manifold $M = \mathcal{M}((m_1,n_1),(m_2,n_2))$, with $p = m_1n_2 + m_2n_1$ and $q = a m_1 + b n_1$, where $a,b\in\mathbb{Z}$ satisfy $a m_2 - b n_2 = 1$.
Expanding $m_1/n_1 = [r_1,\dots,r_l]$ and $m_2/n_2 = [\rho_1,\dots,\rho_k]$, we find that
\begin{align} \label{eq:doubleFibre}
Z_{\cT(M)} = \int \da{-k}&\cdots \da{l}\, 
\cT^{\rho_k}(a_{-k}) \cS_X(a_{-k},a_{-k+1})\cT^{\rho_{-k+1}}(a_{-k+1}) \cdots \cT^{\rho_1}(a_{-1}) \times\nonumber\\
& \times\cS_X(a_{-1},a_0) \cS_X(a_0,a_1) \cT^{r_1}(a_1) \cdots \cS_X(a_{l-1},a_l) \cT^{r_l}(a_l),
\end{align}
which we recognize as the partition function for the Lens space $L(\tilde{p},\tilde{q})$, where $-\tilde{p}/\tilde{q} = [\rho_k,\dots\rho_1,0,r_1,\dots,r_l]$. This is indeed homeomorphic to the Lens space $L(p,q)$ described above.


\subsection{Special Limits and Topological Invariance}

We are currently not able to compute the Seifert manifold partition function for general $\epsilon$.
Nevertheless, in certain limits the general formula \eqref{eq:general-seifert} reduces to a simpler form and we are able to calculate the partition function explicitly. 

By analytic continuation, we will consider the limits $\ep \to0$ and $\ep \to Q$ which are expected to correspond to removing the puncture from $T^2$. For example, in the limit $\ep\to0$, it is straightforward to show that 
\be
\cT(a) \to \exp\left({2\pi i \left(\Delta(\al) - \frac{c}{24}\right)}\right) \, ,
\ee
where $c = (N-1)+N(N^2-1)Q^2$, and up to a numerical factor that~\cite{Bullimore:2014upa}
\be
S_b(\epsilon)^{N-1} \cS_X(a,a')  \longrightarrow \frac{\sum_{\sigma \in S_N} e^{-2\pi \ii\sum_j a_{\sigma(j)} a'_j }}{\prod_{i<j}2\sin(\pi b^\pm(a_i-a_j))} \, .
\ee
 This reproduces the modular $S$- and $T$-matrices for characters of non-degenerate representations of the $W_N$-algebra with momentum $\al=Q\rho-a$~\cite{Drukker:2010jp}, as expected once we remove the puncture from $T^2$.

In this section, we will simply consider the case $\gf=\suf(2)$ and discuss the limits $\epsilon\rightarrow 0$ and $\epsilon\to Q$. In the limit $\ep\to0$, we can fully determine the partition function, while for $\ep \to Q$, we can expresse the integrals in terms of trigonometric functions, which can then be used to get some analytic and numerical results.

In these limits, we test the statement that the partition function of $\cT(M_3)$ on $S_b^3$ is a topological invariant of Seifert manifolds $M_3$.
We have tested in both limits the equality of partition functions of the manifolds satisfying the following homeomorphisms \cite{Freed:1991wd}:
\begin{itemize}
\item $L(p,q) \cong L(p,q')$ if and only if $q' \equiv \pm q^{\pm 1} \mod p$. Note that we had already established invariance when $q q'\equiv 1 \mod p$ in section \ref{sec:Lens}.
\item $L(5,4) \cong \mathcal{M}((-2,1),(3,1),(1,1)).$ 
\item $L(7,2) \cong \mathcal{M}((-2,1),(3,1),(-1,1)).$
\end{itemize}
Furthermore, we will investigate the following homeomorphism: 
\begin{equation}
M = \mathcal{M}((0,1),(-p_1,q_1),\dots,(-p_n,q_n)) \cong \bighash_{j=1}^n L(p_j,q_j)\, ,
\end{equation}
where $\bighash$ denotes the connected sum, and show that the relevant partition functions satisfy
\begin{equation} \label{eq:connsum}
Z_M = \frac{\prod_{j=1}^n Z_{L(p_j,q_j)}}{Z_{S^3}^{n-1}}\, ,
\end{equation}
with $M$ and each $L(p_j,q_j)$ in Seifert framing and $S^3$ in canonical framing.
This suggests that the following formula from regular Chern-Simons theory~\cite{Witten:1988hf}
\begin{equation}
Z_{\bighash_{j=1}^n M_j} = \frac{\prod_{j=1}^n Z_{M_j}}{Z_{S^3}^{n-1}}\, ,
\end{equation}
is valid in our construction.

In these limits, all partition functions become either $0$ or infinite due to the contribution from an adjoint multiplet of $T_R$ and $T_f$ charge $0$. In fact we find that the combination
\begin{equation}
\frac{1}{S_b(\epsilon)}Z_{\cT(M)}(\epsilon)
\end{equation}
is regular, with an overall factor of $S_b(\epsilon)$ in $Z_{\cT(M)}(\epsilon)$ from the contribution of such an adjoint chiral at the central node of the plumbing tree.
In principle, one should first compute the partition function $Z_{\cT(M)}(\epsilon)$ explicitly for general $\epsilon$, and then take a limit after removing the $S_b(\ep)$ factor. However, since we cannot perform the relevant integrals in closed form for general $\epsilon$, we shall assume that we can push the limits through integrals. We find that this leads to consistent results.

\subsubsection{The limit $\epsilon \rightarrow Q$}

Let us first consider the limit $\epsilon\rightarrow Q$. This limit of the partition function $S_X(a,a')$ has been considered previously in \cite{Hosomichi:2010vh}.
We find that
\be\label{eq:ZlimQ}
\frac{1}{S_b(\epsilon)}\nu_X(a) \to\nu(a)^2\, , \qquad S_b(\epsilon) \cS_X(a,a')\to 2 \sqrt{2} \frac{\cos(4\pi aa')}{\nu(a)\nu(a')}.
\ee
Specifically, note that the product $\nu_X(a)\cS_X(a,a')$ is regular.
Evaluating the integrals \eqref{eq:general-seifert} in closed form for a general Seifert manifold is beyond our current capabilities. However, we checked numerically that the expression \eqref{eq:generalLensPart} for Lens spaces is invariant under the homeomorphism $L(p,q) \cong L(p,q')$ whenever $q'=-q \mod p$.

Moreover, we can check exactly that the integrals \eqref{eq:generalLensPart} and \eqref{eq:general-seifert} coincide for the following exceptional pairs of homeomorphic 3-manifolds:
\begin{equation}
L(5,4)\cong \cM((-2,1), (3,1),(1,1))\,, \qquad L(7,2)\cong \cM((-2,1), (3,1),(-1,1)) \,.
\end{equation}

Furthermore, we consider the homeomorphism
\begin{equation}
\mathcal{M}((0,1),(-p_1,q_1),\dots,(-p_n,q_n)) \cong \bighash _{j=1}^n L(p_j,q_j)\, .
\end{equation}
Let $-p_i/q_i = [r^i_1,\dots,r^i_{m_i}]$ in the general formula \eqref{eq:general-seifert}. 
In the limit $\epsilon\to Q$, the partition function simplifies to
\be
\begin{split}
 \frac{1}{S_b(\epsilon)} Z_{\cT(M)} =& \int_\mathbb{R} \frac{\rd x}{2}\, \frac{1}{\nu(\ii x)^{n-1}} \int_\mathbb{R} \frac{\rd t}{2} \, \nu(\ii t)  2\sqrt{2}\cos(4\pi x t) \\
 &\; \times \prod_{j=1}^n \int_\mathbb{R} \frac{\rd x_j}{2}\, \nu(\ii x_j) 2\sqrt{2} \cos(4\pi x x_j) \phi_j(\ii x_j),
\end{split}
\ee

Consider the singular fibre $(0,1)$, represented above by the $t$ integral.
The integration of $\sqrt{2}\, \nu(\ii t)\cos(4\pi x t)$ yields a sum of delta functions
\begin{equation}
\int \rd t\, \sqrt{2}\, \nu(\ii t) \cos(4\pi x t) = \sum_k a_k \delta(x-x^0_k)\, .
\end{equation}
This simplifies the integral to
\begin{align}
\frac{1}{S_b(\epsilon)} Z_{\cT(M)} 
= \frac{1}{\left[-2\sqrt{2}\sin(\pi b^2)\sin(\pi b^{-2})\right]^{n-1}}  \prod_{j=1}^n \int_\mathbb{R} \frac{\rd x_j}{2}\, \nu_X(x_j) \phi_j(\ii x_j)\, .
\end{align}
Now recognize the remaining integrals as the partition functions $Z_{\cT(L(p_j,q_j))}$, and notice the following limit of $Z_{\cT(S^3)}$,
\begin{equation}
\lim_{\epsilon\rightarrow Q}  \frac{1}{S_b(\epsilon)} Z_{\cT(S^3)} (\epsilon) =-2 \sqrt{2}\sin(\pi b^2) \sin(\pi b^{-2}) \,.
\end{equation}
Therefore the connected sum formula \eqref{eq:connsum} holds, with $M$ and $L(p_j,q_j)$ both in Seifert framing and $S^3$ in canonical framing.


\subsubsection{The limit $\epsilon\rightarrow 0$} \label{sec:limeps->0}

In the limit $\ep\to0$, it is straightforward to check that
\be \label{eq:Zlim0}
\frac{1}{S_b(\epsilon)}\nu_X(a) \to 1\, , \qquad S_b(\epsilon)\cS_X(a,a') \to 2\sqrt{2} \cos \left( 4\pi aa'\right).
\ee
Again, we note that the product $\nu_X(a) \cS_X(a,a')$ is regular in the limit.

Now consider a general Seifert manifold $M = \mathcal{M}\left( (p_1,q_1),\dots,(p_n,q_n) \right)$. 
Assume that each $p_i \neq 0$ and, as before, write $p_i/q_i = [r^i_1,\dots,r^i_{m_i}]$.
Each fibre in the plumbing diagram contributes
\begin{align}
\phi_i(a_i) 
&= \int \rd\nu_X(a^i_1) \dots \rd\nu_X(a^i_{m_i}) \, \cS_X\left(a_i,a^i_1\right) \cT^{r^i_1}\left(a^i_1\right)\dots \cS_X\left(a^i_{m_{i-1}}, a^i_{m_i}\right) \cT^{r^i_m}\left(a^i_{m_i}\right) \nonumber\\
&= \int_{\ii \mathbb{R}^{m_i-1}} \frac{\rd a^i_1}{2\ii}\cdots \frac{\rd a^i_{m_i}}{2\ii} \, 2\sqrt{2}\cos(4\pi a a^i_1) \cT^{r^i_1}(a^i_1)\dots 2\sqrt{2}\cos(4\pi a^i_{m_i-1} a^i_{m_i}) \cT^{r^i_{m_i}}(a^i_{m_i})\, .
\end{align}
Unlike the limit $\ep\to Q$, this integral has a nice recursive structure, namely:
\begin{align}
\int_{\ii\mathbb{R}} \frac{\rd a_{j+1}}{2\ii}~ e^{-2\pi \ii r_j a_j^2} 2\sqrt{2} \cos (4\pi a_j a_{j+1}) e^{-2\pi \ii r_{j+1} a_{j+1}^2} 
= \frac{e^{\frac{\pi \ii}{4} \text{sign}(r_{j+1})}}{\sqrt{|r_{j+1}|}} e^{-2\pi \ii [r_j,r_{j+1}] a_j^2}\, ,
\end{align} 
whence
\bea
\widetilde{\phi}_i(a) & := \int \rd \nu_X(a_i) \cS_X\left(a,a_i\right) \phi_i(a_i) 
= e^{\frac{\pi \ii}{4} \sum_{j=1}^{m_i} \text{sign} \left( r^i_j\right) -\frac{\pi \ii}{12} \sum_{j=1}^{m_i} r^i_j} |p_i|^{-1/2} e^{2\pi \ii \frac{q_i}{p_i} a_i^2}\, ,
\eea
where we used that
\begin{equation}
\prod_{j=1}^{m_i} |[r^i_j, \dots, r^i_{m_i}]| = |p_i|\, ,
\end{equation}
and that $\text{sign} \left( [r^i_j,\dots,r^i_{m_i}] \right) = \text{sign}(r^i_j)$.

Performing the final integration over $a$, we then find that
\bea
\frac{1}{S_b(\epsilon)} Z_{\cT(M)} &= \int \frac{\rd \nu_X(a)}{S_b(\epsilon)} \prod_{i=1}^n \widetilde{\phi}_i(a) 
= \frac{e^{-\frac{\pi \ii}{12}   \left(3\text{ sign}\left(\sum_{i=1}^n q_i/p_i\right) +  \sum_{i=1}^n \sum_{j=1}^{m_i}( -3 \text{ sign} \left( r^i_j\right) + r^i_j) \right)} }{2\left| 2 \sum_{j=1}^n \frac{q_j}{p_j}\prod_{i=1}^n p_i\right|^{1/2} }\,  .
\eea
Observe that
\begin{equation}
-\frac{\pi \ii}{12}\left( -3\text{ sign}\left(- \sum_{i=1}^n \frac{q_i}{p_i} \right) +  \sum_{i=1}^n  \sum_{j=1}^{m_i}\left( -3 \text{ sign}  \left( r^i_j\right)  + r^i_j \right)\right) = -\frac{\pi \ii}{12} \phi_L\, ,
\end{equation}
where $\phi_L =  -3\sigma(Q_L) +  \sum_{i=1}^n  \sum_{j=1}^{m_i} r^i_j$ is the framing of the manifold, with $\sigma(Q_L)$ the signature of the linking matrix $Q_L$ of the plumbing tree.
Furthermore, recognize that \cite{Gadde:2013sca}
\begin{equation}
| \det Q_L | =  \left| \sum_{j=1}^n \frac{q_j}{p_j}\prod_{i=1}^n p_i\right|,
\end{equation}
to get the result
\begin{equation}
\lim_{\epsilon\rightarrow 0} \frac{1}{S_b(\epsilon)} Z_{\cT(M)}(\epsilon) = \frac{e^{-\frac{\pi\ii}{12}\phi_L}}{2\sqrt{2|\det Q_L|}}\, .
\end{equation}
This expression gives the partition function in Seifert framing; this suggests that to move to canonical framing we multiply by $\exp(\pi\ii\phi_L/12)$ and find
\begin{equation} \label{eq:generalSeifertZeroEps}
\lim_{\epsilon\rightarrow 0} \frac{1}{S_b(\epsilon)} Z_{\cT(M)}(\epsilon) = \frac{1}{2\sqrt{2|\det Q_L|}}\, ,
\end{equation}
which is a topological invariant.

Finally, consider again the homeomorphism: 
\begin{equation}
M = \mathcal{M}((0,1),(-p_1,q_1),\dots,(-p_n,q_n)) \cong \bighash_{j=1}^n L(p_j,q_j)\, ,
\end{equation}
which is not covered by our previous computation due to the appearance of the $(0,1)$.
Again, let $-p_j/q_j = [r^j_1,\dots,r^j_{m_j}]$.
Then
\begin{align}\label{eq:PfMe01}
\frac{1}{S_b(\epsilon)} Z_{\cT(M)} = \frac{1}{S_b(\epsilon)} \int \rd\nu_X(x) \, \widetilde{\phi}_0(x) \prod_{j=1}^n\rd\nu_X(a_j)\cS_X(a,a_j)\phi_j(a_j)\, ,
\end{align}
where
\begin{equation}
\widetilde{\phi}_0(x)= 
\int \rd \nu_X(t) ~ \cS_X(a,t) = \int \rd t ~ \sqrt{2}\cos(4 \pi a t) =  \frac{1}{\sqrt{2}}\Delta(a),
\end{equation} 
where $\Delta(a) = \frac{1}{2} \left(\delta(a) + \delta(-a) \right)$ is a Weyl-invariant delta function on the Cartan subalgebra of $\mathfrak{su}(2)$.
Furthermore $S_b(\epsilon)\cS_X(0,a') = 2\sqrt{2}$, so that, using the definition of $\widetilde{\phi}_j(a)$ and $\phi_j(a_j)$, \eqref{eq:PfMe01} simplifies to
\begin{align}
\frac{1}{S_b(\epsilon)} Z_{\cT(M)} 
&=( 2\sqrt{2})^{n-1} \prod_{j=1}^n  \frac{1}{S_b(\epsilon)}  \int \rd\nu_X(a_j)~ \phi_j(a_j) \, .
\end{align}
By the definition of $\phi_j(a_j)$, the latter integrals are precisely the partition functions of the Lens spaces $L(p_j,q_j)$ in Seifert framing:
\begin{equation}
Z_{\cT(L(p_j,q_j))} = \int \rd \nu_X(a_j)~ \phi_j(a_j)\, .
\end{equation}
Moreover, using the general result \eqref{eq:generalSeifertZeroEps}, we see that $S^3$ has the following partition function in canonical framing:
\begin{equation}
\lim_{\epsilon\rightarrow 0} \frac{1}{S_b(\epsilon)} Z_{\cT(S^3)}(\epsilon) = \frac{1}{2\sqrt{2}}
\end{equation}
Lastly, observe that $M$ and all $L(p_j,q_j)$ are both in Seifert framing.
Thus it is again true in this limit that the connected sum formula \eqref{eq:connsum} holds. 


\section{Discussion}
\label{sec:Conclusions}

We have given a prescription for computing the partition functions of 3d $\cN=2$ theories $\cT(M_3)$ associated to Seifert manifolds $M_3$ by compactification of a 4d $\cN=2^*$ theory on an interval with appropriate boundary conditions and a set of $SL(2,\Z)$ duality interfaces. Throughout, we have turned on a mass parameter for the distinguished $\uf(1)_f$ flavour symmetry associated to the circle action on Seifert manifolds. This construction is the analogue of Dehn surgery on the supersymmetric side of the 3d-3d correspondence.

We expect the partition functions of 3d $\cN=2$ theories $\cT(M_3)$ to correspond to computations in $SL(N,\C)$ Chern-Simons theory on $M_3$ with a network of defects supporting the mass parameter for the flavour symmetry $\uf(1)_f$. In particular, we recovered an analytic continuation of the $S$-matrix of refined Chern-Simons theory~\cite{Aganagic:2012ne,Aganagic:2011sg} from the study of supersymmetric line defects in $\cT(S^3)$. Our analysis therefore provides an insight into the structure of refined Chern-Simons with complex gauge group $SL(N,\C)$. 

To develop the full 3d-3d correspondence with complex Chern-Simons theory, it is important to consider the complete spectrum of supersymmetric defects of the 6d $\cN=(2,0)$ theory. In the case $\gf=\suf(N)$, we could consider general combinations of codimension-2 and codimension-4 defects of the 6d $\cN=(2,0)$ theory wrapping a curve $C$ in $M_3$ labelled by data $\Lambda = (\rho,\lb,\tilde\lb)$ with
\begin{itemize}
\item An embedding $\rho : \suf(2) \to \gf$.
\item A pair of dominant integral weights $(\lb,\tilde\lb)$ of the stabilizer $\Im(\rho) \subset \gf$.
\end{itemize}
Here, $\lambda$ and $\tilde\lambda$ correspond to codimension-4 defects wrapping respectively the circles $S^1$ and $\tilde S^1$ inside the squashed sphere $S^3_b$ on the supersymmetric side of the correspondence. In terms of $SL(N,\C)$ Chern-Simons theory, $\rho$ specifies a monodromy defect on $C$, while the weights $\lambda$, $\tilde\lambda$ correspond to Wilson loops in irreducible representations of the subgroup of $SL(N,\C)$ left unbroken by the monodromy defect~\cite{Gang:2015wya,Gang:2015bwa}. 

It would be interesting to map out the full dictionary with the supersymmetric side of the correspondence. For example, it seems reasonable to construct an $S$-matrix $S_{\Lambda_1,\Lambda_2}$ element corresponding to the correlation function of any combinations of defects labelled by data $\Lambda_1$ and $\Lambda_2$ supported on Hopf linked circles in $S^3$. Here, we have considered only particular combinations:
\begin{enumerate}
\item $\Lambda = ([1]^N,0,0)$: maximal codimension-2 defects supporting a flavour symmetry $\gf$.
\item $\Lambda = ([N],\omega_r,0)$: codimension-4 defects labelled by the fundamental weights of $\gf$.
 \end{enumerate}
The $S$-matrix for a pair of maximal codimension-2 defects is the normalized partition function $\cS_X(a,a')$ of $T(\gf)$. This should have a natural extension to a pair of general codimension-2 defects $\Lambda = (\rho,0,0)$ and $\Lambda = (\rho',0,0)$: it is the partition function of the theory $T^\rho_{\rho'}(\gf)$ introduced in~\cite{Gaiotto:2008ak}. The $S$-matrix for a pair of codimension-4 defects labelled by fundamental weights $\omega_r$ and $\omega_s$ is an analytic continuation of the $S$-matrix $S_{r,s}$ of refined Chern-Simons theory, constructed as the partition function of the $\cT(S^3)$ in the presence of a pair of supersymmetric loop operators. Extending this computation to general weights $\lambda_1$ and $\lambda_2$ will require a better understanding of monopole bubbling effects for supersymmetric 't Hooft loops. Clearly we have only scratched the surface of the spectrum of such correlation functions.

We should note that the minimal codimension-2 defect with $\Lambda = ([N-1,1],0,0)$ has played a ubiquitous background role in supporting the distinguished flavour symmetry $\uf(1)_f$.

Finally, we have focussed on computing the partition functions of $\cT(M_3)$ on squashed $S^3_b$, which is expected to correspond to $SL(N,\C)$ Chern-Simons theory at level $(k,\sigma)$ with
\be 
k = 1\, , \qquad \sigma = \frac{1-b^2}{1+b^2} \, .
\ee 
It would clearly be very interesting to perform the analogous computations for the superconformal index and Lens space partition functions, which should allow access to complex Chern-Simons theory at other values of the levels~\cite{Dimofte:2014zga}.

\section*{Acknowledgements}

We would like to thank Masahito Yamazaki for discussions at an early stage of this work. 
We would also like to thank Tudor Dimofte, Fabrizio Nieri, James Sparks and Maxim Zabzine for helpful discussions. 
The work of L.F.A., M.B. and M.v.L. was supported by ERC STG grant 306260.
L.F.A. is a Wolfson Royal Society Research Merit Award holder.
P.B.G. was supported by the EPSRC and a Scatcherd European Scholarship. M.v.L. was also supported by the EPSRC.

\appendix

\section{Conventions}
\label{app:conv}

We work with the `double-sine' function
\be
S_b(z) := \frac{1}{S_2\left(z \, |\, b,b^{-1}\right)}\, , 
\ee
where $S_2(x\, | \, \boldsymbol{\omega})$ is defined in \cite{Kharchev:2001rs}. It has the following properties:
\begin{enumerate} 
\item $S_b(z+b^{\pm}) = 2\sin(\pi b^{\pm}z) \, S_b(z)$,
\item $S_b(z)S_b(Q-z)=1$,
\item $S_{b^{-1}}(z)=S_b(z)$,
\item $S_b(z)^*=S_b(z^*)$ ,
\item $S_b(z)$ is pure phase for $z=Q/2+ \ii r$ with $r\in \mathbb{R}$,
\item $S_b\left(\frac{Q}{2}\right) = 1$, $S_b\left(\frac{b}{2}\right) = \frac{1}{\sqrt{2}}$.
\end{enumerate}
where $Q=b+b^{-1}$.

In addition, it has simple zeroes at
\begin{equation}
z = Q + nb + mb^{-1}, \qquad n,m\in \mathbb{Z}_{\geq0},
\end{equation}
and simple poles at
\begin{equation}
z=-nb-mb^{-1}, \qquad n,m\in \mathbb{Z}_{\geq0}
\end{equation}
with residue
\be \label{eq:SbResidue}
R_{n,m} = \text{Res}\left(S_b(z);z= -nb-mb^{-1}\right) = \frac{1}{2\pi}\frac{(-1)^{nm+n+m}}{\prod_{j=1}^n2\sin \pi j b^2\prod_{j=1}^m2\sin\pi jb^{-2}}.
\end{equation}

The following useful formula for any $n,m \in\mathbb{Z}_{\geq 0}$,
\begin{equation}
\label{eq:Rescaling}
S_b(x + nb + mb^{-1}) = (-1)^{nm} S_b(x) \prod_{j=0}^{n-1} 2 \sin \left(\pi b (x + jb) \right) \prod_{l=0}^{m-1} 2 \sin \left(\pi b^{-1}(x + l b^{-1})\right)\, ,
\end{equation}
is a consequence of the functional equations for the double sine function.

The asymptotics of the double sine function are given by
\begin{equation}\label{eq:AsymptoticsSb}
\lim_{z\rightarrow\infty} S_b(z) = \begin{cases}
e^{-\frac{\pi \ii}{2} B(z)} & \text{Im } z > 0 \\
e^{\frac{\pi \ii}{2} B(z)} & \text{Im } z < 0\, ,
\end{cases}
\end{equation}
where
\begin{equation}
B(z) = B_{2,2}\left(z\, |\, b,b^{-1}\right) = z^2 - Q z + \frac{1}{6}(Q^2+1)\, .
\end{equation}

Let us now summarize the contributions to the partition function of three-dimensional theories on $S^3_b$ with these conventions:
\begin{enumerate}
\item $\cN=2$ Chiral multiplet with R-charge $R$ : $S_b\left(\frac{Q \, R}{2}+x\right)$
\item $\cN=2$ $U(N)$ vectormultiplet: $\displaystyle \prod_{\substack{i,j=1  \\ i < j} }^N  \, 2 \sin ( \pi b\, (a_i-a_j) ) 2 \sin ( \pi b^{-1}(a_i-a_j)) $ 
\end{enumerate}


\section{Surgery on three-manifolds}\label{app:3-surgery}

In this appendix we 
review some of the ideas in three-dimensional topology that are relevant to our constructions, specifically those relating to surgery. Excellent reviews are \cite{neumannlectures} and \cite{Saveliev}.

Consider two compact $n$-manifolds with boundary $M_1$ and $M_2$, with homeomorphic boundaries, and a homeomorphism $f: \partial M_2 \rightarrow \partial M_1$ between the latter.
The operation of \textit{surgery} between the two consists in the construction of a new manifold by gluing the boundaries with $f$. More precisely, we define 
\be
M_1 \cup_f M_2 := \left(M_1 \sqcup M_2\right) / \sim,
\ee
where the equivalence relation is between points of the boundaries:
\be
x \sim y \, \Leftrightarrow \,  y = f(x) \qquad \forall x\in \partial M_1,\, \forall y\in \partial M_2 \ .
\ee
Recall that a knot $K$ in a closed orientable 3-manifold $M$ is a smooth embedding of $S^1$ in $M$. 
A link $L$ is a disjoint union of a finite collection of knots in $M$. 

A knot $K \subset M$ can be thickened to a tubular neighbourhood $N(K)$, a smoothly embedded disjoint solid torus ($D^2\times S^1$), whose core $\{0\}\times S^1$ forms the knot $K$.
Consider now the knot exterior $M_1 = M \setminus N(K)$ and tubular neighbourhood $M_2 =  N(K)$, which both have a $T^2$ boundary, and an arbitrary homeomorphism $f: (\partial M_2 \cong T^2) \rightarrow (\partial M_1 \cong T^2)$. We perform surgery between the two using $f$, gluing the knot exterior and the tubular neighbourhood using $f$. This results in a new closed orientable 3-manifold
\be
\widetilde{M} \equiv M_1 \cup_f M_2
\ee
We say that $\widetilde{M}$ is obtained from $M$ via surgery along the knot $K$, and refer to the process as \textit{Dehn surgery}.

The gluing process above depends on the boundary homeomorphism $f$; in fact it is completely determined by the image of a meridian $\partial D^2 \times \{x\} $, with $x\in S^1$, in $\partial M_1$. 
If $M =S^3$, then, after picking bases for $H_1(\partial M_1,\Z) \cong \mathbb{Z} \oplus \mathbb{Z}$, a curve on $\partial M_1$ is given up to isotopy by a pair of relatively prime integers $(p,q)$.
This pair describes in a basis of $H_1(\partial M_1,\Z)$ to what curve the meridian $(1,0)\in \mathbb{Z} \oplus \mathbb{Z} \cong H_1(\partial M_2,\Z)$ gets mapped.
Such surgeries are called \textit{rational} surgery, with a surgery called \textit{integral} if $q=\pm 1$. In the latter case, the surgery along $K$ is determined by both $K$ and the choice of an integer, which is called a \textit{framing} of the knot.

Another way we can describe a Dehn surgery is by determining the knot $K$ along which it is performed and the homeomorphism up to isotopy, that is, by an element of the mapping class group of the boundary. In this specific case, the boundaries are homeomorphic to tori, and the mapping class group is
\be
{\rm Homeo}(T^2) / {\rm Homeo}_0(T^2) \cong SL(2,\Z)\, ,
\ee
therefore we can decompose the homeomorphism in terms of the generators $S,T$ of $SL(2,\Z)$.

The \textit{Lickorish-Wallace theorem} states that any closed orientable connected 3-manifold can be obtained from $S^3$ through an integral Dehn surgery on a link in $S^3$ \cite{Lickorish:1962, Wallace:1960}.

Seifert manifolds are a special class of $3$-manifolds that are $S^1$-bundles over two-dimensional orbifolds. They can also be described using surgery in the following way. Let $M = F\times S^{1}$, where $F= S^2\setminus \text{int}\left( D_1^2\cup\dots D_n^2\right)$ is a two-sphere with $n$ discs removed. 
Then $\partial M = \bigcup_{i=1}^n N_i$ is a disjoint union of $n$ solid tori. 
We can glue in solid tori by identifying the meridian on the $i$-th solid torus boundary to a curve on $N_i$, whose isotopy class is described by $(p_i,q_i)\in\mathbb{Z}\oplus \mathbb{Z}$. 
This forms the Seifert manifold $M(0;(p_1,q_1),\dots,(p_n,q_n)) \equiv M((p_1,q_1),\dots,(p_n,q_n))$, where the $0$ refers to the fact that the construction used $S^2$, a genus $0$ surface. 
The construction can be generalized by using $F_g = \Sigma_g \setminus \text{int}\left( D_1^2\cup\dots D_n^2\right)$ instead of $F$, where $\Sigma_g$ is a closed orientable surface of genus $g$. 

Lens spaces are Seifert manifolds with $2$ singular fibres. 
Specifically, $\cM(0;(q,p)) \cong L(p,q)$, and $\mathcal{M}(0;(a_1,b_1),(a_2,b_2)) \cong L(p,q)$,with $p = a_1b_2 + a_2b_1$ and $q = m a_1 + nb_1$, where $m,n\in\mathbb{Z}$ satisfy $m a_2 - n b_2 = 1$  \cite{neumannlectures}\footnote{Note an early edition of \cite{Saveliev} claims in section 1.6 that $\mathcal{M}(0;(a_1,b_1),(a_2,b_2)) = L(a_1b_2 + a_2b_1, a_1a_2)$. This seems to be an error and has been removed in later editions.}.

For a general Seifert manifold $ M((p_1,q_1),\dots,(p_n,q_n))$, obtained as above, the $i$-th component of the link was glued back in after twisting the boundary using $M_i\in SL(2,\mathbb{Z})$, where $M_i=\begin{pmatrix}
p_i&r_i\\q_i&s_i \end{pmatrix}$. 
Such an $M_i$ is not unique: the choice of $r_i,s_i$ determines the framing of the manifold \cite{Freed:1991wd}. 
We would like to obtain $M_i$ from the $SL(2,\mathbb{Z})$-generators, which in our conventions are taken to be
\begin{equation}
S = \begin{pmatrix}
0&-1\\1&0
\end{pmatrix}\, \qquad T = \begin{pmatrix} 1&1\\0&1\end{pmatrix}\ .
\end{equation}
This can be achieved by noting that $T^n S = \begin{pmatrix}
n & -1 \\ 1&0
\end{pmatrix}$
and that if $A = \begin{pmatrix}
p & r\\q &s\end{pmatrix} \in SL(2,\mathbb{Z})$ then
\begin{equation}
T^nS A = \begin{pmatrix}
np -q & nr - s\\
p & r
\end{pmatrix}\in SL(2,\mathbb{Z}).
\end{equation}
Hence, by induction on $m$:
\begin{equation}
M_i = T^{a^i_1} S\dots T^{a^i_m} S
\end{equation}
where 
\be
p_i/q_i = [a^i_1,\dots,a^i_m] = a^i_1 - \frac{1}{a^i_2- \frac{1}{a^i_3 - \dots}}\, .
\ee

Seifert manifolds can also be described in terms of surgery diagrams, which encode how the surgery on links in $S^3$ has taken place. The simplest such diagram is that of a Lens space $L(p,1)$: 
\begin{center}
\begin{tikzpicture}
\node at (1.3,0.7) {$-p$};
\centerarc[black](0,0)(0,360, 1cm, 1cm);
\end{tikzpicture}
\end{center}
\noindent This indicates the surgery happened over a single unknot, with framing $-p$.
For a general Lens space $L(p,q)$ with $-p/q = [a_1,\dots,a_m]$, we have the following diagram:
\begin{center}
\begin{tikzpicture}
\centerarc[black](0,0)(60, 400, 1cm, 1cm);
\node at (0,1.3) {$a_1$};
\centerarc[black](1.3,0)(60, 220, 1cm, 1cm);
\centerarc[black](1.3,0)(240, 400, 1cm, 1cm);
\node at (1.3,1.3) {$a_2$};
\centerarc[black](2.6,0)(60, 220, 1cm, 1cm);
\centerarc[black](2.6,0)(240, 400, 1cm, 1cm);
\node at (2.6,1.3) {$a_3$}; 
\centerarc[black](3.9,0)(100, 220, 1cm, 1cm);
\centerarc[black](3.9,0)(240, 260, 1cm, 1cm);
\draw[loosely dotted] (4.4,0) -- (6.4,0);
\centerarc[black](6.6,0)(60, 80, 1cm, 1cm);
\centerarc[black](6.6,0)(280, 400, 1cm, 1cm);
\centerarc[black](7.9,0)(240, 580, 1cm, 1cm);
\node at (7.9,1.3) {$a_m$}; 
\end{tikzpicture}
\end{center}

Alternatively, we can represent these as \textit{plumbing graphs}, or plumbing trees, which are weighted graphs with each vertex representing an unknot, and each edge representing that two unknots are linked. For example, the diagram above translates into the plumbing tree
\begin{center}
\begin{tikzpicture}
\node at (0,0) (a1)[circle, draw, minimum size=1mm, inner sep=0.5mm, fill=black, label={[xshift=0cm, yshift=-0.1cm]\small{$a_1$}}] {};
\node at (1.5,0) (a2)[circle, draw, minimum size=1mm, inner sep=0.5mm, fill=black, label={[xshift=0cm, yshift=-0.1cm]\small{$a_2$}}] {};
\node at (3,0) (a3)[circle, draw, minimum size=1mm, inner sep=0.5mm, fill=black, label={[xshift=0cm, yshift=-0.1cm]\small{$a_3$}}] {};
\node at (7.5,0) (am)[circle, draw, minimum size=1mm, inner sep=0.5mm, fill=black, label={[xshift=0cm, yshift=-0.1cm]\small{$a_m$}}] {};
\draw (a1) -- (a2) -- (a3) -- (3.8,0);
\draw[dotted] (4.5,0) -- (6,0);
\draw (6.7,0) -- (am);
\end{tikzpicture}
\end{center}

A general Seifert manifold $M((p_1,q_1),\dots,(p_n,q_n))$ with the rational surgery coefficients $p_i/q_i = [a^i_{1},\dots, a^i_{m_i}]$ can be described diagrammatically as
\begin{center}
\begin{tikzpicture}
\centerarc[black](0,0)(42, 382, 0.4cm, 1cm);
\node at (-0.7,-0.3) {$a_1$};
\centerarc[black](1,-0.25)(45, 385, 0.4cm, 1cm);
\node at (0.5,-1.15) {$a_2$};
\centerarc[black](4.5,0)(157, 500, 0.4cm, 1cm);
\node at (5.2,-0.3) {$a_k$};
\centerarc[loosely dotted](2.25,0)(258, 310, 2.6cm, 0.7cm);
\centerarc[black](2.25,1)(-10, 192, 2.6cm, 0.7cm);
\centerarc[black](2.25,1)(202, 231, 2.6cm, 0.7cm);
\centerarc[black](2.25,1)(237, 337, 2.6cm, 0.7cm);
\node at (2.7,1.2) {{\small $0$}};
\end{tikzpicture}
\end{center}
Alternatively, one can draw this as a plumbing tree
\begin{center}
\begin{tikzpicture}
\node at (-2,0) (0) [circle, draw, minimum size=1mm, inner sep=0.5mm, fill=black, label={left:\small{$0$}}]{};
\node at (0,1.5) (11) [circle, draw, minimum size=1mm, inner sep=0.5mm, fill=black, label={[yshift=-0.1cm]\small{$a_1^1$}}] {};
\node at (0,0.6) (21) [circle, draw, minimum size=1mm, inner sep=0.5mm, fill=black, label={[yshift=-0.1cm]\small{$a_1^2$}}] {};
\node at (0,-1.3) (n1) [circle, draw, minimum size=1mm, inner sep=0.5mm, fill=black, label={[yshift=-0.1cm]\small{$a_1^n$}}] {};
\node at (1.5,1.5) (12) [circle, draw, minimum size=1mm, inner sep=0.5mm, fill=black, label={[yshift=-0.1cm]\small{$a_2^1$}}] {};
\node at (1.5,0.6) (22) [circle, draw, minimum size=1mm, inner sep=0.5mm, fill=black, label={[yshift=-0.1cm]\small{$a_2^2$}}] {};
\node at (1.5,-1.3) (n2) [circle, draw, minimum size=1mm, inner sep=0.5mm, fill=black, label={[yshift=-0.1cm]\small{$a_2^n$}}] {};
\node at (6,1.5) (1m1) [circle, draw, minimum size=1mm, inner sep=0.5mm, fill=black, label={[yshift=-0.1cm]\small{$a_{m_1}^1$}}] {};
\node at (6,0.6) (2m2) [circle, draw, minimum size=1mm, inner sep=0.5mm, fill=black, label={[yshift=-0.1cm]\small{$a_{m_2}^2$}}] {};
\node at (6,-1.3) (nmn) [circle, draw, minimum size=1mm, inner sep=0.5mm, fill=black, label={[yshift=-0.1cm]\small{$a_{m_n}^n$}}] {};
\draw (0) -- (11) -- (12) -- (2.3,1.5);
\draw (0) -- (21) -- (22) -- (2.3,0.6);
\draw (0) -- (n1) -- (n2) -- (2.3,-1.3);
\draw (5.2,1.5) -- (1m1);
\draw (5.2,0.6) -- (2m2);
\draw (5.2,-1.3) -- (nmn);
\draw[dotted] (0,0.3) -- (0,-0.5);
\draw[dotted] (1.5,0.3) -- (1.5,-0.5);
\draw[dotted] (6,0.3) -- (6,-0.5);
\draw[dotted] (3,1.5) -- (4.5,1.5);
\draw[dotted] (3,0.6) -- (4.5,0.6);
\draw[dotted] (3,-1.3) -- (4.5,-1.3);
\end{tikzpicture}
\end{center}

To a manifold $M$ described by a surgery diagram with knots $\{ L_i\}$ and surgery coefficients $\{ a_i\}$, we can associate the intersection form, or linking matrix, $Q$ defined by 
\begin{equation}
Q_{ij} = \begin{cases}
a_i & \text{ if } i=j\\
\mathrm{lk}(K_i,K_j) & \text{ if } i\neq j
\end{cases}
\end{equation}
where $\mathrm{lk}(K_i,K_j)$ is the linking number of knots $K_i$ and $K_j$.
The intersection form is particularly simple given a plumbing graph with weights $a_i$ at vertex $i$, namely
\begin{equation}
Q_{ij} = \begin{cases}
a_i & \text{ if } i=j,\\
1 & \text{ if } i \text{ and } j \text{ are connected by an edge},\\
0 & \text{else}.
\end{cases}
\end{equation}

The above prescription constructs a Seifert manifold as a framed manifold in `Seifert framing', which differs from the `canonical framing' by $\phi$ units, where
\begin{equation}
\phi_L =  -3\sigma(Q) +  \sum_{i=1}^n  \sum_{j=1}^{m_i} r^i_j \, ,
\end{equation}
with $\sigma(Q_L)$ the signature of the linking matrix $Q$ \cite{Freed:1991wd}:
\begin{equation}
\sigma(Q) =   \text{sign}\left( - \sum_{i=1}^n \frac{q_i}{p_i}\right) + \sum_{i=1}^n \sum_{j=1}^{m_i}\text{sign}  \left( r^i_j\right)\, . 
\end{equation}

Finally, we use the following formula for the determinant of the intersection form describing a Seifert manifold $M((p_1,q_1),\dots,(p_n,q_n))$ \cite{Gadde:2013sca}:
\begin{equation}
\det Q = \left(\sum_i \frac{q_i}{p_i}\right) \prod_j p_j\, .
\end{equation}

\section{Details of $\cT\left(S^3\right)$}\label{app:DetailsTwoDefects}

In this appendix we provide some details on computations used in section \ref{sec:CaseStudyTS3}.

\subsection*{$S^3$ Integral}\label{sec:S3integral}

In our conventions explained in the main body of the paper, the partition function for $\cT_{A_N}\left(S^3\right)$ is given by \eqref{eq:ZS3NTinvN}. For the first non-trivial case $A_1$ we have 
\be
Z^{\suf(2)}_{S^3} = \int_{\gamma_{\ep}}\frac{\rd x}{2\ii}\, S_b(\ep) \frac{S_b(\ep + 2x)S_b(\ep-2x)}{S_b(2x)S_b(-2x)}\, e^{-2\pi \ii x^2} \equiv I(\ep)  \, 
\ee
where $\gamma_{\ep}$ is a suitably deformed contour coming from the localization computation, such that $I(\ep)$ with $\Im(\ep)<0$ is the analytic continuation of $I(\ep)$ in the physical region, where $\Re(\ep)>0, \Im(\ep)>0$ and $\gamma_{\ep}=\ii \R$.
Due to the asymptotics of the $S_b$ functions \eqref{eq:AsymptoticsSb}, the contour needs to close in either the second or fourth quadrant of the complex plane, and as such our integral is only defined for $\text{arg}(\ep) \in [0,\pi)$, i.e. we can think of our integral defined on the half-open disk in $\mathbb{C}\mathbb{P}^1$.

We claim that the integral above evaluates to
\be\label{eq:S3intconjecture}
I(\ep) = \frac{1}{\sqrt{2}} e^{\frac{\pi\ii}{4}}e^{\frac{\pi\ii}{2}\epsilon(Q+\epsilon)} S_b(2\epsilon)\, .
\ee
We will check that the asymptotics and the analytic structures match as functions of $\ep$. The poles are all located at $\text{arg }(\ep) = \pi$, and as such the residues should be interpreted as the coefficient of $\ep^{-1}$ of the Laurent expansion of $I(\ep)$, for $\ep$ near the pole with $\text{arg } (\ep) <\pi$.

We start by considering the asymptotics at $\ep \to \infty$ of the two sides of the equation in the region $\Im(\ep)>0$. For the right-hand side, we immediately find that
\begin{equation}
\frac{1}{\sqrt{2}} e^{\frac{\pi\ii}{4} +\frac{\pi\ii}{2} \ep(Q+\ep) } S_b(2\ep) \sim \frac{1}{\sqrt{2}}\exp\left[\frac{\pi\ii}{12}\left(5+2Q^2-18 B(\ep) \right) \right]\, .
\end{equation}
On the other hand, we assume that we can exchange limit and integral in $I(\ep)$ and thus obtain the following expression for the integrand
\begin{equation}
S_b(\ep)S_b(\ep+2x)S_b(\ep-2x) \sim \exp\left(-4\pi \ii x^2 - \frac{3\pi \ii}{2}B(\ep) \right)\, .
\end{equation}
By closing a contour in the second quadrant, it is then easy to see that
\be
\begin{split}
I(\ep) &\sim -2\ii e^{-\frac{3\pi \ii}{2}B(\ep)}\int_{\R}\rd x\, \sin(2\pi bx)\sin(2\pi b^{-1}x) e^{-6\pi \ii x^2} \\
&= \frac{1}{\sqrt{2}}\exp\left[\frac{\pi\ii}{12}\left(5+2Q^2-18 B(\ep) \right) \right]\, ,
\end{split}
\ee
which matches the behaviour of the left-hand side in the physical region of $\Im(\ep)>0$.

Another immediate check that we can perform is considering the behaviour near $\ep = 0$. Both sides of the equality have a simple pole there, and we can compute the residues, which should match. For the right-hand side, using \eqref{eq:SbResidue}, we immediately find
\be
\text{Res}\left[ \frac{1}{\sqrt{2}} e^{\frac{\pi\ii}{4} +\frac{\pi\ii}{2} \ep(Q+\ep) } S_b(2\ep)\, ; \ep = 0 \right] = \frac{e^{\frac{\pi\ii}{4}}}{4\pi\sqrt{2}}\, .
\ee
On the other side, we have
\be
\text{Res}\left[I(\ep); \ep = 0 \right] = \frac{1}{4\pi \ii} \int_{\gamma_{\ep}}\rd x\, e^{-2\pi \ii x^2} = \frac{e^{\frac{\pi\ii}{4}}}{4\pi \sqrt{2}} \, ,
\ee
thus obtaining a match.

In the same way, performing the integrals on the left-hand side using a computer, we can check consistency of equation \eqref{eq:ZS3NTinvN}. More specifically we check the matching of the asymptotic behaviour for $\ep\to\infty$ in the region $\Im\ep>0$ of both sides of
\be
\int \rd\nu_X(a)\, e^{-\pi\ii \sum_ja_j^2} = \frac{1}{\sqrt{N}}\exp\left[ \frac{\pi\ii}{6}N(N^2-1)\ep^2 + \frac{\pi\ii}{4}(N-1)(1+N\ep^*\ep)\right]\prod_{k=2}^NS_b(k\ep)\, ,
\ee
which is equivalent to \eqref{eq:ZS3NTinvN}. This was done for $N=3, 4$, as for larger $N$ it becomes too challenging from the computational point of view.

\subsection*{Value of $\cS_X(\rho\ep,a')$} \label{sec:Satt^rho}

We would like to show that $\cS_X(\rho\ep,a') = \cS_X(a',\rho\ep)$ is independent of $a'$, and that it is proportional to $\left[\prod_{k=2}^N S_b(k\epsilon) \right]^{-1}$.

To show independence of $\cS_X(\rho\ep,a')$ from $a'$, recall equation \eqref{eq-diffeq}, evaluated at $a'=\rho\ep$\footnote{The operators with the tilde indicate that we are considering loops of length $2\pi b^{-1}$.}:
\begin{align}
H^{(r)}(a) \cdot Z(a,\rho\ep,\epsilon) & = W^{(r)}(\rho\ep) Z(a,\rho\ep,\epsilon)\, , \\ 
\widetilde{H}^{(r)}(a) \cdot Z(a,\rho\ep,\epsilon)  & =\widetilde {W}^{(r)}(\rho\ep) Z(a,\rho\ep,\epsilon)\, .
\end{align}
Furthermore, recall equation \eqref{eq:PartFuncGaugedQM}, so that
\begin{align}
H^{(r)}(a) \cdot 1 & = W^{(r)}(\rho\ep)\, , \\ 
\widetilde{H}^{(r)}(a) \cdot 1  & = \widetilde{W}^{(r)}(\rho\ep)\, .
\end{align}
Since the space of simultaneous eigenfunctions of the difference operators $\left(H^{(r)}(a), \widetilde{H}^{(r)}(a)\right)$ with the respective eigenvalues $\left(W^{(r)}(\rho\ep), \widetilde{W}^{(r)}(\rho\ep)\right)$ is one-dimensional \cite{Beem:2012mb}, this shows that $Z(a,\rho\ep,\epsilon)$ is constant, so independent of $a$.

To find the value of $S_X(\rho\ep,a')$, we shall evaluate $Z(a,a',\epsilon)$ at the symmetric point $a=a'= \rho\ep$ and use the explicit evaluation of $Z(a,a',\epsilon)$ in \cite{Bullimore:2014awa}, equation (3.37)\footnote{After correcting a small error in this formula to restore Weyl invariance in the variable $t$.}.

Let $\sigma\in\cW\cong {\rm Sym}(N)$. Firstly, we find
\begin{align}
K_{Y}(\rho\ep) = K_{Y}(\epsilon\sigma(\rho))
&= \left[S_b(\epsilon)^{N-1} \prod_{1\leq i\neq j \leq N} S_b(\epsilon(1+\rho_{\sigma_i} - \rho_{\sigma_j}))\right]^{-1}\nonumber\\
&= \left[S_b(\epsilon)^{N-1} \prod_{1\leq i\neq j \leq N} S_b(\epsilon(1+\sigma_j - \sigma_i))\right]^{-1}\, .
\end{align}
Then we use mirror symmetry to write $Z(\rho\ep,\rho\ep,\epsilon) = Z(\rho\ep,\rho\ep,\epsilon^*)$.
Using the explicit form in \cite{Bullimore:2014awa}, we see that this contains the following product:
\begin{align}
A_\sigma := \prod_{1\leq i < j \leq N} \frac{S_b(\epsilon(\rho_{\sigma_i} - \rho_{\sigma_j}))}{S_b(\epsilon^* - \epsilon(\rho_{\sigma_i} - \rho_{\sigma_j}))} 
= \prod_{1\leq i < j \leq N}S_b(\epsilon(\sigma_j-\sigma_i))S_b( \epsilon(1+\sigma_i-\sigma_j))\,,
\end{align}
using properties of the $S_b$ function.

Consider now the combination
\begin{align}
A_\sigma K_{Y}(\rho\ep) &=  \frac{1}{S_b(\epsilon)^{N-1}}\frac{\prod_{1\leq i < j \leq N}S_b(\epsilon(\sigma_j-\sigma_i))S_b( \epsilon(1+\sigma_i-\sigma_j))
}{ \prod_{1\leq i\neq j \leq N} S_b(\epsilon(1+\sigma_i - \sigma_j))}\nonumber\\
&= \frac{1}{S_b(\epsilon)^{N-1}} \prod_{1\leq i < j \leq N} \frac{S_b(\epsilon(\sigma_j-\sigma_i))}{S_b(\epsilon(1+\sigma_j - \sigma_i))} \, .
\end{align}
Note that the numerator in the product is never $0$ or infinite. 
The denominator however causes the expression to vanish whenever $1+\sigma_j - \sigma_i = 0$ for some $i<j$, by introducing a factor of $\tfrac{1}{S_b(0)} = 0$.
This happens for every $\sigma\in \text{Sym}(N)$ except $\sigma = {\rm id}$, in which case this product simplifies to give
\begin{equation}
A_{\rm id} K_{Y}(\rho\ep)  = \frac{1}{\prod_{k=2}^N S_b(k\epsilon)}\, .
\end{equation}

Furthermore, when $\ii m = \rho\ep$, and $\sigma ={\rm  id}$, the vortex and anti-vortex partition functions become $1$.
To see this, firstly note that (in our language) $\eta^2 = t$, so that we can rewrite in our language
\begin{equation}
\eta^2 \frac{\mu_{\sigma(i)}}{\mu_{\sigma(j)}} = t^{1+\rho_{\sigma(j)} - \rho_{\sigma(i)}} =  t^{1+\sigma_i-\sigma_j} \,.
\end{equation}
In the Weyl sum over $\sigma\in\text{Sym}(N)$, the only contribution to $Z(a,a',\epsilon)$ is from $\sigma = {\rm id}$, and there is always a pair $(i,j)$ with $1\leq i \leq n , 1\leq j \leq n+1$ with $1\leq n \leq N-1$ such that $1+i - j = 0$.
Therefore, there is always such a pair $(i,j)$ for which
\be 
\left(\eta^2 \tfrac{\mu_i}{\mu_j};q\right)_{k_i^{(n)} - k_j^{(n+1)}} = (1;q)_{k_i^{(n)} - k_j^{(n+1)}} = 0\, ,
\ee 
making any contribution to the vortex partition function with $n\geq 1$ vanish, as claimed.
An isomorphic calculation shows that the anti-vortex partition function becomes 1.

Putting this together, we find the result
\be
\begin{split}
S_X(\rho\ep,\rho\ep) 
= \frac{ e^{-\pi \ii \epsilon N (\rho\ep_N) +2\pi \sum_{j=1}^N (\rho\ep_j)(-\ii \rho\ep_j)} }{\prod_{k=2}^N S_b(k\epsilon)} & =\frac{ e^{-\pi \ii \epsilon^2\left( N \rho_N + 2\sum_{j=1}^N \rho_j \rho_j\right)}}{\prod_{k=2}^N S_b(k\epsilon)} \\
&=\frac{ e^{-\frac{\pi\ii}{6} \epsilon^2 N(N-1)(N-2)}}{\prod_{k=2}^N S_b(k\epsilon)},
\end{split}
\ee
which is the required form. Multiplying by $\sqrt{N}$ gives $\cS_X(\rho\ep,\rho\ep)$.

\bibliographystyle{JHEP}
\bibliography{refined}

\end{document}